\documentclass[11pt,pdf,twoside]{article}
\usepackage{graphicx,color}
\usepackage{epsfig}
\usepackage{color}
\usepackage{cite}

\newcommand{\mrm}[1]{\mathrm{#1}}
\newcommand{\tsc}[1]{\textsc{#1}}
\newcommand{\ttt}[1]{\texttt{#1}}
\newcommand{\pT}[1]{\ensuremath{p_{\perp#1}}}
\newcommand{\TeV}{\,\mbox{Te\kern-0.2exV}}
\newcommand{\GeV}{\,\mbox{Ge\kern-0.2exV}}
\newcommand{\MeV}{\,\mbox{Me\kern-0.2exV}}
\newcommand{\keV}{\,\mbox{ke\kern-0.2exV}}
\newcommand{\eV}{\,\mbox{e\kern-0.2exV}}

\newcommand{\Py}{\tsc{Pythia}}

\begin{document}
\vspace*{-2.0cm}\noindent\begin{minipage}{\textwidth}
\flushright
\end{minipage}\vspace*{0.75cm}
\begin{center}
\Large{\bf Tuning Monte Carlo Generators: The Perugia Tunes}\\[8mm]
{\normalsize {\bf P.~Z.~Skands} (peter.skands@cern.ch)\\
{\sl CERN PH-TH, Case 01600, CH-1211 Geneva 23, Switzerland}}\\[3mm]
\end{center}

\begin{abstract}
We present 9 new tunes of the $p_\perp$-ordered shower and
underlying-event model in \textsc{Pythia} 6.4. 
These ``Perugia'' tunes 
update and supersede the older ``S0'' family. The data sets used to
constrain the models include hadronic $Z^0$ decays at LEP, Tevatron
min-bias data at 630, 1800, and 1960 GeV, Tevatron Drell-Yan data at
1800 and 1960 GeV, and SPS min-bias data at 200, 546, and 900 GeV. 
In addition to the central parameter set, called ``Perugia 0'', 
we introduce a set of 8 related ``Perugia variations'' that attempt to
systematically explore soft, hard, parton density, and colour structure
variations in the theoretical parameters. 
Based on these variations, a best-guess prediction of 
the charged track multiplicity in inelastic, non-diffractive
minimum-bias events at the LHC is made. Note that these tunes can
only be used with \textsc{Pythia} 6, not with \textsc{Pythia} 8. 
\textbf{Note:} this report was updated in March 2011 with a new set
of variations, collectively labelled ``Perugia 2011'', 
that are optimized for matching applications and which also 
take into account some lessons from the early LHC data. In order not
to break the original text, these are
described separately in Appendix \ref{app:2011}. \textbf{Note 2:} 
a subsequent ``Perugia 2012'' update is described in Appendix
\ref{app:2012}.  
\end{abstract}

\tableofcontents

\section{Introduction}

Perturbative calculations of collider observables (see
e.g.~\cite{Skands:2012ts} for an introduction) rely
on two important prerequisites: factorization and
infrared (IR) safety. These are the tools that permit us to relate
theoretical calculations to detector-level measured quantities, up to
corrections of 
known dimensionality, which can then be suppressed (or enhanced) by 
appropriate choices of the dimensionful scales appearing in
the observable and process under study. However, in the context of
the underlying event (UE), say, 
we are faced with the fact that we do not (yet) have formal
factorization theorems for this component --- in fact 
the most naive attempts at factorization 
can easily be shown to fail \cite{Snigirev:2003cq,Korotkikh:2004bz}. 
At the same time, not all collider measurements can be made
insensitive to the UE at a level comparable to the achievable experimental
precision, and hence the extraction of parameters from such
measurements acquires an implicit
dependence on our modelling of the UE. 
Further, when considering observables such as track
multiplicities, hadronization corrections, or even short-distance
quantities if the precision required is very high, we are confronted
with observables which may be experimentally well measured, but which
are explicitly sensitive to infrared physics. 

\paragraph{The Role of Factorization: }
Let us
begin with factorization. When applicable, factorization allows us to
subdivide the calculation of an observable (regardless of whether it
is IR safe or not) into a perturbatively calculable
short-distance part and an approximately universal long-distance part,
the latter of 
which may be modelled and constrained by fits to data. However, in the
context of hadron collisions, the possibilities of multiple
perturbative parton-parton interactions and parton rescattering
processes explicitly go beyond the factorization theorems so far
developed. Part of the problem is that the underlying event may
contain short-distance physics of its own, that can be as hard as, or
even harder than, the bremsstrahlung emissions associated with the
scattering that triggered the event. Hence 
the conceptual 
separation into what we think of as ``hard-scattering'' and ``underlying-event''
components is not necessarily equivalent to a clean separation in
terms of ``short-distance'' and ``long-distance'' physics. 
Indeed, from ISR energies \cite{Akesson:1986iv} through the SPS
\cite{ua1minijets,Alitti:1991rd} to  
the Tevatron
\cite{Abe:1993rv,Abe:1997bp,Abe:1997xk,Abazov:2002mr,Abazov:2009gc}, and also
in photoproduction at HERA
\cite{Gwenlan:2002st}, we see evidence of (perturbative)
``minijets'' in the underlying event, beyond what bremsstrahlung alone
appears to be able to account for. It therefore appears plausible 
that a universal modelling of the underlying event 
must take into account that the hard-scattering and underlying-event
components can involve similar time scales and have a common,
correlated evolution. It is in this spirit
that the concept of ``interleaved evolution'' \cite{Sjostrand:2004ef} 
was developed as the cornerstone of the $\pT{}$-ordered models
\cite{Sjostrand:2004pf,Sjostrand:2004ef} in both 
\textsc{Pythia}~6~\cite {Sjostrand:2006za} and, more recently,
\textsc{Pythia}~8~\cite{Sjostrand:2007gs}, the latter of which now also
incorporates a model of parton rescattering \cite{Corke:2009tk}. 

\paragraph{The Role of Infrared Safety:}
The second tool, infrared safety\footnote{By ``infrared'' we here mean any 
non-UV limit, without regard to whether it is collinear or soft.}, provides us
with a class of observables which are insensitive to the details of
the long-distance physics. This works up to corrections of order the
long-distance scale divided by the short-distance scale to some
(observable-dependent) power, typically
\begin{equation}
\mbox{IR Safe Corrections~~~$\propto$~~~} \frac{Q_{\mathrm{IR}}^2}{Q_{\mathrm{UV}}^2} 
\end{equation}
where $Q_\mathrm{UV}$ denotes
a generic hard scale in the problem, and $Q_\mrm{IR} \sim
\Lambda_\mrm{QCD} \sim \mathcal{O}(\mrm{1\ GeV})$. Of course, in 
minimum-bias, we typically have $Q_{\mathrm{UV}}^2 \sim Q_{\mathrm{IR}}^2$, wherefore
\emph{all} observables depend significantly on the IR
physics (or in other words, when IR physics is all there is, then any
observable, no matter how carefully defined, depends on it). 

Even when a high scale is present, as in resonance decays, jet
fragmentation, or underlying-event-type studies, infrared safety only
guarantees us that infrared corrections are small, not that they are zero. 
Thus, ultimately, we run into 
a precision barrier even for IR safe observables, 
which only a reliable understanding of the
long-distance physics itself can address. 

Finally, there are the non-infrared-safe observables.
Instead of the suppressed corrections above, 
such observables contain logarithms
\begin{equation}
\mbox{IR Sensitive Corrections~~~$\propto$~~~} 
\alpha_s^n\log^{m}\left(\frac{Q_{\mathrm{UV}}^2}{Q_{\mathrm{IR}}^2}\right)~~~,~~~m
\le 2n ~~~,
\end{equation}
 which grow
increasingly large as $Q_{\mathrm{IR}}/Q_{\mathrm{UV}}\to 0$.   
As an example, consider such a fundamental quantity as particle 
multiplicities; in the absence of nontrivial infrared
effects, the number of partons that would be
mapped to hadrons in a na\"ive local-parton-hadron-duality
\cite{Azimov:1984np} picture would tend logarithmically to infinity 
as the IR cutoff is lowered. Similarly, the distinction between
a charged and a neutral pion only occurs in the very last phase of
hadronization, and hence observables that only include charged tracks
are always IR sensitive. 

\paragraph{Minimum-Bias and the Underlying Event:}
Minimum-bias (MB) and Underlying-Event (UE) physics can therefore be
perceived of as offering  
an ideal lab for \emph{studying} non-factorized and nonperturbative
phenomena, with the added benefit of having
access to the highest possible  
statistics in the case of min-bias. In this context 
there is no strong preference for IR safe over IR sensitive
observables; they merely represent two different lenses through which
we can view the infrared physics, each revealing different
aspects. By far the most important point is that it is in
their \emph{combination} that we achieve a sort of stereo vision, in
which infrared safe observables measuring the overall energy flow
are simply the slightly averaged progenitors of the spectra and correlations 
that appear at the level of individual particles.  
A systematic programme of such 
studies can give crucial tests of our ability to model and understand
these ubiquitous components, and the resulting improved physics models 
 can then be fed back into the modelling of 
high-$\pT{}$ physics.  

Starting from early notions such as ``KNO scaling'' of
multiplicity distributions \cite{Koba:1972ng}, a large number of 
theoretical and experimental
investigations have been brought to bear on 
what the physics of a generic, unbiased sample of hadron collisions
looks like (for a recent review, see, e.g.,
\cite{GrosseOetringhaus:2009kz} and references therein).  
However, in step with the gradual shift in focus
over the last two decades, towards 
higher-$\pT{}$ (``maximum-bias'') physics, the field of 
QCD entered a golden age of perturbative calculations and infrared
safety, during which 
time the unsafe ``soft'' physics became viewed increasingly
as a non-perturbative quagmire, into the depths of 
which ventured only fools and old men. 

From the perspective of the author's generation, it was chiefly
with a comprehensive set of measurements carried out
by Rick Field using the CDF detector at the Tevatron
\cite{Field:2000dy,Affolder:2001xt,Field:2002vt,Acosta:2004wqa,Field:2005qt,Kar:2009kc},  
that this perception began to change back towards one of 
a definable region of particle production
that can be subjected to rigorous scrutiny in a largely 
model-independent way, and an ambitious programme of such measurements
is now being drawn up for the LHC experiments. 
In other words, a well-defined experimental laboratory has
been prepared, and is now ready for the testing of theoretical models. 

Simultaneously with the LHC efforts, it is important to remember that 
interesting connections are also being explored towards other,
related, fields, such as cosmic ray fragmentation (related to forward
fragmentation at the LHC) and heavy-ion physics (related to collective
phenomena in hadron-hadron interactions). A nice example of this
interplay is given, for instance, by the \textsc{Epos} model~\cite{Werner:2008zza}, 
which originated in the heavy-ion community, but uses a parton-based model
as input and whose properties in the  context of
ultra-high-energy cosmic ray fragmentation are currently being
explored~\cite{Apel:2009sv,Pierog:2009zt}. 
Also methods from the field of numerical optimization 
are being applied to Monte Carlo tuning (cf., e.g., 
the Professor \cite{Buckley:2009bj} and Profit \cite{Bacchetta:2010hh}
frameworks), and there are tempting connections back to perturbative
QCD. Along the latter vein, we believe that by bringing the logarithmic 
accuracy of perturbative parton shower calculations 
under better control, there would be less room for playing out
ambiguities in the non-perturbative physics against ambiguities on the shower
side, and hence the genuine soft physics could also be revealed 
more clearly. This is one of the main motivations behind the
\textsc{Vincia}  project \cite{Giele:2007di,Giele:2011cb}.

For the present, as part of the effort to prepare for the LHC era and 
spur more interplay between theorists and
experimentalists, we shall here report on a new set of tunes 
of the $\pT{}$-ordered \textsc{Pythia} framework, 
which update and supersede the older ``S0''
family \cite{Sandhoff:2005jh,Skands:2007zg,Wicke:2008iz,Skands:2007zz}.
We have focused in particular on the 
scaling from lower energies towards the LHC (see also
\cite{Alekhin:2005dx,Albrow:2006rt,Buttar:2008jx,Bartalini:2008zz}) 
and on attempting to provide
at least some form of theoretical uncertainty estimates, represented
by a small number of alternate parameter sets that systematically
explore variations in some of the main tune parameters. 
The full set of new tunes have been made available 
starting from \textsc{Pythia} version 6.4.23 (though some have been available
longer; see the \textsc{Pythia} update notes
\cite{updatenotes} for details). 

This concludes a several-year long
effort to present the community with an optimized set of parameters
that can be used as default settings for 
the so-called ``new'' interleaved shower and underlying-event model in
\textsc{Pythia} 6. The author's intention is to now move
fully to the development of \textsc{Pythia} 8. We note that the 
Perugia tunes can unfortunately not be used directly in
\textsc{Pythia} 8, 
since it uses slightly different parton-shower and
colour-reconnection models. A separate set of tunes 
for \textsc{Pythia} 8 are therefore under development, with
several already included in the current version 8.1.42 of that
generator. 

We also present a few distributions that carry
interesting information about the underlying
physics, updating and complementing those contained in
\cite{Skands:2007zz,Skands:2009zm}. 
For brevity, this text only includes a
representative selection, with more results available on the web
\cite{lhplots,Karneyeu:2013aha}.  

The main point is that, while any plot of an infrared sensitive
quantity represents a
complicated cocktail of physics effects, such that any sufficiently
general model presumably could be tuned to give an acceptable
description observable by observable, it is very difficult 
to simultaneously describe the entire set. The real game 
is therefore not to study one distribution in detail, for which a
simple fit would in principle suffice, but to study the
degree of simultaneous agreement or disagreement over many, mutually
complementary, distributions. 

\section{Procedure}

\subsection{Manual vs Automated Tuning}
Although Monte Carlo models may appear to have a
bewildering array of independently adjustable parameters, it is worth
keeping at the front of one's mind that most  of these parameters only control
relatively small (exclusive) details of the event generation. The majority of the
(inclusive) physics is determined by only a few, very important ones, 
such as, e.g., the value of the strong coupling, in the perturbative
domain, and the form of the fragmentation function for massless
partons, in the non-perturbative one. 

\paragraph{Manual Tuning:}
Armed with a good understanding of the underlying model, and using only
the generator itself as a tool, a generator expert would
therefore normally take a highly factorized approach to constraining
the parameters, 
first constraining the perturbative ones and thereafter the
non-perturbative ones, each ordered in a measure of their relative
significance to the overall modelling. This factorization, and carefully chosen
experimental distributions corresponding to each step,  allows the expert
to concentrate on just a few parameters and distributions 
at a time, reducing the full parameter space to manageable-sized
chunks. Still, each step will often involve more than one single
parameter, and non-factorizable 
corrections still imply that changes made in 
subsequent steps can change the agreement obtained in previous ones 
by a non-negligible amount, requiring additional iterations from
the beginning to properly tune the entire generator framework. 

Due to the large and varied data sets available, and the high statistics 
required to properly explore tails of distributions,  mounting a proper
tuning effort can therefore be 
quite intensive --- 
often involving testing the 
generator against the measured data 
for thousands of observables, collider energies, and
generator settings. 
Although we have not kept a detailed record, an
approximate guess is that the generator runs involved in producing the
particular tunes reported on here consumed on the order of 1.000.000 CPU
hours, to which can be added an unknown number of man-hours. While
some of these man-hours were undoubtedly productive, teaching the
author more about his model and resulting in some of the conclusions
reported on in this paper, most of them were merely 
tedious, while still disruptive enough to prevent getting much other work
done. 

The main steps followed in the tuning procedure for the Perugia
tunes are described in more detail in section \ref{sec:steps} below. 

\paragraph{Automated Tuning:}
As mentioned in the introduction, recent years have seen the emergence
of automated tools that attempt to reduce the amount of both computer
and manpower required. The number of machine hours can, for instance, 
be substantially reduced by making full generator runs only for a
limited set of parameter points, and then interpolating between
these  to obtain approximations to what the true generator result
would have been for any intermediate parameter point. In the Professor 
tool  \cite{Buckley:2009vk,Buckley:2009bj}, which we rely on for our
LEP tuning here, this optimization technique is used heavily, so that
after an initial (intensive) initialization period, approximate
generator results for \emph{any} set of generator parameters within
the sampled space can be obtained without any need of
further generator runs. Taken by itself, such optimization techniques 
 could in principle also be used as an aid to manual tuning, but Professor,
 and other tools such as Profit \cite{Bacchetta:2010hh}, attempt to
  go a step
 further. 

Automating the human expert input is of course more difficult (so the
experts believe). What parameters to include, in what order, and
which ranges for them to consider ``physical''? 
What distributions to include, over which regions, how to treat
correlations between them, and how to judge the relative
importance, for instance, between getting the right average of an 
observable versus getting the right asymptotic slope? 
In the tools currently on the market,
these questions are addressed by a combination of input solicited from
the generator authors (e.g., which parameters and ranges to consider,
which observables constitute a complete set, etc)
and the elaborate construction of non-trivial weighting
functions that determine how much weight is assigned to each 
individual bin and to each distribution. The field is still
burgeoning, however, and future sophistications are to be
expected. Nevertheless, at this point the overall quality of the tunes
obtained with automated methods appear to  
the author to at least be competitive with the manual ones.

\subsection{Sequence of Tuning Steps \label{sec:steps}}

We have tuned the Monte Carlo in five consecutive steps (abbreviations
which we use often below are highlighted in boldface):
\begin{enumerate}
\item Final-State Radiation ({\bf FSR}) and Hadronization ({\bf HAD}): 
 using LEP data \cite{Ackerstaff:1998hz,Amsler:2008zzb}. 
For most of the Perugia tunes, 
 we take the LEP parameters given by the
 Professor collaboration \cite{Buckley:2009vk,Buckley:2009bj}. This
 improves several event shapes and fragmentation spectra as compared
 to the default settings. For hadronic yields, especially $\phi^0$ was
 previously wrong by more than a factor of 2, and $\eta$ and $\eta'$
 yields have likewise been improved. For a ``HARD'' and a
 ``SOFT'' tune variation, we deliberately change 
 the re-normalization scale for FSR 
 slightly away from the central Professor value. Also, since 
 the Professor parameters were originally optimized for the $Q^2$-ordered parton
 shower in \textsc{Pythia}, the newest (2010) Perugia tune goes 
 slightly further, by changing the other fragmentation parameters (by
 order of 5-10\% relative to their Professor values) in an attempt to 
 improve the description of high-$z$ fragmentation 
 and strangeness yields reported at LEP \cite{Ackerstaff:1998hz,Amsler:2008zzb} 
and at RHIC \cite{Adams:2006nd,Abelev:2006cs}, relative to the
Professor \pT{}-ordered tuning. The amount of ISR jet
broadening (i.e., FSR off ISR) in hadron collisions has also been increased 
in Perugia 2010, relative to Perugia 0, in an attempt to improve 
hadron collider jet shapes and rates \cite{Acosta:2005ix,Banfi:2010xy}. 
\item Initial-State Radiation ({\bf ISR}) and Primordial $k_T$: using 
  the Drell-Yan $\pT{}$ spectrum at 1800 and
  1960 GeV, as measured by CDF \cite{Affolder:1999jh} and D\O\ 
  \cite{:2007nt}, respectively. Note that we treat the data as fully corrected
  for photon bremsstrahlung effects in this case, i.e., we compare the
  measured points to the Monte Carlo distribution of the ``original $Z$
  boson''. We are aware that this is not a physically meaningful observable 
  definition, but believe it is the closest we can come to the definition
  actually used for the data points in both
 the CDF and D\O\ studies. See \cite{hesketh} for a more detailed
 discussion of this issue. Again, we deliberately change
  the renormalization scale for ISR away from its best fit value for
  the HARD and SOFT variations, by about a factor of 2 in either
  direction, which does not appear to lead to serious conflict with
  the data (see distributions below). 
\item Underlying Event ({\bf UE}), Beam Remnants ({\bf BR}), and Colour
  Reconnections ({\bf CR}): using 
  $N_\mrm{ch}$ \cite{Acosta:2001rm,moggi}, 
$dN_\mrm{ch}/d\pT{}$ \cite{Abe:1988yu,Aaltonen:2009ne}, and
  $\left<\pT{}\right>(N_\mrm{ch})$ \cite{Aaltonen:2009ne} in min-bias events  
at 1800 and 1960 GeV, as measured by CDF. Note that the $N_\mrm{ch}$ spectrum 
extending down to zero $\pT{}$ measured by the E735 Collaboration at
1800 GeV \cite{Alexopoulos:1998bi} was left out of the tuning, 
since we were not able to consolidate this measurement with the rest
of the data. We do not know
whether this is due to intrinsic limitations in the modelling (e.g., mismodeling of 
the low-$\pT{}$ and/or high-$\eta$ regions, which are included in the
E735 result but not in the CDF one) or to a
misinterpretation on our part of the measured observable. Note, however,
that the E735 collaboration itself remarks
\cite{Alexopoulos:1998bi} that its results are
inconsistent with those reported by UA5
\cite{Alner:1987wb,Ansorge:1988kn} over the entire range of
energies where both experiments have data.  So
far, the early LHC results at 900 GeV appear to be consistent with UA5, 
within the limited $\eta$ regions accessible to the experiments
\cite{Collaboration:2009dt,Collaboration:2010xs,Aad:2010rd},
but it remains important to check the 
high-multiplicity tail in detail, in as large a phase space region as
possible. We also note that there are
some discrepancies between the CDF Run-1 \cite{Acosta:2001rm} and Run-2
\cite{moggi} measurements at very low multiplicities, presumably due
to ambiguities in the procedure used to correct for diffraction. We
have here focused on the high-multiplicity tail, which is consistent
between the two. Hopefully, this question can also be addressed by 
 comparisons to early low-energy LHC data. Although the 4 main LHC
 experiments are not ideal for  diffractive studies and cannot
 identify forward protons, it is likely that
 a good sensitivity can still be obtained by requiring events with
 large rapidity gaps, where the gap definition would essentially be 
 limited by the noise levels achievable in the electromagnetic
 calorimeters. 
\item Energy Scaling: using $N_\mrm{ch}$ in min-bias events 
at 200, 546, and 900 GeV, as measured by
UA5~\cite{Alner:1987wb,Ansorge:1988kn}, and at 630 and 1800 GeV, 
as measured by CDF~\cite{Acosta:2001rm}.
\item The last two steps were 
iterated a few times.
\end{enumerate}
\paragraph{Remarks on Jet Universality:}
Note that the clean separation between the first and second points in
the list above assumes jet universality, i.e., that a $Z^0$, for instance,
fragments in the same way at a hadron collider as it did at LEP. This
is not an unreasonable first assumption  \cite{Field:1976ve}, but since the infrared
environment in hadron collisions is characterized by a different
(hadronic) initial-state vacuum, by a larger final-state gluon
component, and also by simply having a lot more
colour flowing around in general, it is still important to check to
what precision it holds explicitly, e.g., by measuring multiplicity
and $\pT{}$ spectra of 
identified particles, particle-particle correlations, and particle
production ratios (e.g.,
strange to unstrange, vector to pseudoscalar, baryon to meson, etc.) 
\emph{in situ} at hadron colliders. We
therefore very much encourage the LHC experiments not to blindly rely
on the constraints implied by LEP, but to construct and publish their own
full-fledged sets of fragmentation constraints using identified
particles. This is the only way to verify explicitly to what extent the models
extrapolate correctly to the LHC environment, and gives the
possibility to highlight and address any discrepancies. 

\paragraph{Remarks on Diffraction:}
Note also that the modelling of diffraction in
\textsc{Pythia 6} lacks a dedicated modelling of diffractive jet
production, and hence we include neither elastic nor diffractive Monte
Carlo events in any of our comparisons. This affects 
the validity of the modelling for the first few
bins in multiplicity. Due also to the discrepancy noted above
between the two CDF measurements in this region
\cite{Acosta:2001rm,moggi}, 
we therefore assigned less importance to these bins when doing
the tunes\footnote{To ensure an apples-to-apples comparison for the
low-multiplicity bins between these models and present measurements, 
one must take care
 to include any relevant diffractive components using a 
 (separate) state-of-the-art modelling of diffraction.}. We emphasize
that widespread use of ill-defined terminologies such as ``Non-Single
Diffractive'' (NSD) events without an accompanying
definition of what is meant by that terminology at the level of physical
observables contributes to the ambiguities surrounding diffractive
corrections in present data sets. Since different diffraction models
produce different 
spectra at the observable level, an intrinsic ambiguity is introduced
which was not present in the raw data. We strongly encourage future
measurements if not to avoid such terminologies entirely then to at least 
\emph{also} make data  available in a form which is defined only in terms of
physical observables, i.e., using explicit cuts, weighting functions, and/or trigger
conditions to emphasize the role of one component over another.

\paragraph{Remarks on Observables:}
Finally, note that we did not include any explicit ``underlying-event''
observables in the tuning. 
Instead, we rely on the large-multiplicity tail of
minimum-bias events to mimic the underlying event. 
A similar procedure was followed for the older ``S0'' tune
\cite{Sandhoff:2005jh,Skands:2007zg}, which 
gave a very good simultaneous description of 
underlying-event physics at the Tevatron\footnote{Note: when
extrapolating to lower energies, the alternative scaling represented
by ``S0A'' appears to be preferred over the default scaling used in
``S0''.}.  Conversely, Rick Field's ``Tune A'' 
\cite{tunea,Albrow:2006rt} gave a 
good simultaneous description of minimum-bias data, despite only
having been tuned on underlying-event data. Tuning to one and
predicting the other is therefore not only feasible but simultaneously
a powerful cross-check on the universality properties of the
modelling. 

Additional important quantities to consider for further model tests
and tuning would be event shapes at hadron colliders 
\cite{Banfi:2004nk,Banfi:2010xy}, 
observables involving explicit jet reconstruction --- 
 including so-called ``charged jets'' \cite{Affolder:2001xt} 
(a jet algorithm run on a set of charged tracks, omitting neutral energy), 
which will have fluctuations in the charged-to-neutral 
ratio overlaid on the energy flow and therefore will be more
IR than full jets, but still less so than individual
particles, and ``EM jets'' (a jet algorithm run on a set of charged
tracks plus photons), which basically adds back the $\pi^0$ component
to the charged jets and hence is less IR sensitive than pure charged jets 
while still remaining free of the noisy environment of
hadron calorimeters --- explicit
underlying-event, fragmentation, and jet structure (e.g., jet mass, jet
shape, jet-jet separation) observables in events with jets
\cite{Abe:1993rv,Field:2000dy,Affolder:2001xt,Field:2002vt,Acosta:2004wqa,D0jets,Acosta:2005ix,Field:2005sa,Field:2005yw,Field:2005qt,Aaltonen:2008yn,Cacciari:2009dp,Aaltonen:2010rm},   
photon + jet(s) events (including the important $\gamma$ +
3-jet signature for double-parton interactions
\cite{Abe:1997xk,Abazov:2009gc}), Drell-Yan events
\cite{Field:2000dy,Kar:2008zza,Aaltonen:2010rm}, and
observables sensitive to the 
initial-state shower evolution in DIS (see, e.g.,
\cite{Alekhin:2005dx,Carli:2010cg}). 

As mentioned above, it is also important that fragmentation models
tuned at LEP be tested \emph{in situ} at hadron colliders. 
To this effect, single-particle
multiplicities and momentum spectra for identified particles such as
$K^0_S$, vector mesons, protons, and hyperons (in units of GeV and/or 
normalized to a global measure of transverse energy, such as, e.g., the $\pT{}$
of a jet when the event is clustered back to a dijet topology) 
are the first order of business, 
and particle-particle
correlations the second (e.g., how charge, strangeness,
baryon number, etc., are compensated
as a function of a distance measure and how the correlation 
strength of 
particle production varies over the measured phase space region). Again, 
these should be considered at the same time as less
infrared sensitive variables measuring the overall energy flow. 
We expect a programme of such measurements to gradually develop as it
becomes possible to extract more detailed information from the LHC data and 
note that some such observables, from earlier experiments, 
have already been included, e.g., in the Rivet
framework, see \cite{Buckley:2009bj}, most notably underlying-event
observables from the Tevatron, but also recently some 
fragmentation spectra 
from RHIC  \cite{Adams:2006nd,Abelev:2006cs}. See also the  
underlying-event sections in the HERA-and-the-LHC
\cite{Alekhin:2005dx}, Tevatron-for-LHC \cite{Albrow:2006rt}, and Les
Houches write-ups \cite{Buttar:2008jx}. A complementary and useful
guide to tuning has been produced by the ATLAS collaboration in the
context of their MC09 tuning efforts \cite{atlasmc09}. 

\section{Main Features of the Perugia Tunes}

Let us first describe the overall features common to all the Perugia
tunes, divided into the same main steps as  in the outline of the
tuning procedure given in the preceding section: {\sl 1)} final-state
radiation and hadronization, {\sl 2)} initial-state radiation and
primordial $k_T$, {\sl 3)} underlying event, beam remnants, and colour
reconnections, and {\sl 4)} energy scaling. 
Each step will be accompanied by plots to illustrate salient points and 
by a summary table in appendix \ref{sec:tables} giving the
Perugia parameters relevant to that step, as compared to the older Tune S0A-Pro,
which serves as our reference. We shall then turn to 
the properties of the individual tunes in the following section, and
finally to extrapolations to the LHC in the last section.

\subsection{Final-State Radiation and Hadronization (Table \ref{tab:fsrhad})}
As mentioned above, we have taken the LEP tune obtained
by the Professor group  \cite{Buckley:2009vk,Buckley:2009bj} 
as our starting point for the FSR and HAD
parameters for the Perugia tunes. Since we did not perform this part of the tuning
ourselves, we treat these parameters almost as fixed inputs, and 
only a very crude first attempt at varying them was
originally made for the Perugia HARD and SOFT variations. 
This is reflected in the relatively
small differences between the FSR and HAD parameters listed in table
\ref{tab:fsrhad}, compared to S0A-Pro which uses the
original Professor parameters. (E.g., most of the tunes use the same 
parameters for the longitudinal fragmentation function applied in the 
string hadronization process, including the same Lund functions
\cite{LundFrag} for light quarks and Bowler 
functions \cite{Bowler} for heavy quarks.) With the most recent
Perugia 2010 tune, an effort was made to manually improve jet shapes,
strangeness yields, and high-$z$ fragmentation, which is the reason
several of the hadronization parameters differ in this tune as well as
in its sister tune Perugia K.
A more systematic exploration of
variations in the fragmentation parameters is certainly a point to
return to in the future, especially in the light of the new
identified-particle spectra and jet shape data 
that will hopefully soon be available from
the LHC experiments. For the present, we have focused on the 
the uncertainties in the hadron-collider-specific parameters, as
follows. 

\subsection{Initial-State Radiation and Primordial $\mathbf{k_T}$  (Table \ref{tab:isrkt})}

\paragraph{Evolution Variable, Kinematics, and Renormalization Scale:}
One of the most significant changes when going from the old
(virtuality-ordered) to the new (\pT{}-ordered)
ISR/FSR model concerns the Drell-Yan \pT{} spectrum. 
In the old model, when an originally massless 
ISR parton evolves to become a jet with a timelike invariant mass, 
then that original parton is pushed off its mass shell by
reducing its momentum components. In particular the transverse
momentum components are reduced, and hence each final-state emission
off an ISR parton effectively removes $\pT{}$ from that parton, and by
momentum conservation also from the recoiling Drell-Yan pair. Via this
mechanism, the \pT{} distribution generated for the Drell-Yan
pair 
is shifted towards lower values than what was initially produced. 

\begin{figure}[t!]
\vspace*{15mm}\begin{center}\hspace*{-2mm}
\scalebox{1.3}{
\begin{tabular}{ll}
\rotatebox{90}{\sl CDF}\hspace*{-10mm}& 
\hspace*{7mm}\includegraphics*[scale=0.26]{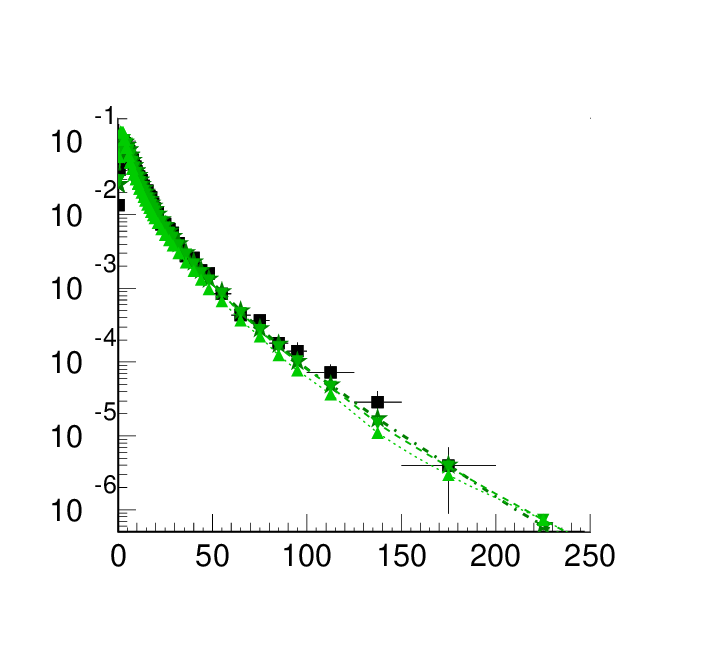} \\[-5.0cm]
\hspace*{-15mm}& \includegraphics*[scale=0.34]{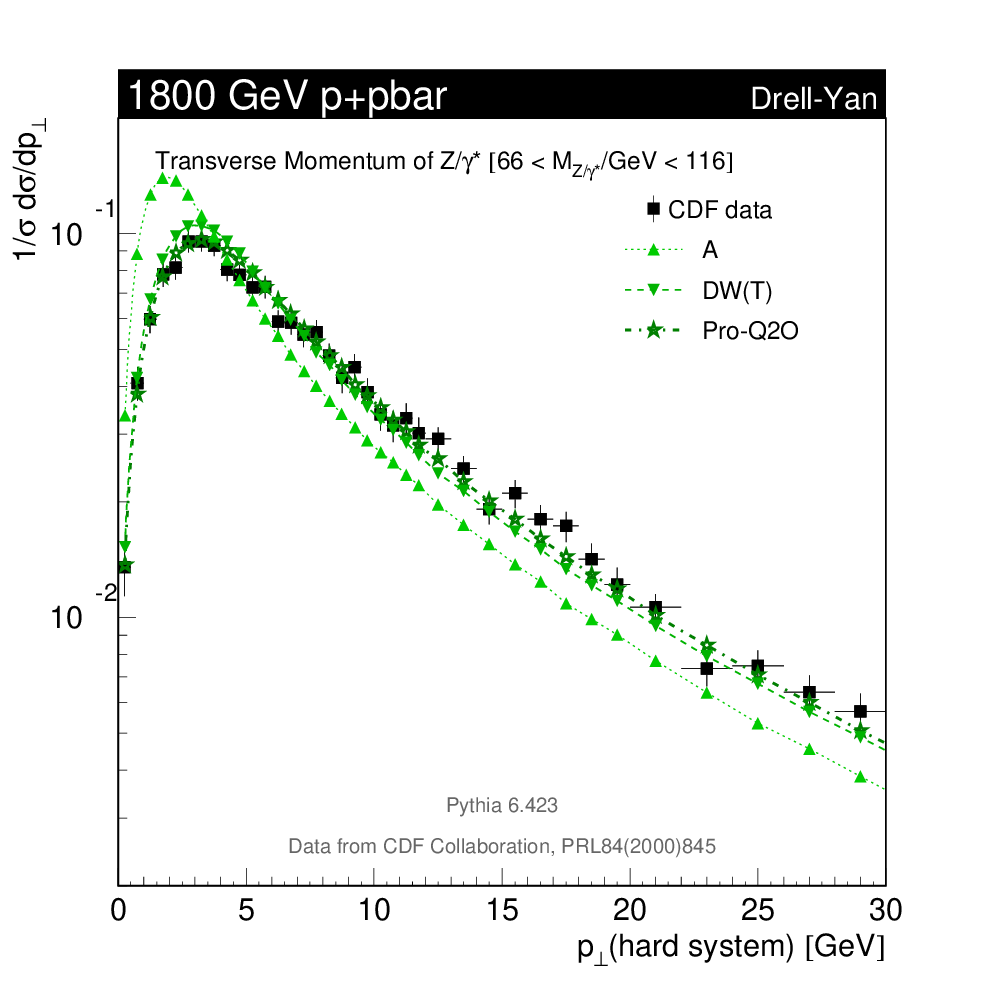}\hspace*{-5mm} \\[1.6cm]
\rotatebox{90}{\sl D\O}\hspace*{-8mm}& 
\hspace*{7mm}\includegraphics*[scale=0.26]{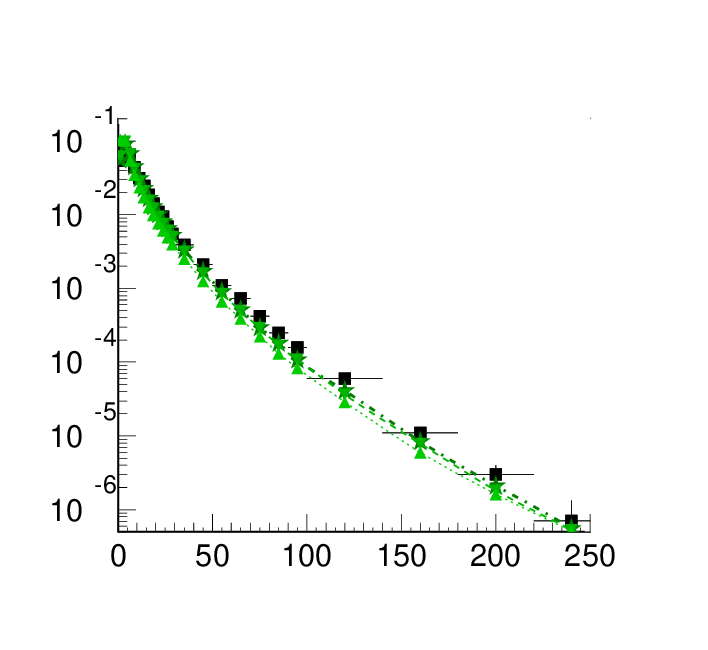} \\[-5.0cm]
\hspace*{-15mm}& \includegraphics*[scale=0.34]{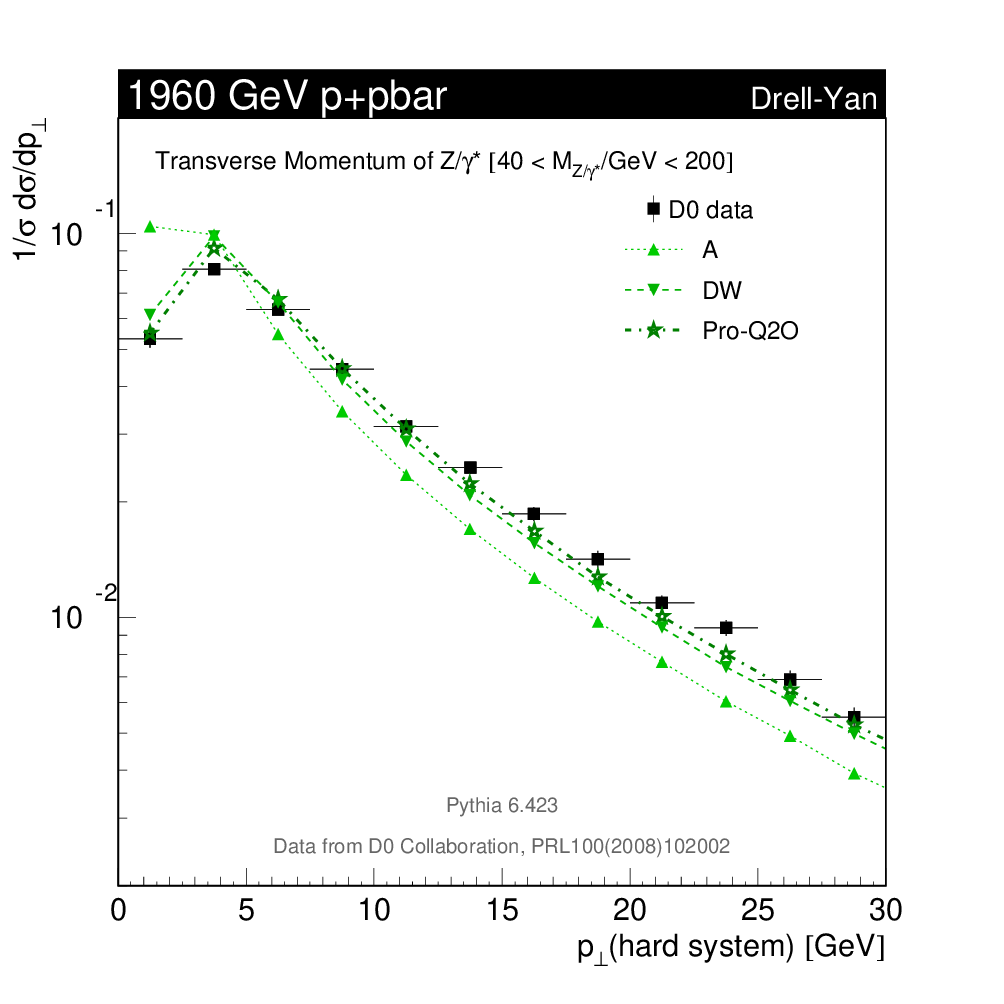}\hspace*{-5mm} 
\end{tabular}}
\scalebox{1.3}{
\begin{tabular}{ll}
\rotatebox{90}{\sl CDF}\hspace*{-10mm}& 
\hspace*{7mm}\includegraphics*[scale=0.26]{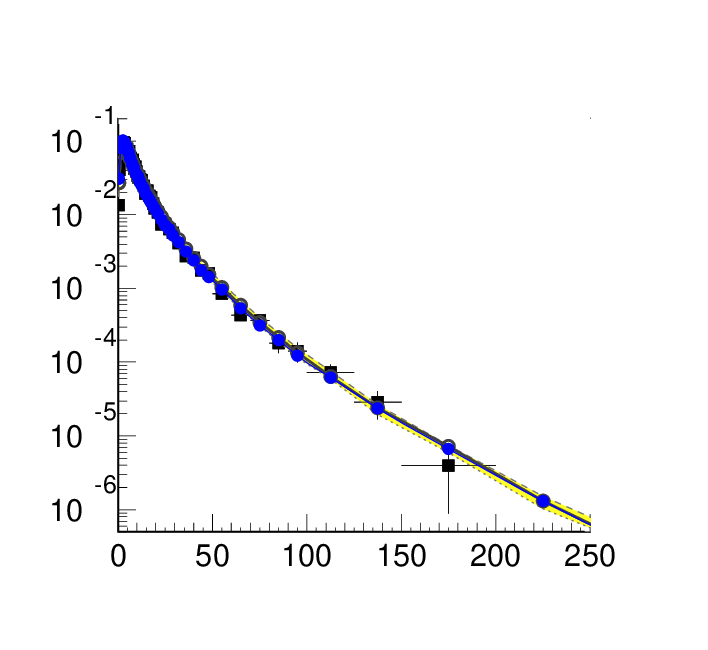} \\[-4.9cm]
\hspace*{-15mm}& \includegraphics*[scale=0.34]{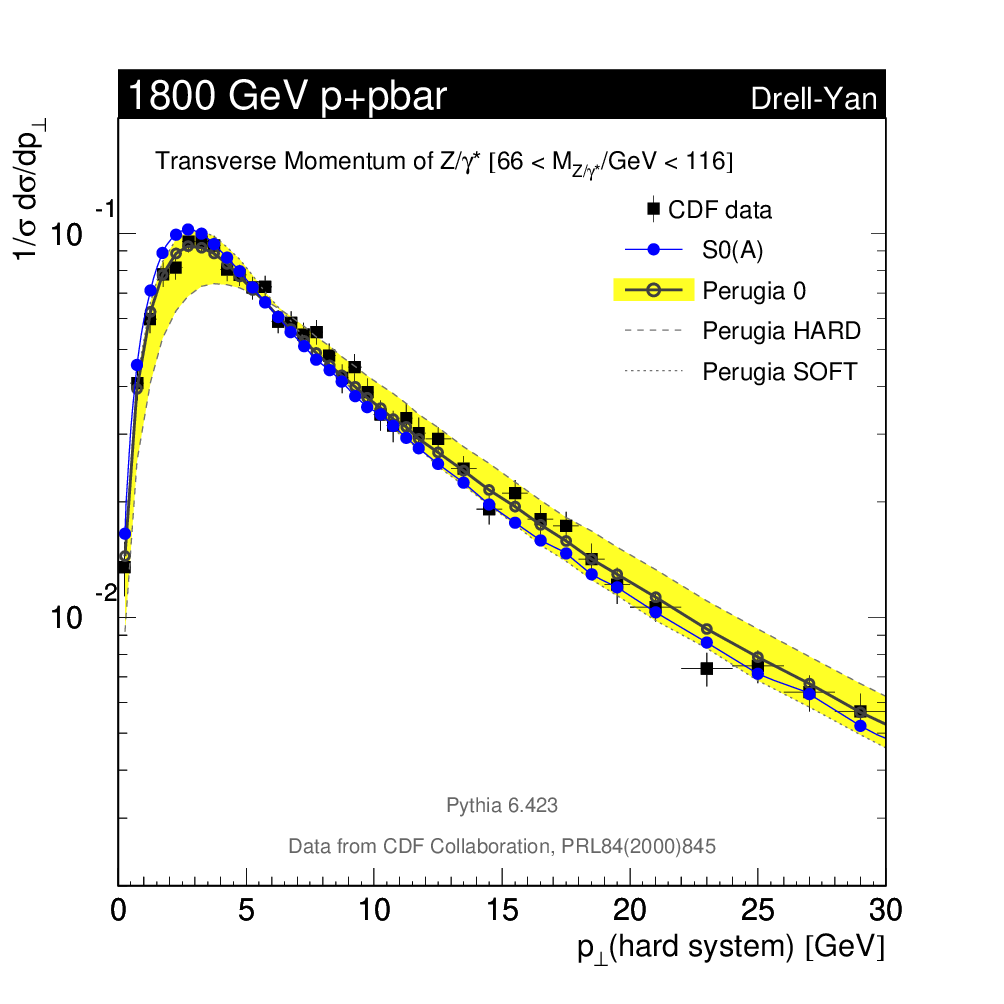}\hspace*{-5mm} \\[1.6cm]
\rotatebox{90}{\sl D\O}\hspace*{-8mm}&
\hspace*{7mm}\includegraphics*[scale=0.26]{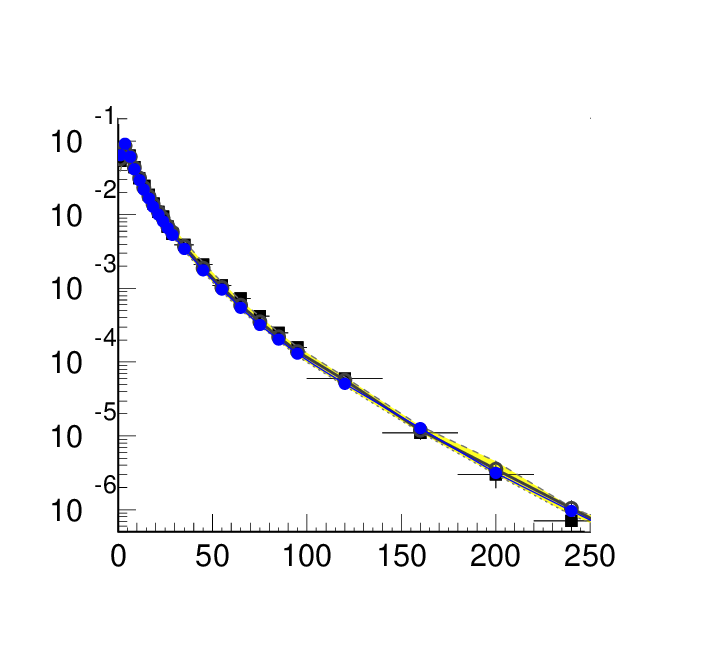} \\[-4.9cm]
\hspace*{-15mm}& \includegraphics*[scale=0.34]{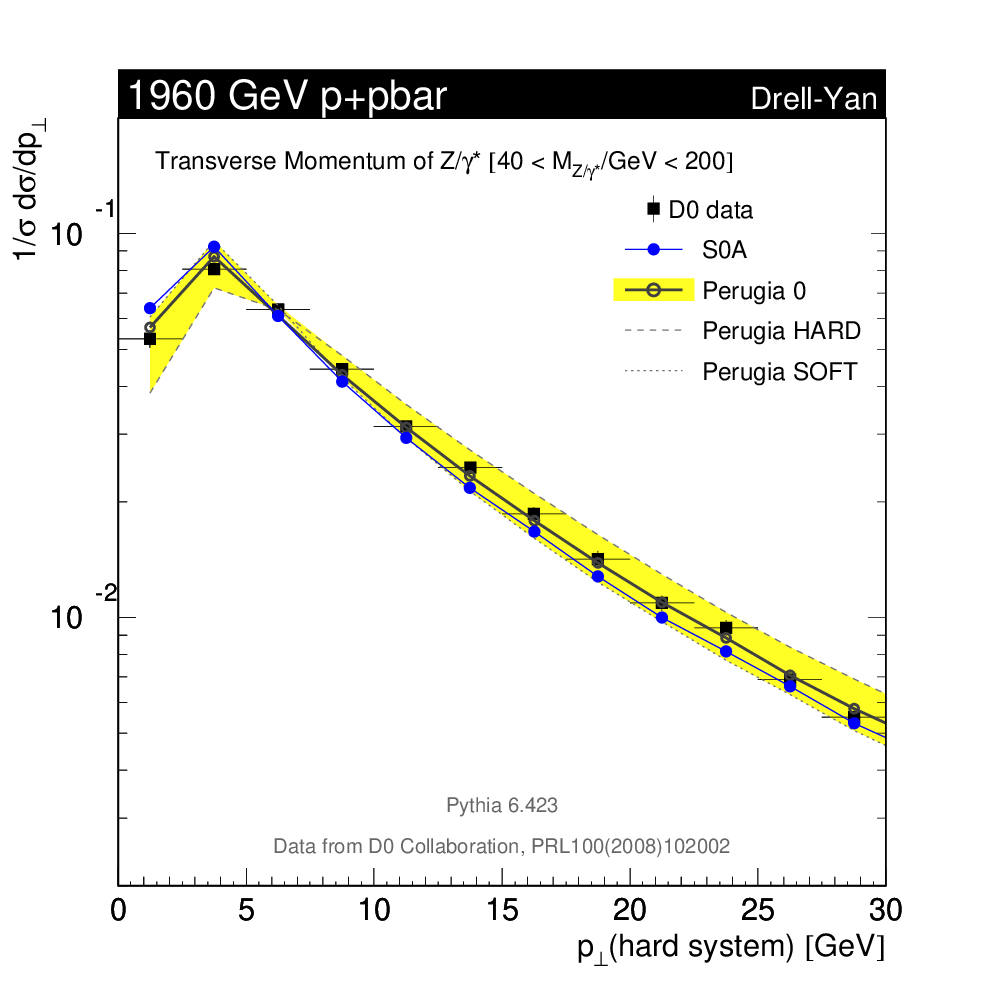}\hspace*{-5mm} 
\end{tabular}}\hspace*{-5mm}
\vspace*{-3mm}
\caption{\small Comparisons to the CDF and D\O\ measurements of
 the $\pT{}$ of Drell-Yan pairs \cite{Affolder:1999jh,:2007nt}. Insets
 show the high-$\pT{}$ tails.
 {\sl Left:} virtuality-ordered showers.  {\sl
   Right:} $\pT{}$-ordered showers. 
See
  \cite{lhplots} for 
  other tunes and collider energies. 
\label{fig:tevatronDY}}
\end{center}
\end{figure}

Compared to data, this appears to effectively cause any tune of
the old \textsc{Pythia} framework with default ISR settings --- such as
Tune A or the ATLAS DC2/``Rome'' tune --- 
to predict a too narrow spectrum for the Drell-Yan $\pT{}$
distribution, as illustrated by the comparison of Tune A to CDF and
D\O\ data in fig.~\ref{fig:tevatronDY}
(left column). (The inset shows the high-\pT{}\ tail which in all
cases is matched to $Z+$jet matrix elements, the default in
\textsc{Pythia} for both the virtuality- and \pT{}-ordered 
shower models.)
We note that a recent theoretical study \cite{Nagy:2009vg} using
virtuality-ordering with a different kinematics map did not find this
problem, consistent with our suspicion that it is not the virtuality ordering
\emph{per se} which results in the narrow shape, but the specific
recoil kinematics of FSR off FSR in the old shower model. 

To re-establish agreement with the measured spectrum without changing
the recoil kinematics, the total amount of ISR in the old model 
had to be increased. This can be accomplished, e.g., by 
choosing very low values of the
renormalization scale (and hence large $\alpha_s$ values) for ISR, as
illustrated by tunes DW-Pro and Pro-Q2O in fig.~\ref{fig:tevatronDY}
(left column). To summarize, the $\alpha_s$ choices corresponding to
each of the three tunes of the old shower shown in the left pane of 
fig.~\ref{fig:tevatronDY} are,
\begin{equation}
\begin{array}{c}
\mbox{{\bf ISR}}\\
\mbox{$Q^2$-ordering}
\end{array}~~~\left\{\begin{array}{rcclll}
\mbox{Tune A~(100)} & : & \alpha_s(\pT{}^2) &
~\overline{\mrm{MS}},&\mbox{1-loop}, &\Lambda_{\mrm{CTEQ5L}}\\
\mbox{Tune DW~(103)} & : &
\alpha_s(0.2\pT{}^2)&~\overline{\mrm{MS}},&\mbox{1-loop},&\Lambda_{\mrm{CTEQ5L}}\\ 
\mbox{Tune Pro-Q2O~(129)} & : & \alpha_s(0.14 \pT{}^2)&~\overline{\mrm{MS}},&\mbox{1-loop},&\Lambda_{\mrm{CTEQ5L}}\\
\end{array} \right.~,
\end{equation}
where, for completeness, we have given also the renormalization
scheme, loop order, and choice of $\Lambda_{\mrm{QCD}}$, which are the
same for all the tunes. 

While the increase of $\alpha_s$ 
nominally reestablishes a good agreement with the Drell-Yan
\pT{} spectrum, the whole business does smell faintly of
fixing one problem by introducing another and hence 
the defaults in \textsc{Pythia} for these parameters 
have remained the Tune A ones, at the price of retaining the poor
agreement with the Drell-Yan spectrum.

In the new $\pT{}$-ordered
showers~\cite{Sjostrand:2004ef}, however, FSR off ISR is treated
within individual QCD dipoles and does not affect the Drell-Yan
$\pT{}$. This appears to make the spectrum come out generically much
closer to the data, as illustrated by the S0(A) curves in 
fig.~\ref{fig:tevatronDY} (right column), which use
$\alpha_s(\pT{})$. The only change going to Perugia 0 --- which can be
seen to be slightly harder --- was 
implementing a translation from the $\overline{\mrm{MS}}$ definition
of $\Lambda$ used previously, to the so-called CMW choice \cite{Catani:1990rr} 
for $\Lambda$, similarly to what is done in
\textsc{Herwig} \cite{Corcella:2000bw,Bahr:2008pv}. 

For both CTEQ5L and CTEQ6L1, the 
$\Lambda^{\overline{\mrm{MS}}}_{\mrm{QCD}}$ 
value in the PDF set 
is derived with an LO (1-loop) running of $\alpha_s$, which is also
what we use in the backwards evolution algorithm in our ISR model. In
the Perugia tunes (and also in \textsc{Pythia} by default) we
therefore let the $\alpha_s$ value for the ISR evolution be determined
by the PDF set. The MRST LO* set \cite{Sherstnev:2007nd}, however,
uses an NLO (2-loop) running 
for $\alpha_s$, which gives a roughly 50\% larger value for $\Lambda$. Since we
do not change the loop order of our ISR evolution, this higher
$\Lambda$ value would lead to an increase in, e.g., the mean Drell-Yan 
$\pT{}$ at the Tevatron. In practice, however, this point is obscured
by the fact that the LHAPDF interface, used in our code (v.5.8.1), 
does not return the correct $\Lambda_{QCD}$ value for each PDF
set. Instead, a constant value of 0.192 (corresponding to CTEQ5L) is
returned. Since we were not aware of this bug in the interface when
performing the Perugia tunes, we therefore note that all the tunes
are effectively using the CTEQ5L value of $\Lambda$.
The pace of evolution with the LO* PDF set is still slightly higher
than for CTEQ5L, however. To compensate for this, 
the renormalization scale was chosen slightly higher for the LO* tune, cf.\ the
\ttt{PARP(64)} values in table \ref{tab:isrkt}. 

We note that a similar issue afflicted the original CTEQ6L set, which used an NLO
$\alpha_s$ (with a correspondingly larger value of $\Lambda$). We
here use the revised CTEQ6L1 set for our Perugia 6 tune, which uses an
LO running and hence should be more consistent with the 
evolution performed by the shower. Similarly, the LO* set used here 
could be replaced by the newer LO** one, which uses \pT{} instead of
$Q^2$ as the renormalization scale in $\alpha_s$, similarly to what is
done in the shower evolution, but this 
was not yet available at the time our LO* tune was performed. 
The main reason for sticking to CTEQ5L for Perugia 0 was the
desire that this tune can be run with standalone \Py\ 6. We note that 
in \textsc{Pythia} 8, several more recent sets have already been implemented in
the standalone version \cite{Kasemets:2010bx}, hence removing this
restriction from corresponding tuning efforts for \Py\ 8. Note also
that, since these sets are implemented \emph{internally} in \Py\ 8,
the bug in the LHAPDF interface mentioned above does not affect
\Py\ 8\footnote{Note therefore that one has to be  careful when
  linking LHAPDF. If an internally implemented PDF set is replaced by
  its LHAPDF equivalent, there is unfortunately at present no
  guarantee that identical results will be obtained. We therefore strongly
  advise MC tunes to specify exactly which implementation was used to
  perform the tune, and users to regard the implementation as part of
  the tune. We hope this unfortunate situation may be rectified in the
  future.}.

Finally, the HARD and SOFT variations shown by the yellow band in the 
right pane of fig.~\ref{fig:tevatronDY} are obtained by making a
variation of roughly a factor of 2 in either direction from the
central tune (in the case of
the SOFT tune, this is obtained by a combination of reverting to the
$\overline{\mbox{MS}}$ value for $\Lambda$ and using $\sqrt{2}\pT{}$ as the
renormalization scale). In the low-$\pT{}$ peak, 
the HARD variation generates a slightly too broad distribution, but
given the large sensitivity of this peak to subleading corrections
(see below), we
consider this to be consistent with the expected theoretical
precision. The $p_\perp$ spectrum of the other Perugia tunes will be
covered in the section on the individual tunes below. 

\paragraph{Phase Space:}
A further point concerning ISR that deserves discussion is the phase
space over which ISR emissions are allowed. Here, Drell-Yan is a
 special case, since this process is matched to $Z$+jet matrix
 elements in \textsc{Pythia} \cite{Bengtsson:1986hr,Bengtsson:1986et}, 
  and hence the hardest jet is always
 described by the matrix element over all of phase space. For
 unmatched processes which do not contain jets at leading
 order, the fact that we start the parton shower off from the
 factorization scale can, however, produce an illusion of almost zero
 jet activity above that scale. This was studied in
 \cite{Plehn:2005cq,Skands:2005bj}, where also the consequences of
 dropping the phase space cutoff at the factorization scale were
 investigated, so-called power showers. Our current best understanding 
 is that the conventional (``wimpy'') showers with a cutoff at the
 factorization scale certainly underestimate the tail of ultra-hard
 emissions while the power showers are likely to overestimate it,
 hence making the difference between the two a useful measure of
 uncertainty. 
Since other event generators usually provide wimpy
 showers by default, we have chosen to give the power variants as
 the default in \textsc{Pythia} 6 --- not because the power shower
 approximation is necessarily better, but simply to minimize the risk
 that an accidental agreement between two generators is taken as a
 sign of a small overall uncertainty, and also to give a conservative
 estimate of the amount of hard additional jets that can be
 expected. Note that a more systematic description of hard radiation that
 interpolates between the power and wimpy behaviours has
 recently been implemented in \textsc{Pythia} 8 \cite{Corke:2010zj}.

For the Perugia models, we have  implemented a simpler 
possibility to smoothly dampen the tail of
ultra-hard radiation, using a scale determined from the colour flow 
as reference. This is
done by nominally applying a power shower, but dampening it by a
factor
\begin{equation}
P_{\mrm{accept}} = P_{67} ~ \frac{s_{D}}{4\pT{\mrm{evol}}^2}~,
\end{equation}
where $P_{67}$ corresponds to the parameter \ttt{PARP(67)} in the
code, $\pT{\mrm{evol}}^2$ is the evolution scale for the trial
  splitting, and $s_D$ is the invariant mass of the radiating parton
  with its colour neighbour, with all momenta crossed into the final
  state (i.e., it is $\hat{s}$ for annihilation-type colour flows and
  $-\hat{t}$ for an initial-final connection). This is motivated 
  partly by the desire to give an intermediate possibility between the
  pure power and pure wimpy options but also partly from
  findings that similar factors can substantially improve the
  agreement with final-state matrix elements in the context of the
  \textsc{Vincia} shower \cite{Giele:2011cb}. By default, the Perugia tunes
  use a value of 1 for this parameter, with the SOFT and HARD tunes exploring
  systematic variations, see table \ref{tab:isrkt}.

\begin{figure}[t]
\begin{center}\vspace*{-5mm}
\scalebox{1.41}{
\includegraphics*[scale=0.34]{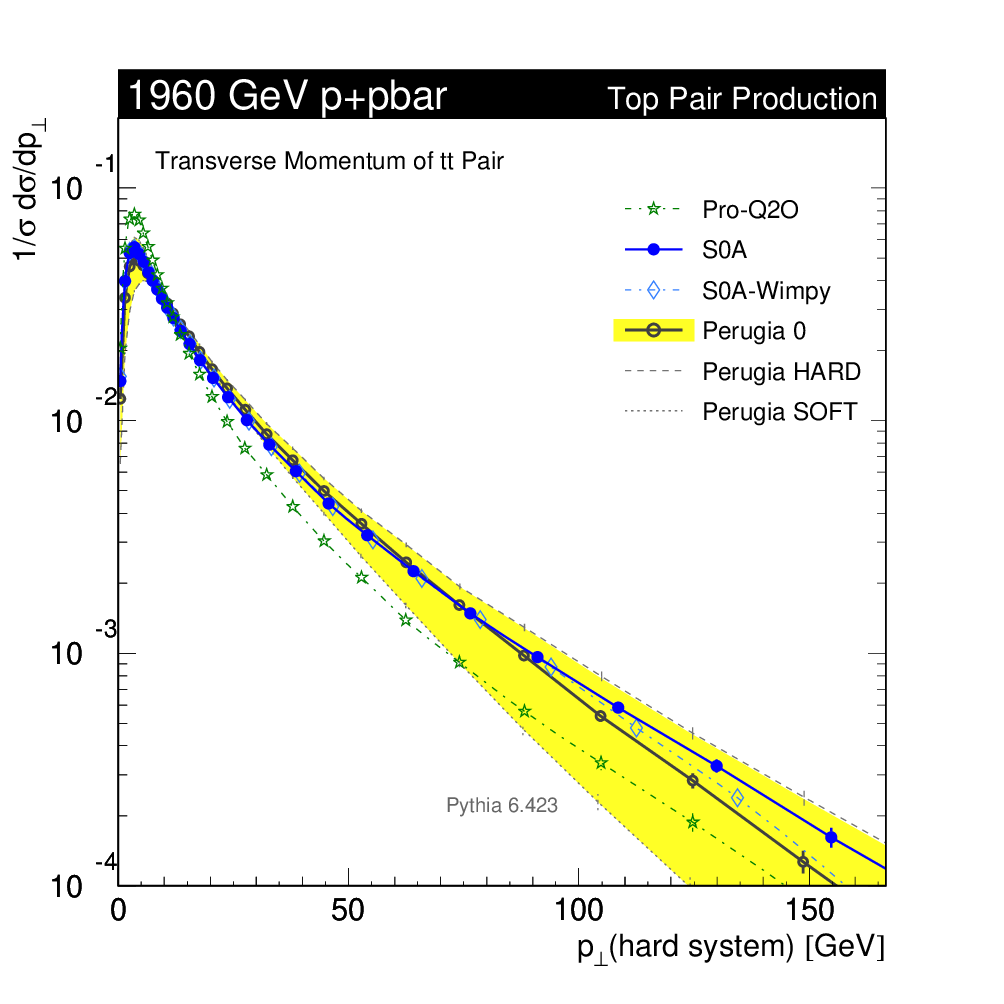}\hspace*{-3mm}
\includegraphics*[scale=0.34]{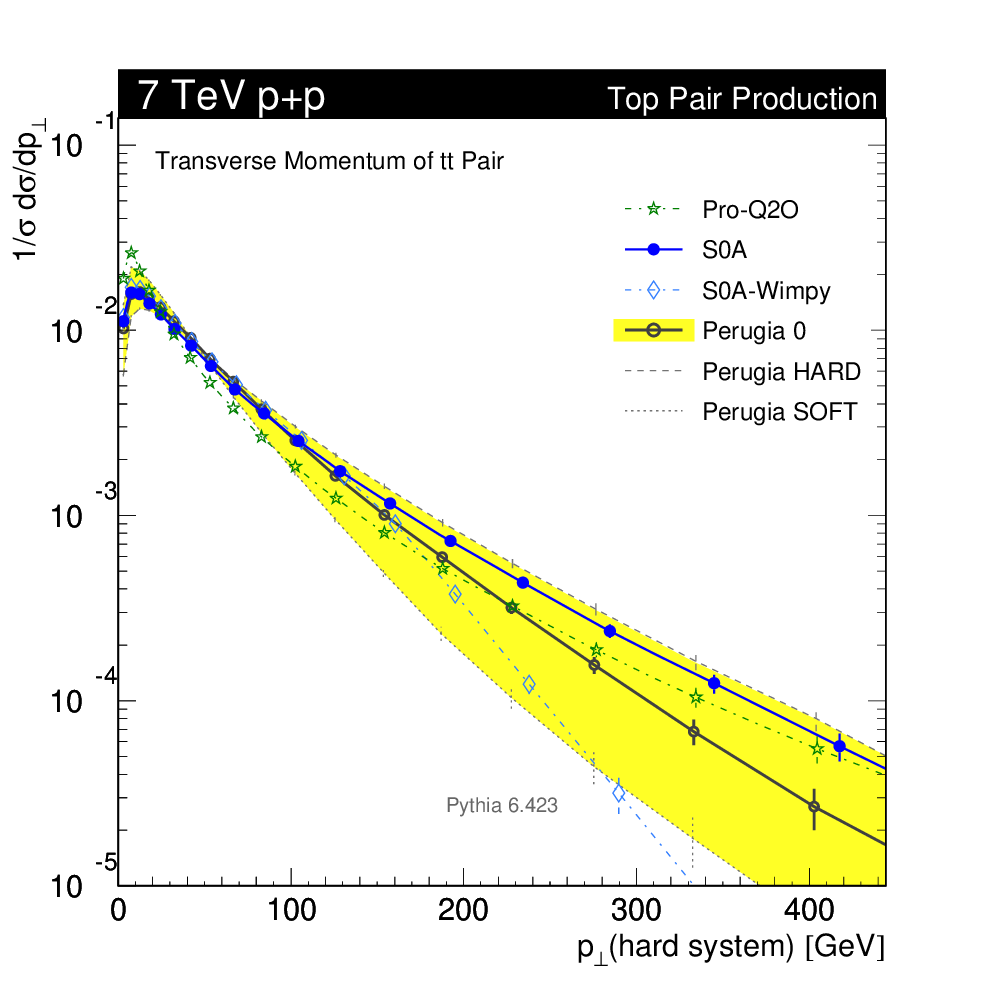}}\vspace*{-3mm}
\caption{\small Comparison of new and old tunes for the $\pT{}$ of $t\bar{t}$ pairs
  at the Tevatron ({\sl left}) and at the LHC at 7 TeV ({\sl right}). 
See
  \cite{lhplots} for
  other tunes and collider energies. 
\label{fig:tt}}
\end{center}
\end{figure}
At the Tevatron, the question of power vs.~wimpy showers is actually not 
 much of an issue, since $H/V$+jets is already matched to matrix elements in
default \textsc{Pythia} and most other interesting
 processes either contain QCD jets already at leading order
 ($\gamma$+jets, dijets, WBF) or have very little phase space for
 radiation above the factorization scale anyway ($t\bar{t}$,
 dibosons). 
This is illustrated by the curves labelled S0A (solid blue) and
 S0A-Wimpy (dash-dotted cyan) in the left pane of
 fig.~\ref{fig:tt}, which shows the  $\pT{}$ spectrum of the
 $t\bar{t}$ system (equivalent to the Drell-Yan \pT{} shown earlier). 
The two curves do begin to diverge around
 the top mass scale, but in light of the limited statistics available at
 the Tevatron, matching to higher-order matrix elements to control
 this  ambiguity does not appear to be of crucial importance.
In contrast, when we extrapolate to $pp$
 collisions at 7 TeV, shown in the right pane of
 fig.~\ref{fig:tt}, the increased phase space makes the ambiguity
 larger. 
 Matching to the proper matrix elements describing the region of jet
 emissions above $\pT{} \sim$ $m_t$ may therefore be
correspondingly more important, see, e.g., \cite{Alwall:2008qv}. Note
that the extremal Perugia variations span most of the full power/wimpy
difference, as desired, while the central ones fall inbetween. Note
also that this only concerns the $\pT{}$ spectrum of the hard jets --- 
power showers cannot in general be expected to properly capture jet-jet
correlations, which are partly generated by polarization effects not
accounted for in this treatment. 

\paragraph{Primordial $\mathbf{k_T}$:}
Finally, it is worth remarking that the peak region of the
Drell-Yan $p_\perp$ spectrum is extremely sensitive to infrared effects. On the
experimental side, this means, e.g., that the treatment of QED
corrections can have significant effects and that care must be taken
to deal with them in a consistent and model-independent manner
\cite{hesketh}. On the theoretical side, relevant infrared effects
include whether the low-$\pT{}$ divergences in the 
parton shower are regulated by a sharp cutoff or by a smooth suppression (and in what
variable), how $\alpha_s$ is treated close to the cutoff, and how much
``Fermi motion'' is given to each of the shower-initiating partons
extracted from the protons. A full exploration of these effects
probably goes beyond what can meaningfully be studied at the current
level of precision. Our models therefore only contain one infrared
parameter (in addition to the infrared regularization scale of the shower), 
called  ``primordial $k_T$'', which should
 be perceived of as lumping together an inclusive sum of unresolved
effects below the shower cutoff. Since the cutoff is typically in the
range 1--2 GeV, we do not expect the
primordial $k_T$ to be much larger than this number, but
there is also no fundamental reason to believe it should be significantly
smaller. This is in contrast to previous lines of thought, which
drew a much closer connection between this parameter and
 Fermi motion, which is expected to be only a few hundred MeV. 
In Tune A, the value of primordial $k_T$, corresponding to
\ttt{PARP(91)} in the code, 
 was originally 1 GeV, whereas it was increased  to 2.1 GeV
in Tune DW. In the Perugia tunes, it varies in the same range, 
cf.~table \ref{tab:isrkt}. Its distribution is assumed to be
Gaussian in all the models. Explicit  attempts exploring alternative
  distributions in connection with the write-up of this paper
  ($1/k_T^6$ tails and even a flat distribution with a cutoff, see
  \cite[\ttt{MSTP(91)}]{Sjostrand:2006za}) did not
  lead to significant differences.

\subsection{Underlying Event, Beam Remnants, and Colour Reconnections  (Table \ref{tab:uebrcr})}

\begin{figure}[t]
\vspace*{2.2cm}
\scalebox{1.41}{
\parbox{12cm}{\hspace*{0.7cm}
\includegraphics*[scale=0.26]{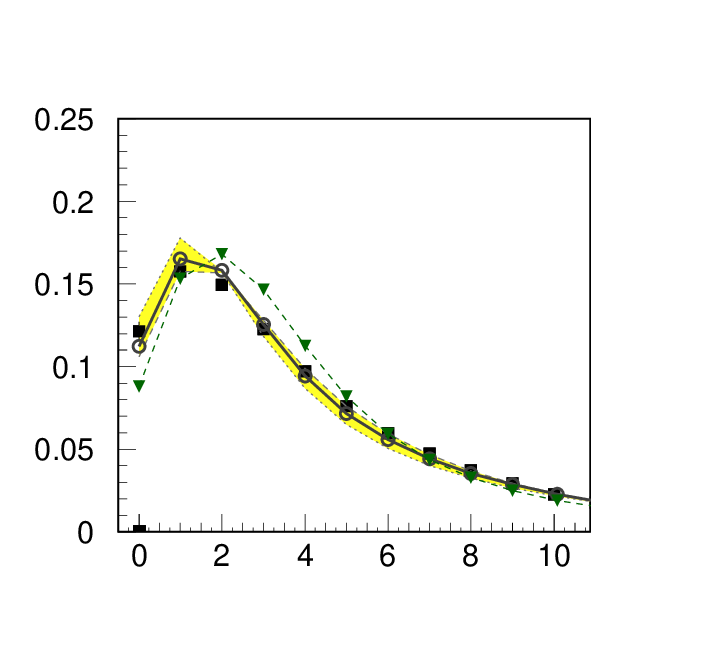} \hspace*{2.37cm}\includegraphics*[scale=0.26]{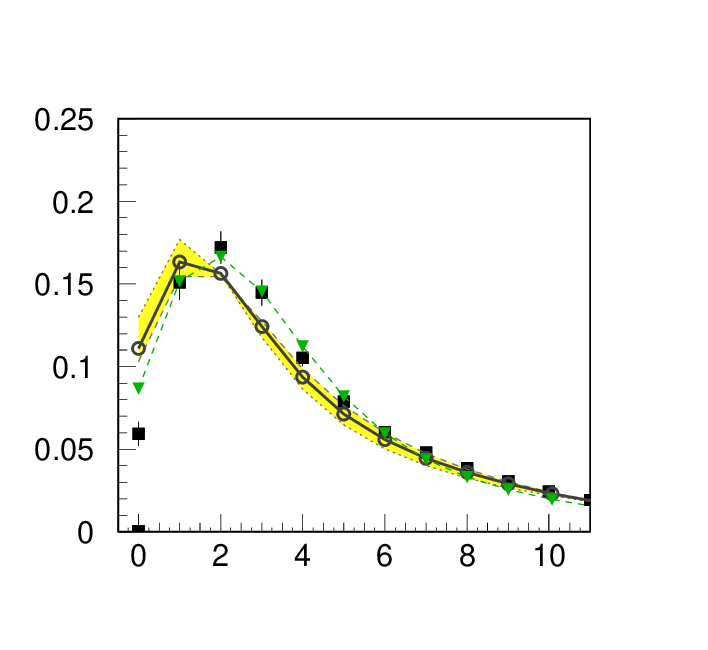} \\[-4.9cm]
\hspace*{1mm}\includegraphics*[scale=0.34]{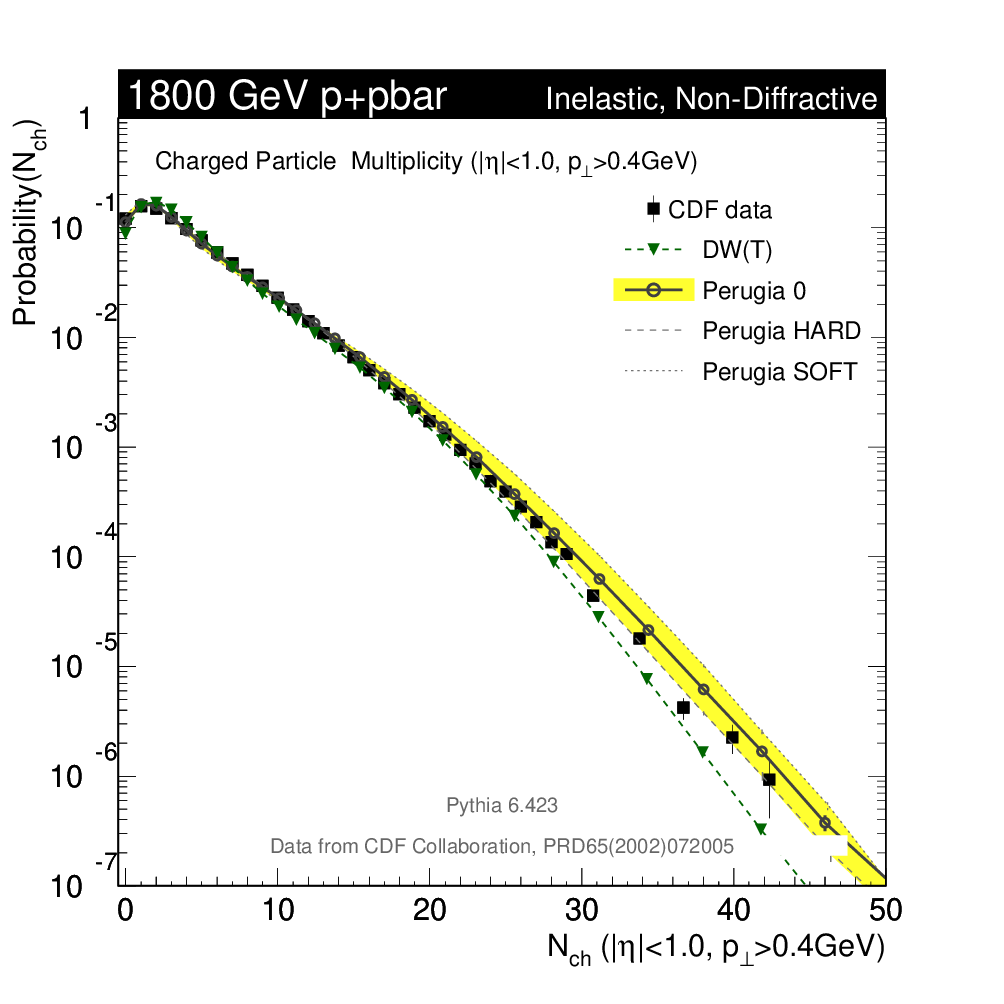}\hspace*{-3mm}
\includegraphics*[scale=0.34]{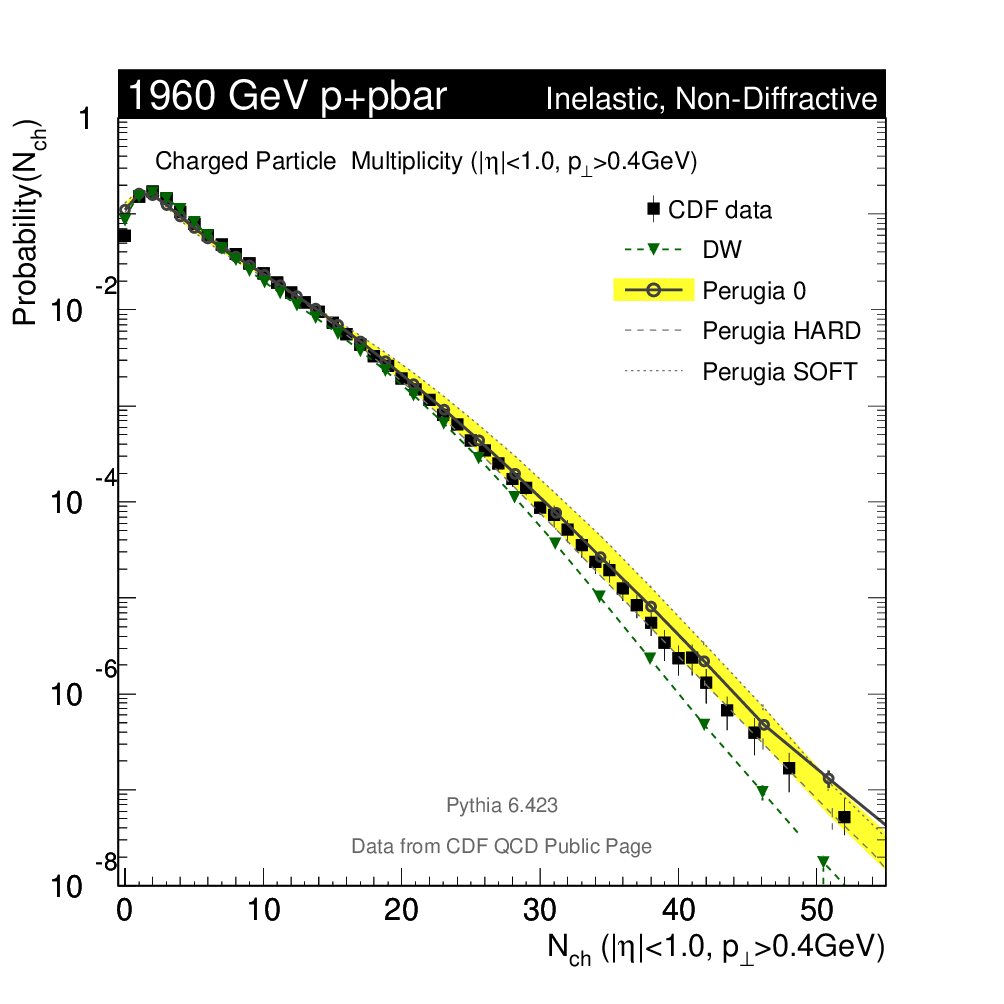}\hspace*{-12mm}\\[-5mm]}}
\caption{\small Comparisons to the CDF measurements of
 the charged track multiplicity at 1800 ({\sl left}) and 1960 GeV ({\sl right}) 
in minimum-bias $p\bar{p}$ collisions. 
See
  \cite{lhplots} for 
  other tunes and collider energies. 
\label{fig:tevatronNCH}}
\end{figure}

\paragraph{Charged Multiplicity}
The charged particle multiplicity ($N_{\mrm{ch}}$) distributions
for minimum-bias events  at 1800 and 1960 GeV at the Tevatron are
shown in fig.~\ref{fig:tevatronNCH}. Particles with $c\tau \ge 10$
mm ($\mu^\pm$, $\pi^\pm$, $K^0_S$, $K^0_L$, $n^0$, $\Lambda^0$,
$\Sigma^\pm$, $\Xi^0$, $\Xi^\pm$, and $\Omega^\pm$) are treated as
stable. Models include the inelastic, non-diffractive component only.
Note that the Perugia tunes included this data in the tuning, while
DW was only tuned to underlying-event data at the same energies. The
overall agreement over the many orders of magnitude spanned by these
measurements is quite good. On the large-multiplicity tails, 
DW appears to give a slightly too narrow distribution. In the
low-multiplicity peak (see insets), the Perugia tunes fit the 1800 GeV
data set better while DW fits the 1960 GeV data set better. As
mentioned above, however, diffractive topologies give large corrections in this
region, and so the points shown in the insets were 
not used to constrain the Perugia tunes. 

\begin{figure}[t]
\begin{center}\vspace*{-5mm}
\scalebox{1.41}{
\includegraphics*[scale=0.34]{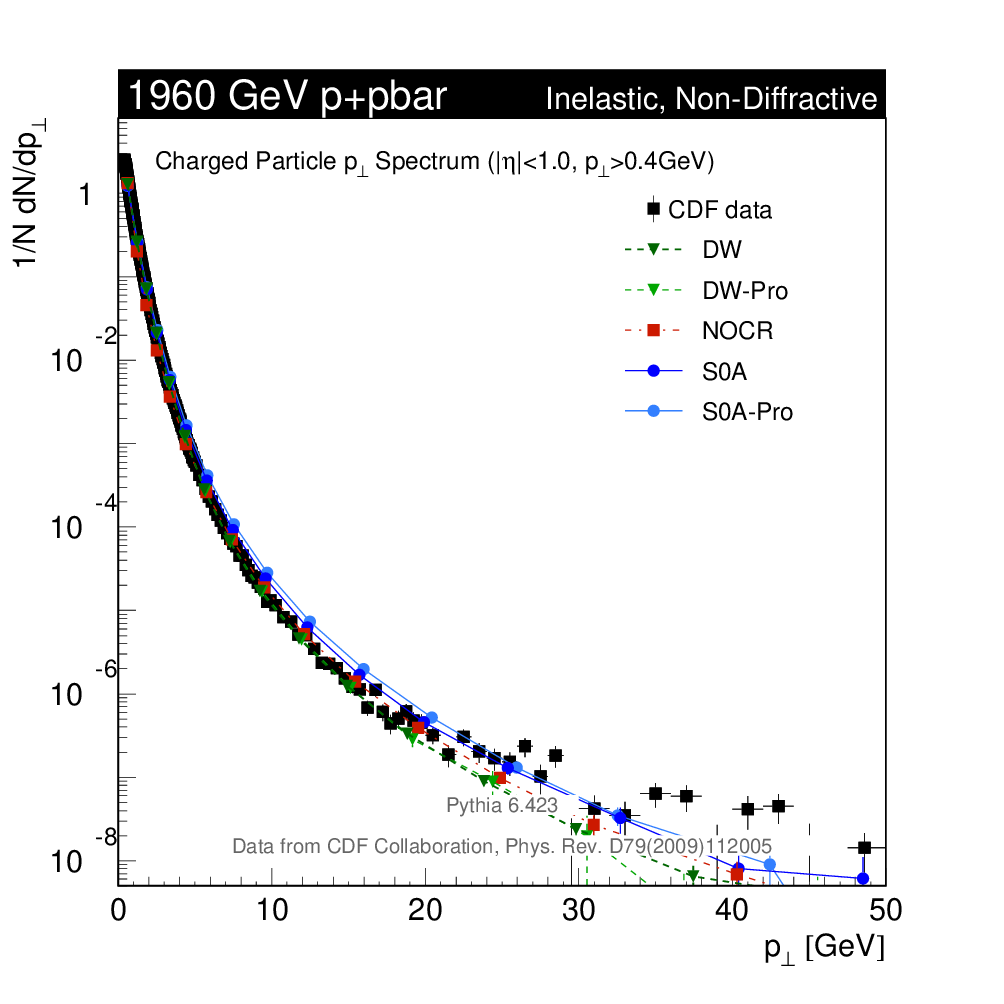}\hspace*{-3mm}
\includegraphics*[scale=0.34]{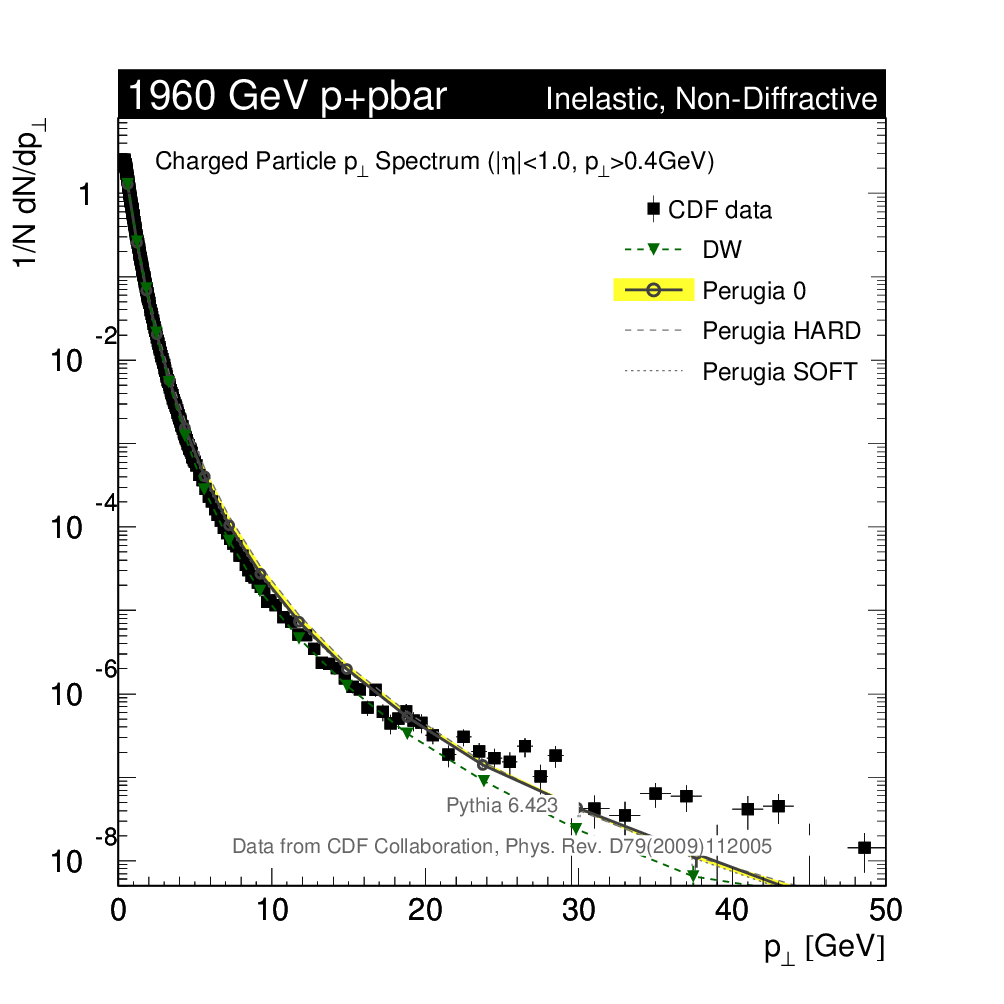}\hspace*{-3mm}}
\vspace*{-8mm}
\caption{\small \small Comparisons to the CDF measurement of
 the charged particle \pT{} spectrum in minimum-bias $p\bar{p}$ collisions 
at 1960 GeV for two sets of models. 
See
  \cite{lhplots} for 
  other tunes and collider energies. 
\label{fig:tevatronPT}}
\end{center}
\end{figure}

\paragraph{Transverse Momentum Spectrum}
The \pT{} spectrum of charged particles at 1960 GeV is shown in
fig.~\ref{fig:tevatronPT}. Note that both plots in the figure 
show the same data; only the model comparisons are different. 

The plot in the left-hand pane illustrates a qualitative difference
between the $Q^2$- and $\pT{}$-ordered models. Comparing DW to NOCR
(a tune of the \pT{}-ordered model which does not employ colour
reconnections) we see that the \pT{} spectrum is generically
slightly harder in the new model than in the old one. Colour
reconnections, introduced in S0A, then act to harden this spectrum 
slightly more, to the point of marginal disagreement with
the data. Finally, when we include the Professor tunes to LEP data,
nothing much happens to this spectrum in the old model --- compare DW
with DW-Pro --- whereas the spectrum becomes yet harder in the new one,
cf.\ S0A-Pro, now reaching a level of disagreement with the data that
we have to take seriously. Since the original spectrum out of the box
--- represented by NOCR --- was originally quite similar to that of
DW and DW-Pro, our tentative conclusion
is that either the revised LEP parameters for the \pT{}-ordered shower 
have some hidden problem and/or the colour reconnection 
model is hardening the spectrum too much. For the Perugia tunes, we
took the latter interpretation, since we did not wish to alter the LEP
tuning. Using a modified colour-reconnection model that suppresses
reconnections among high-\pT{} string pieces (to be described below),
the plot in the right-hand pane illustrates that an acceptable level
of agreement with the data has been restored in the Perugia tunes, 
without modifying the Professor LEP parameters.

For completeness we should also note that there are indications of
a significant discrepancy developing in the extreme tail of particles with 
$\pT{} > 30$ GeV, where all the models fall below the data, a trend that  
was confirmed with higher statistics in \cite{Arleo:2010kw}. This 
discrepancy also appears in the context of 
NLO calculations folded with fragmentation functions 
\cite{Albino:2010em}, 
so is not a feature unique to the \textsc{Pythia} modelling. 
Though we shall not comment on possible causes for this
behaviour here (see \cite{Cacciari:2010yd,Yoon:2010fa} for a critical
assessment), the extreme tail of the \pT{} distribution should 
therefore be especially interesting to check when high-statistics data
from the LHC become available. 

\begin{figure}[t]
\begin{center}\hspace*{-2mm}
\scalebox{1.41}{\includegraphics*[scale=0.34]{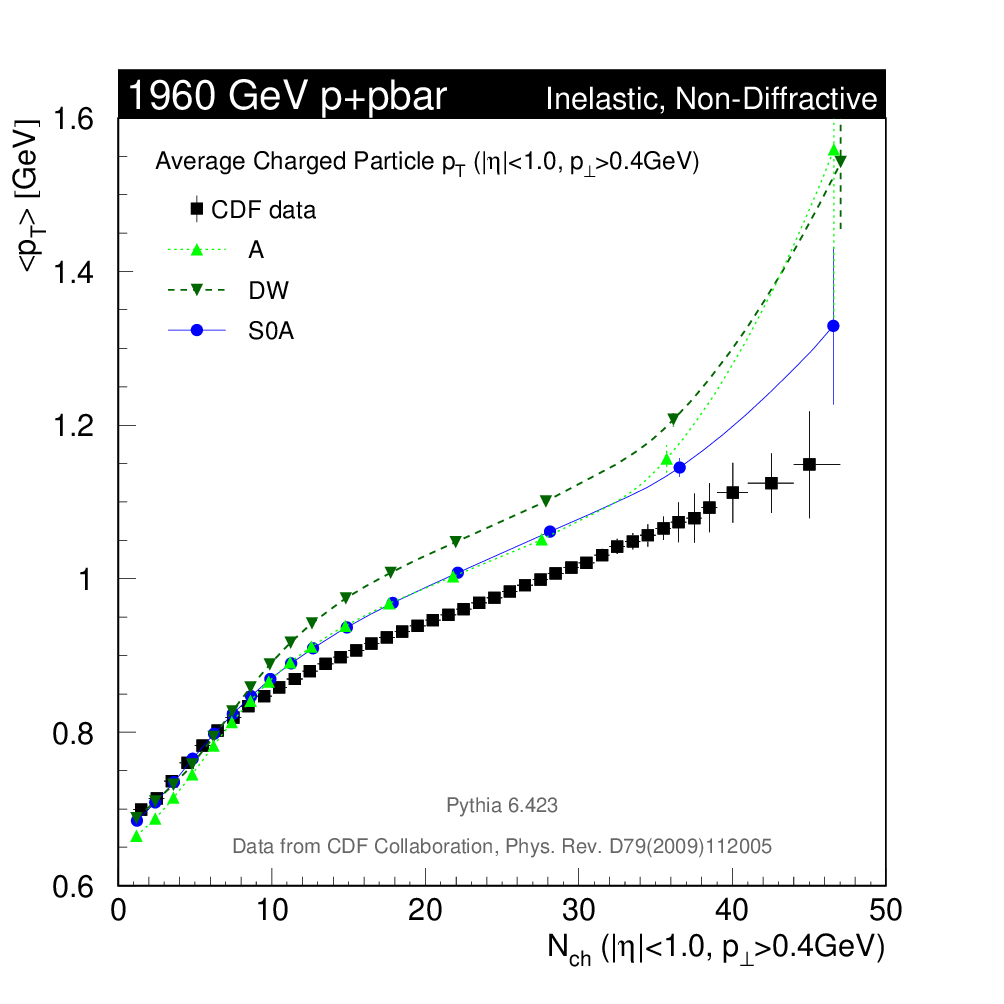}\hspace*{-5mm}
\includegraphics*[scale=0.34]{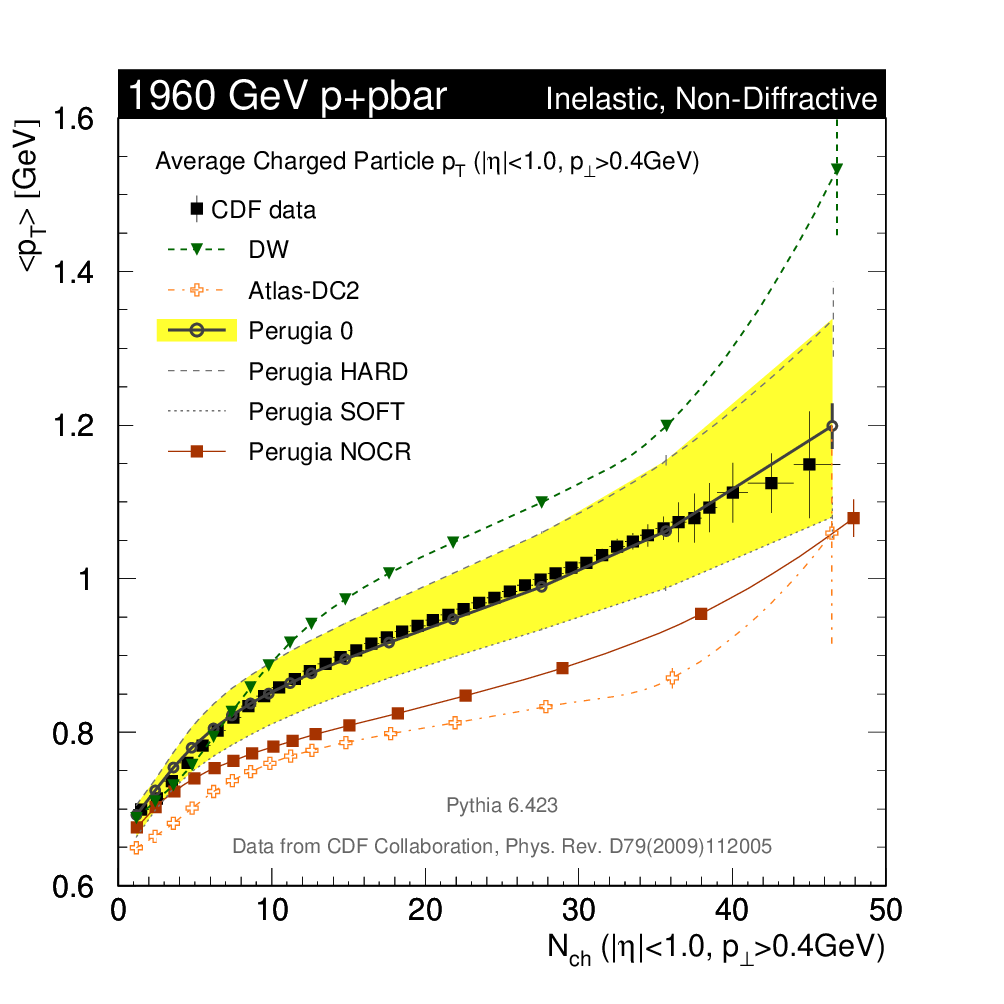}}\hspace*{-5mm}
\caption{\small Comparisons to the CDF Run II measurement of
 the average track $\pT{}$ as a function of track multiplicity 
in min-bias $p\bar{p}$ collisions.  {\sl Left:} The older generation
of tunes. {\sl Right:} the Perugia variations compared to two 
older tunes. See
  \cite{lhplots} for 
  other tunes and collider energies. 
\label{fig:tevatronAVG}}
\end{center}
\end{figure}

\paragraph{$\left<\pT{}\right>(N_{\mrm{ch}})$ and Colour Reconnections}
While the multiplicity and \pT{} spectra are thus, separately, well
described by Tune DW, it 
does less well on their correlation, 
$\left<\pT{}\right>(N_{\mrm{ch}})$, as illustrated by the plot in the
left-hand pane of fig.~\ref{fig:tevatronAVG}. 
Since the S0 family of tunes were initially tuned to Tune A, in the absence of
published data, the slightly smaller discrepancy exhibited by Tune A 
carried over to the S0 set of tunes, as illustrated by the same plot.
Fortunately, CDF Run-2 data has now been made publicly
available \cite{Aaltonen:2009ne}, corrected to the particle level, 
and hence it was possible to take the actual
data into consideration for the Perugia tunes, resulting in somewhat softer
particle spectra in high-multiplicity events, cf.\ the right-hand pane
in fig.~\ref{fig:tevatronAVG}. 

What is more interesting is how this correlation is achieved by the models. Also
shown in the right-hand pane of fig.~\ref{fig:tevatronAVG} 
are comparisons to an older ATLAS tune which did not use the enhanced
final-state colour connections  that  Tunes A and DW employ. A special Perugia 
variation without colour reconnections, Perugia NOCR, is also
shown, and one sees that both this and the ATLAS tune predict too little correlation
between $\left<\pT{}\right>$ and $N_\mrm{ch}$.

This distribution therefore appears to be sensitive to the colour
structure of the events, at least within the framework of the
\textsc{Pythia} modelling
\cite{Sjostrand:1987su,Sandhoff:2005jh,Skands:2007zg,Wicke:2008iz}.  
The Perugia tunes all (with the exception of NOCR) 
rely on an infrared toy model of string interactions
\cite{Sandhoff:2005jh} to drive 
the increase of $\left<\pT{}\right>$ with $N_{\mrm{ch}}$. 
The motivation for a model of this type comes from arguing that, in
the leading-colour limit used by Monte Carlo event generators, and in
the limit of many perturbative parton-parton interactions, 
the central rapidity region in hadron-hadron collisions would be
criss-crossed by a very large number of QCD strings; na\"\i vely one
string per perturbative $t$-channel quark exchange, and two per gluon
exchange. However, since the actual number of colours is only three, and since
the strings would have to be rather closely packed in spacetime, it is not
unreasonable to suppose either that the colour field collapses in a more
economical configuration already from the start, or that the strings
undergo interactions among themselves, before the fragmentation process is complete, 
that tend to minimize their total potential energy, as given by the
 area law of classical strings. 
The toy models used by both
the S0 and Perugia tunes do not address the detailed dynamics of this
process, but instead employ an annealing-like minimization of
the total potential energy, where the string-string interaction
strength was originally the only variable parameter
\cite{Sandhoff:2005jh}. While this gave a reasonable agreement with
$\left<\pT{}\right>(N_\mrm{ch})$, it still tended to give slightly too
hard a tail on the single-particle \pT{} distribution, as compared to
the Tevatron Run 2 measurement. Therefore, a suppression of
reconnections among very high-$\pT{}$ string pieces was introduced,
reasoning that very fast-moving string systems should be able to more
easily ``escape'' the mayhem in the central region. (Similarly, one
could argue that string systems produced in the decay of massive
particles with finite lifetimes, such as narrow BSM or Higgs
resonances, or even possibly hadronic $t$ or $W$ decays, 
should be able to 
escape more easily. We have not so far built in such a suppression,
however.) 

The switch \ttt{MSTP(95)} controls the choice of
colour-reconnection model. 
In the ``S0'' model corresponding to \ttt{MSTP(95)=6} (and
\ttt{=7} to apply it also in lepton
collisions), the total probability for a string piece to survive the
annealing and preserve its original colour connections is
\begin{equation}
\mbox{\ttt{MSTP(95)~=~6,~7~~~:~~~}} P_{\mrm{keep}} = (1-\zeta P_{78})^{n_{\mrm{int}}}~,
\end{equation}
where $P_{78}$ corresponds to the parameter \ttt{PARP(78)} in the
code and sets the overall colour-reconnection strength and 
$n_{\mrm{int}}$ is the number of parton-parton interactions in the 
current event, giving a rough first estimate of the number of strings
spanned between the remnants. (It is thus more likely for a string
piece to suffer ``colour amnesia'' in a busy event, than in a quiet
one.) $\zeta$ was introduced together with the Perugia tunes and 
gives a possibility to suppress reconnections among high-$\pT{}$
string pieces, 
\begin{equation}
\zeta = \frac{1}{1+P_{77}^2\left<\pT{}\right>^2}~,
\end{equation}
with $P_{77}$ corresponding to \ttt{PARP(77)} in the code and
$\left<\pT{}\right>$ being a measure of the average 
transverse momentum per
pion that the string piece would produce, $n_\pi \propto
\ln(s/m_\pi^2)$, with a normalization factor absorbed into $P_{77}$.

Starting from \textsc{Pythia} 6.4.23, a slightly more sophisticated
version of colour annealing was
introduced, via \ttt{MSTP(95)=8} (and 
\ttt{=9}  to apply it also in lepton
collisions), as follows. Instead of using the number of multiple parton-parton
interactions to give an average idea of the total number of strings
between the remnants, the
algorithm instead starts by finding a thrust axis for the event (which
normally will coincide with the $z$ axis for hadron-hadron
collisions).  It then computes the density of string pieces along that axis, 
rapidity-interval by rapidity-interval, with a relatively fine binning
in rapidity. Finally, it calculates the reconnection 
probability for each individual string piece by 
using the average string density in the region spanned by that string
piece, instead of the number of multiple interactions, in the exponent
in the above equation: 
\begin{equation}
\mbox{\ttt{MSTP(95)~=~8,~9~~~:~~~}} P = (1-\zeta P_{78})^{\left<n_s\right>(y_1,y_2)}~,
\end{equation}
where $\left<n_s\right>(y_1,y_2)$ is the average number of other string pieces, 
not counting the piece under consideration, in the rapidity range
spanned by the two endpoints of the piece, $y_1$ and $y_2$. 
Obviously, the resulting model is still relatively crude
--- it still has no explicit space-time picture and hence will not
generate more subtle effects such as (elliptical) flow, no detailed dynamics
model, and no suppression mechanism for reconnections involving
long-lived resonances ---
but at least the reconnection probability has been made a more local
function of the actual string environment, which also provides 
a qualitative variation on the previous models that can be used to 
explore uncertainties. In the code, the ``S0'' type is also referred
to as the ``Seattle'' model, since it was written while on a visit
there. The newer one is referred to as the ``Paquis'' type, for
similar reasons. 

\paragraph{Underlying Event}

\begin{figure}[t]
\begin{center}\vspace*{-5mm}
\includegraphics*[scale=0.34]{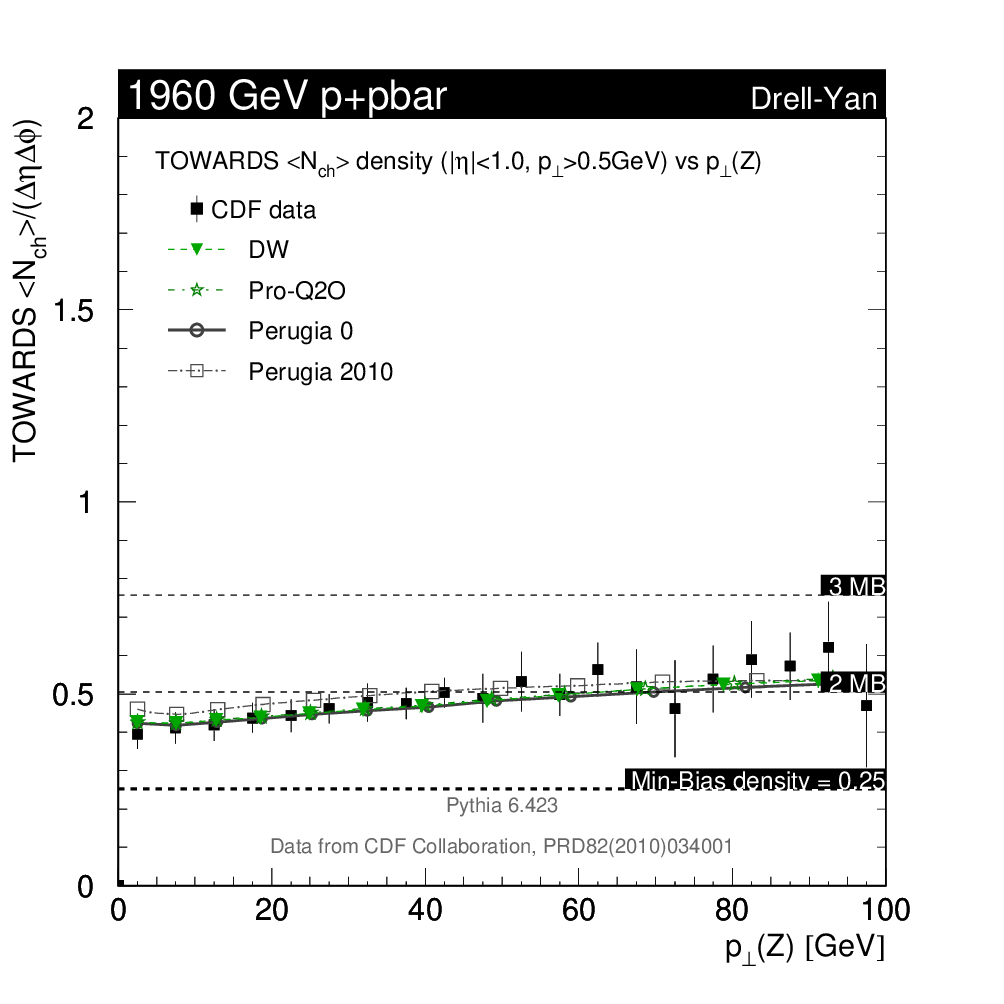}\hspace*{-5mm}
\includegraphics*[scale=0.34]{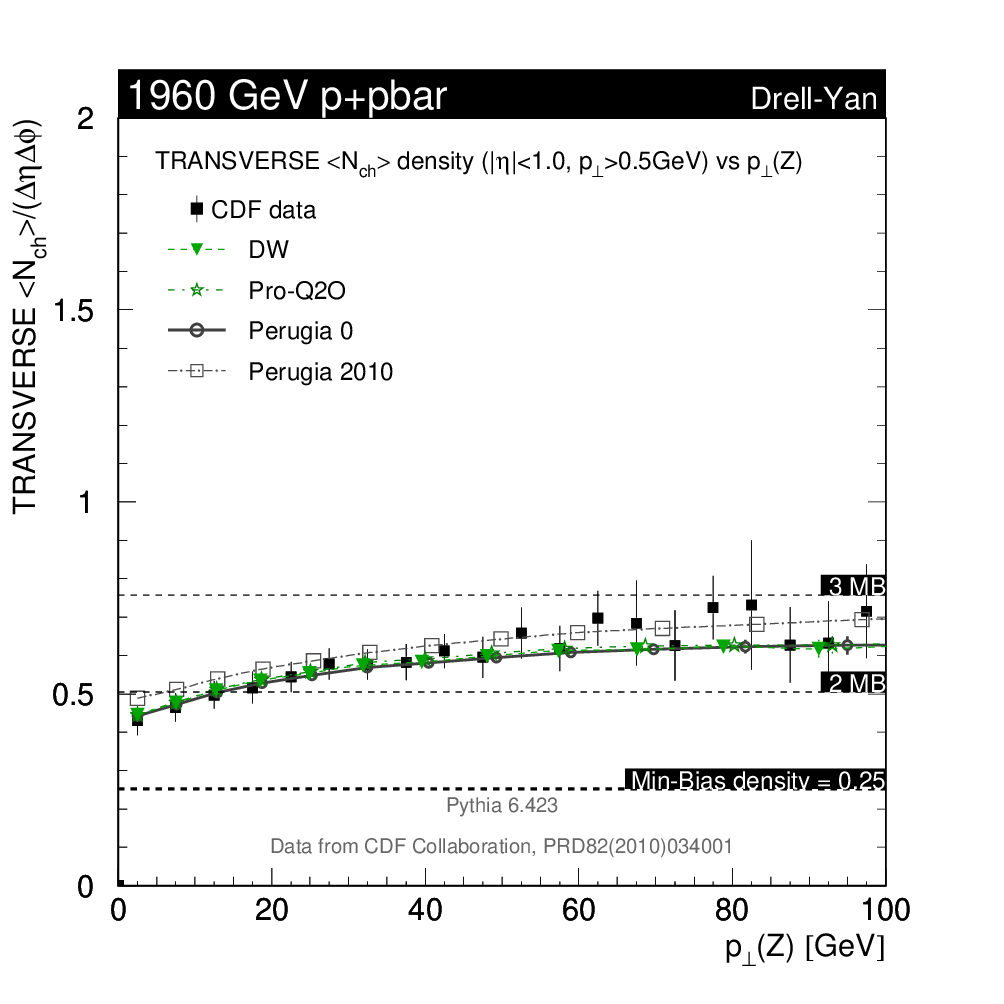}\hspace*{-5mm}
\includegraphics*[scale=0.34]{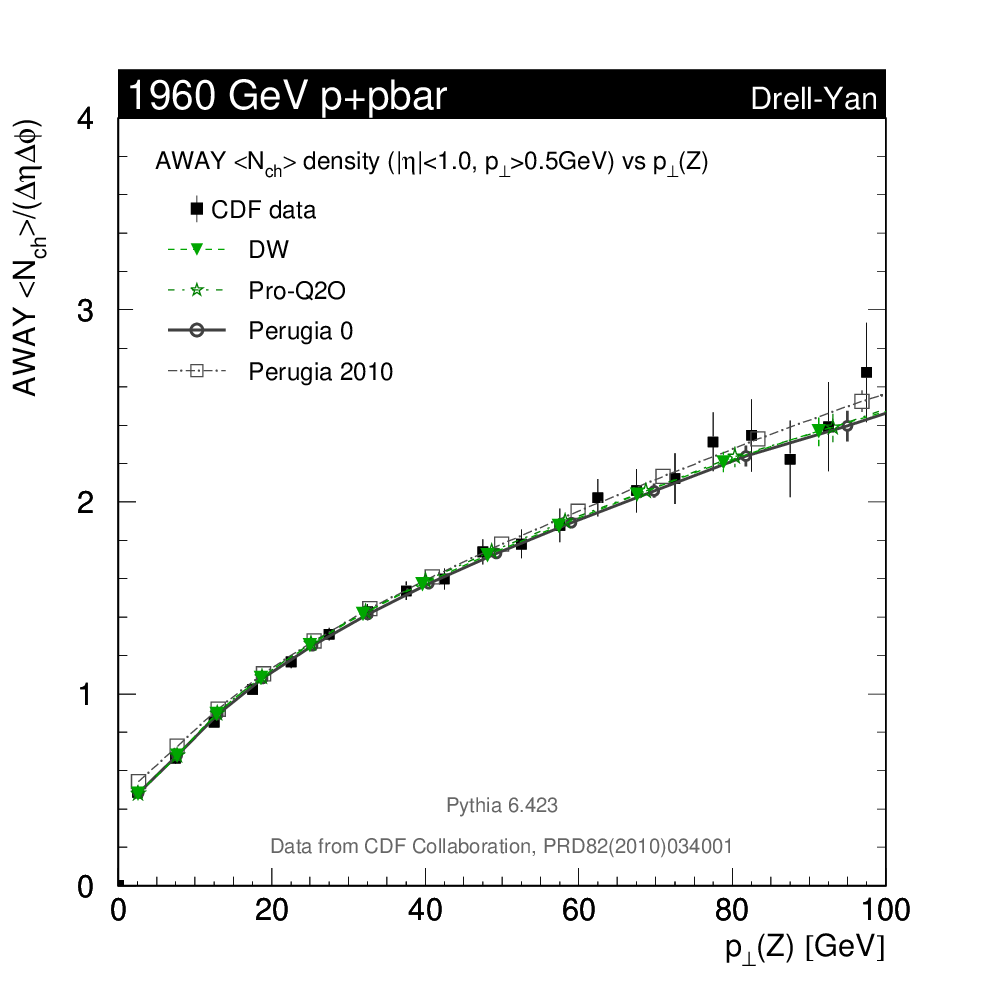}\hspace*{-8mm}\\[-5mm]
\includegraphics*[scale=0.34]{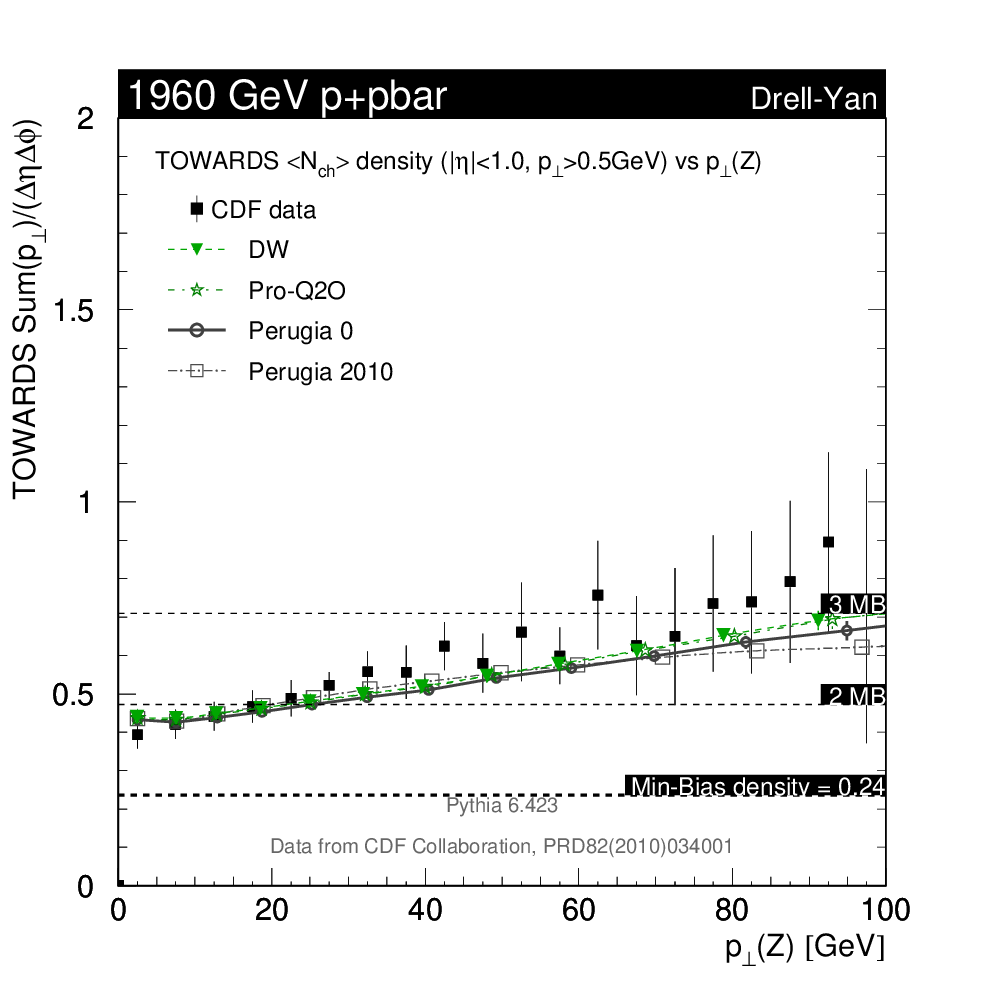}\hspace*{-5mm}
\includegraphics*[scale=0.34]{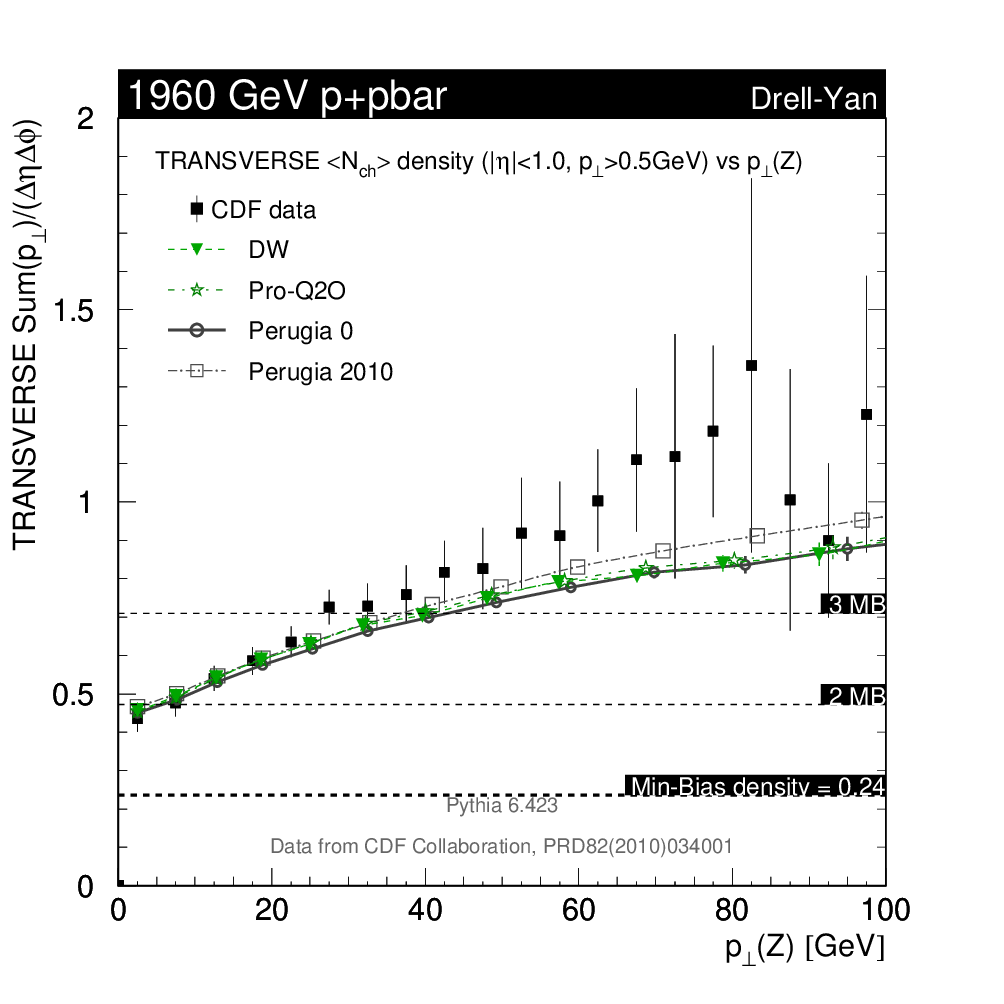}\hspace*{-5mm}
\includegraphics*[scale=0.34]{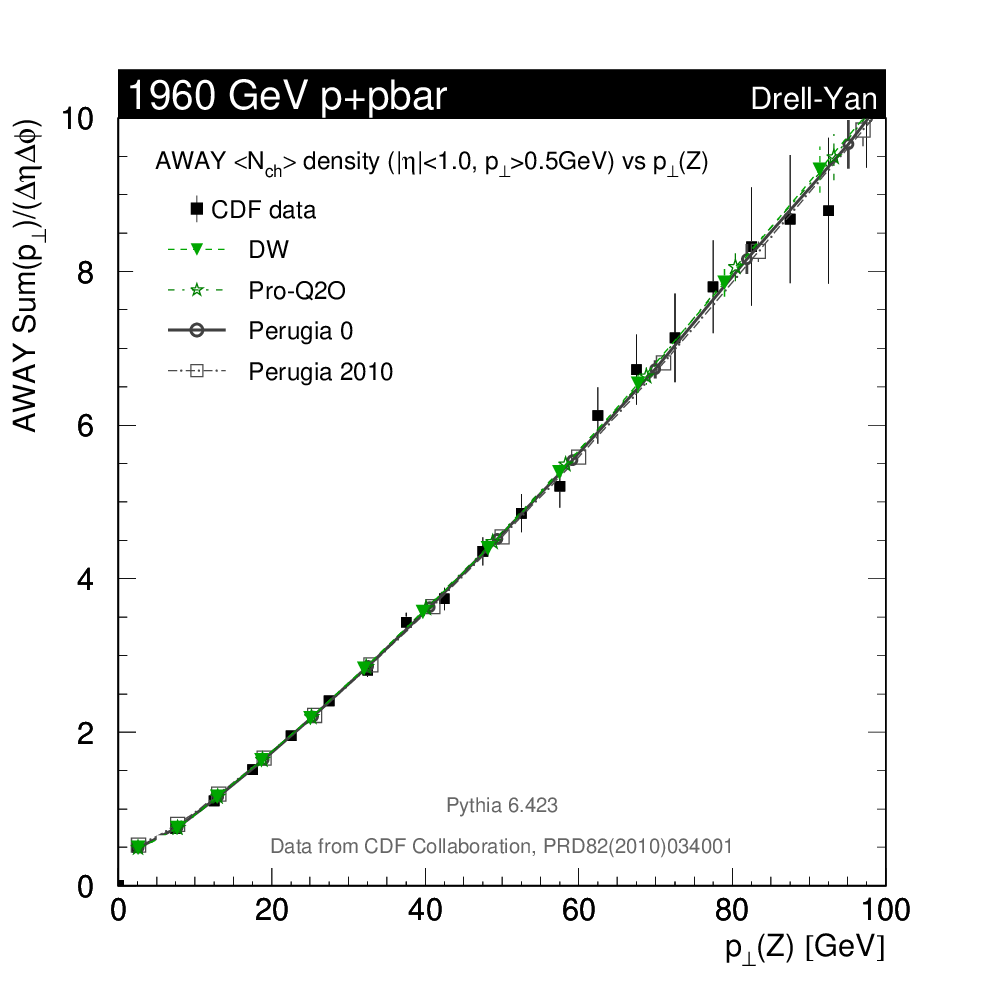}\hspace*{-8mm}\\[-3mm]
\caption{\small Comparisons to the CDF measurements \cite{Kar:2008zza,Aaltonen:2010rm} of
 the charged particle multiplicity (top row) and \pT{} (bottom row) densities in the
 ``TOWARDS'' (left), ``TRANSVERSE'' (middle), and ``AWAY'' (right) 
regions of Drell-Yan production at 1960 GeV, as a function 
of the Drell-Yan $p_\perp$. 
\label{fig:ue}}
\end{center}
\end{figure}

In fig.~\ref{fig:ue}, we show  the $\left<N_{ch}\right>$ density\footnote{The
  $\left<N_{ch}\right>$  density is defined as the average number of tracks
  per unit $\Delta\eta\Delta\phi$ in the relevant region.} 
(top row) and the 
$\left<p_{\perp\mathrm{Sum}}\right>$ density\footnote{The
  $\left<p_{\perp\mathrm{Sum}}\right>$ density is defined as the
    average scalar sum of track $p_\perp$ per unit
    $\Delta\eta\Delta\phi$.} (bottom row) in each of the TOWARDS,
  TRANSVERSE, and AWAY regions, for  
  Drell-Yan production at the Tevatron, compared to CDF data
  \cite{Kar:2008zza,Aaltonen:2010rm}. The invariant mass window for the lepton
  pair for this measurement is $70<m_{\ell^+\ell^-}<110$, in GeV. Tracks with
  $p_T > 0.5$ GeV inside $|\eta|<1$ were included, with the same
  definition of stable charged tracks as above. The leptons from the
  decaying boson were not included. 

The agreement between the Perugia
min-bias tunes and data is at the same level as that of more
dedicated UE tunes, here represented by DW and Pro-Q2, 
supporting  the assertion made earlier concerning the good
universality properties of the \Py\ modelling.  
We note also that the Perugia 2010 variation agrees slightly better with
the data in the TRANSVSERSE region, where it has a bit more activity
than Perugia 0 does. 

\begin{figure}
\begin{center}\vspace*{-5mm}
\scalebox{1.41}{
\includegraphics*[scale=0.34]{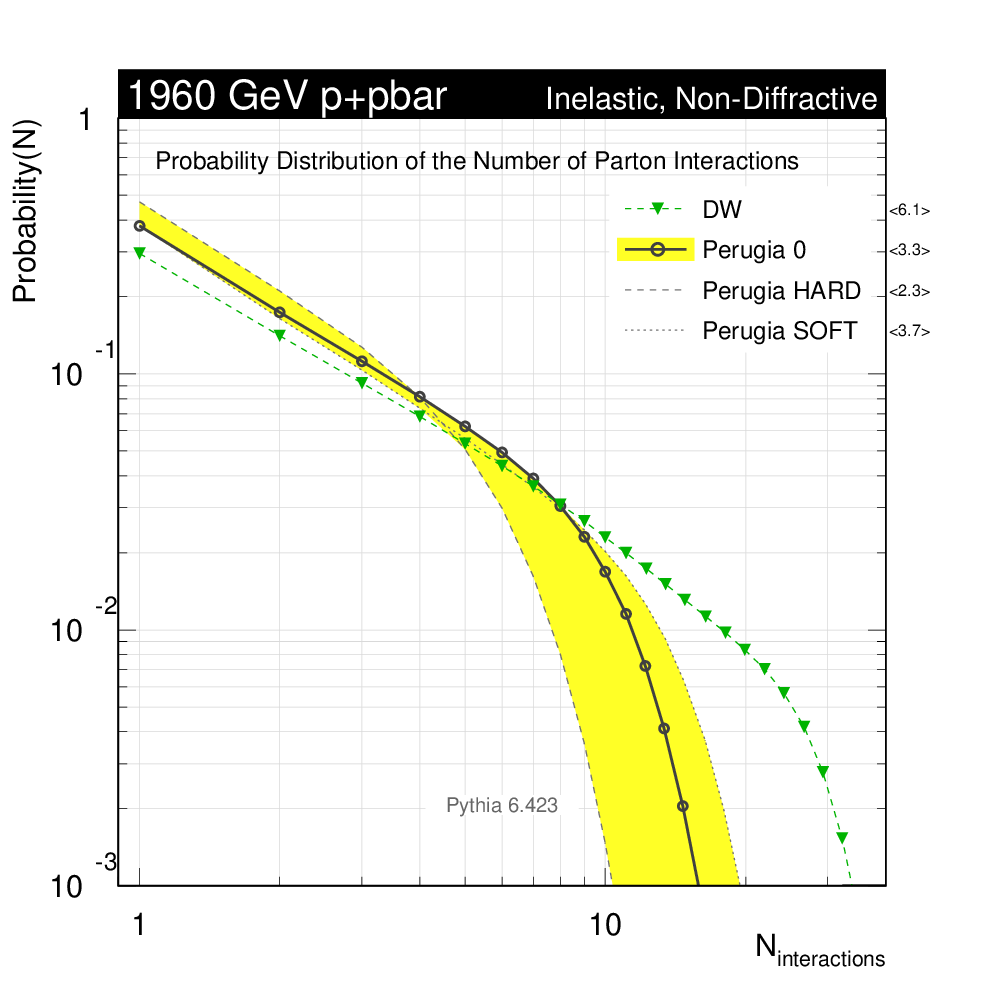}}\vspace*{-3mm}
\caption{\small Double-logarithmic plot of the probability distribution of the number of
  parton-parton interactions in min-bias collisions at the Tevatron,
  showing that the Perugia tunes obtain the same multiplicity 
  distribution, fig.~\ref{fig:tevatronNCH}, with fewer MPI than Tune
  A. 
See
  \cite{lhplots} for 
  other tunes and collider energies. 
\label{fig:theoryplots}}
\end{center}
\end{figure}

\paragraph{Transverse Mass Distribution and MPI Showers}
Finally, the old
framework did not include showering off the MPI
in- and out-states\footnote{It did, of course, include showers off the
  primary interaction. An option to include FSR off the MPI also in
 that framework has since been implemented by S.~Mrenna, see \cite{updatenotes}, 
but tunes using that option have not yet been
made.}. The new framework does include such showers, which furnish
an additional fluctuating physics component. Relatively speaking, 
the new framework therefore  needs \emph{less}
fluctuations from other sources in order to describe the
same data. This is reflected in the tunes of the new framework
generally having a less lumpy proton (smoother proton transverse
density distributions) 
and fewer total numbers of MPI than the old one. This is illustrated
in fig.~\ref{fig:theoryplots}, where a
double-logarithmic scale has been chosen in order 
to reveal the asymptotic behaviour
more clearly. 
Note that, e.g., for Tune A, the plot shows that more than
a per mille of min-bias events have over 30 perturbative parton-parton
interactions per event  at the Tevatron. This
number is reduced by a factor of 2 to 3 in the new models, while the
average number of interactions, indicated on the r.h.s.~of the plot,
goes down by slightly less.

The showers off the MPI also lead to a greater
degree of decorrelation and $\pT{}$ imbalance 
between the minijets produced by the
underlying event, in contrast to the old framework where these 
remained almost exactly balanced and back-to-back. This should show up
in minijet $\Delta\phi_{jj}$ and/or $\Delta R_{jj}$ distributions sensitive to the
underlying event, such as in $Z/W$+multijets with low $\pT{}$ cuts on the
additional jets. It should also show up as a relative enhancement in
the odd components of Fourier transforms of $\phi$ distributions \`a
la \cite{Campanelli:2009hc}. 

\paragraph{Long-Range Correlations}
Further, 
since showers tend to produce shorter-range correlations than MPI, the
new tunes also exhibit smaller long-range correlations than did the old
models. That is, if 
there is a large fluctuation in one end of the detector, it is
\emph{less} likely in the new models that there is a large fluctuation in
the same direction in the other end of the detector. The impact of
this on the overall modelling, and on correction procedures derived from
it, has not yet been studied in great detail. One variable which can 
give direct experimental information on the 
correlation strength over both short and long 
distances is the so-called forward-backward correlation, $b$, defined as in
\cite{Sjostrand:1987su,Alexopoulos:1995ft} 
\begin{equation}
b(\eta_F) = \frac{\left<n_Fn_B\right> - \left<n_F\right>^2}{
\left<n_F^2\right> - \left<n_F\right>^2~}~,
\end{equation}
where $n_F$ and $n_B$ are the 
number of tracks (or a calorimetric measure of energy deposition) in a
pseudorapidity bin centred at $\eta_F$ and $\eta_B = -\eta_F$,
respectively, for a given event. The averages indicate averaging
over the number of recorded events. The resulting
correlation strength, 
$b$, can be plotted either as a function of $\eta_F$ or as a function
of the distance, $\Delta\eta$, between the bins. 
\begin{figure}
\begin{center}\vspace*{-5mm}
\scalebox{1.41}{
\includegraphics*[scale=0.34]{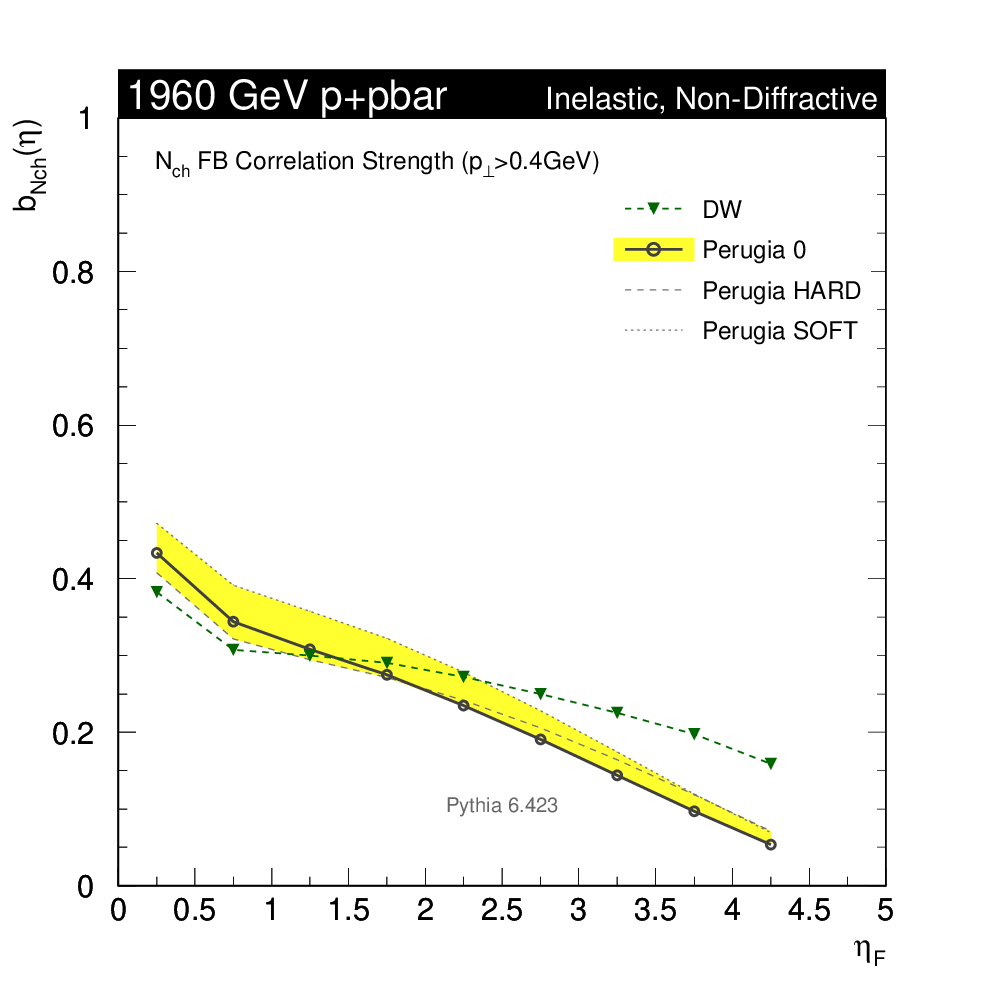}\hspace*{-3mm}
\includegraphics*[scale=0.34]{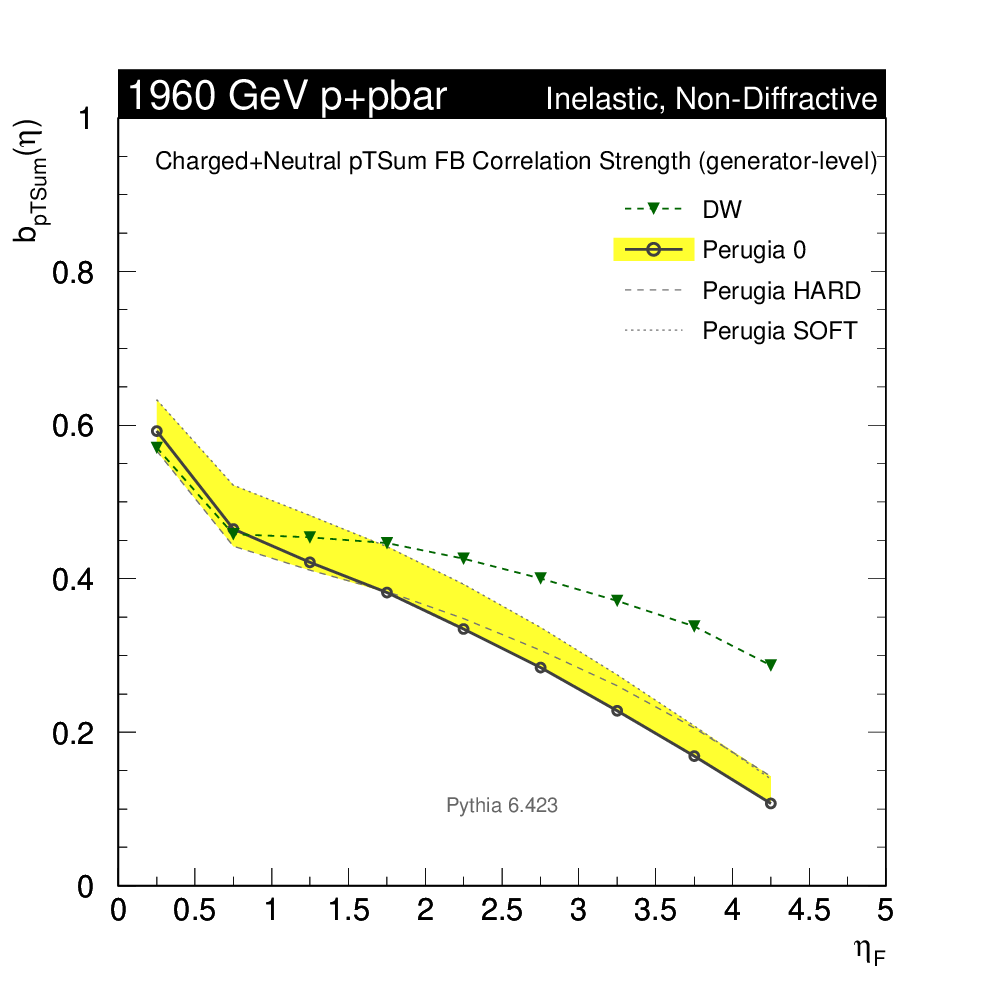}}\vspace*{-3mm}
\caption{\small Forward-Backward correlation strengths at the Tevatron in
  ({\sl left}) charged particles and ({\sl right}) charged plus neutral
  transverse-momentum sum at generator level.
See
  \cite{lhplots} for 
  other tunes and collider energies. 
\label{fig:bFB}}
\end{center}
\end{figure} 
A comparison of the main Perugia tunes to Tune DW
is shown in fig.~\ref{fig:bFB}, for two different
variants of the correlation strength: the plot on the left only
includes charged particles with $\pT{}>0.4$ GeV and the other (right)
includes all energy depositions  (charged plus neutral) 
that would be recorded by an idealized calorimeter. 
Since estimating the impact on the latter of a real (noisy)
calorimeter environment would go beyond the scope of this paper, we
here present the correlation at generator level. For the former, we
show the behaviour out to $\eta=5$ although the CDF and
D\O\ detectors would of course be limited to measuring it
 inside the region $|\eta|<1.0$. Note that a measurement of this
 variable would also be a prerequisite for combining the $dN/d\eta$
 measurements from negative and positive $\eta$ regions to form $dN/d|\eta|$, 
 with the proper correlations taken into account. This particular
 application of the $b$ measurement would require a measurement of $b$
 with the same bin sizes as used for $dN/d\eta$. Since the
 amount of correlation 
 depends on the bin size used (smaller bin sizes are more sensitive to 
 uncorrelated fluctuations), we would advise to perform the $b$
 measurement using several different bin sizes, ranging from a very 
 fine binning (e.g., paralleling that of the $dN/d\eta$ measurement), to
 very wide bins (e.g., one unit in pseudorapidity as used in
 \cite{Sjostrand:1987su}). For our plots here, we used an intermediate-sized 
 binning of 0.5 units in pseudorapidity. 

\subsection{Energy Scaling  (Table \ref{tab:uebrcr})}

A final difference with respect to the older S0(A) family of tunes 
is that we here include data from different
colliders at different energies, in an attempt to fix the energy
scaling better. 

The energy scaling of min-bias and underlying-event phenomena, in
both the old and new \textsc{Pythia} models, 
is driven largely by a single parameter,
the scaling power of the infrared regularization scale for the
multiple parton interactions, $\pT{0}$, see, e.g.,
\cite{Sjostrand:1987su,Sjostrand:2004pf,Sjostrand:2006za}. This
parameter is assumed to scale with the collider CM energy squared, 
$s$, in the following way,
\begin{equation}
\pT{0}^2(s) = \pT{0}^2(s_{\mrm{ref}}) \left(\frac{s}{s_{\mrm{ref}}}\right)^{P_{90}}~,
\end{equation}
where $\pT{0}^2(s_{\mrm{ref}})$ is the IR regularization scale
given at a specific reference $s=s_{\mrm{ref}}$,  
and $P_{90}$ sets the scaling away
from $s=s_{\mrm{ref}}$. In the code, $\pT{0}^2(s_{\mrm{ref}})$ is
represented by \ttt{PARP(82)}, $\sqrt{s}_{\mrm{ref}}$ by
\ttt{PARP(89)}, and $P_{90}$ by \ttt{PARP(90)}. Note that large values
of $P_{90}$ produce a \emph{slower} rate of increase in the overall
activity with collider energy 
than low values, since the generation of additional parton-parton
interactions in the underlying event is suppressed below
$\pT{0}$.

\begin{figure}[t]
\begin{center}
\includegraphics*[scale=0.34]{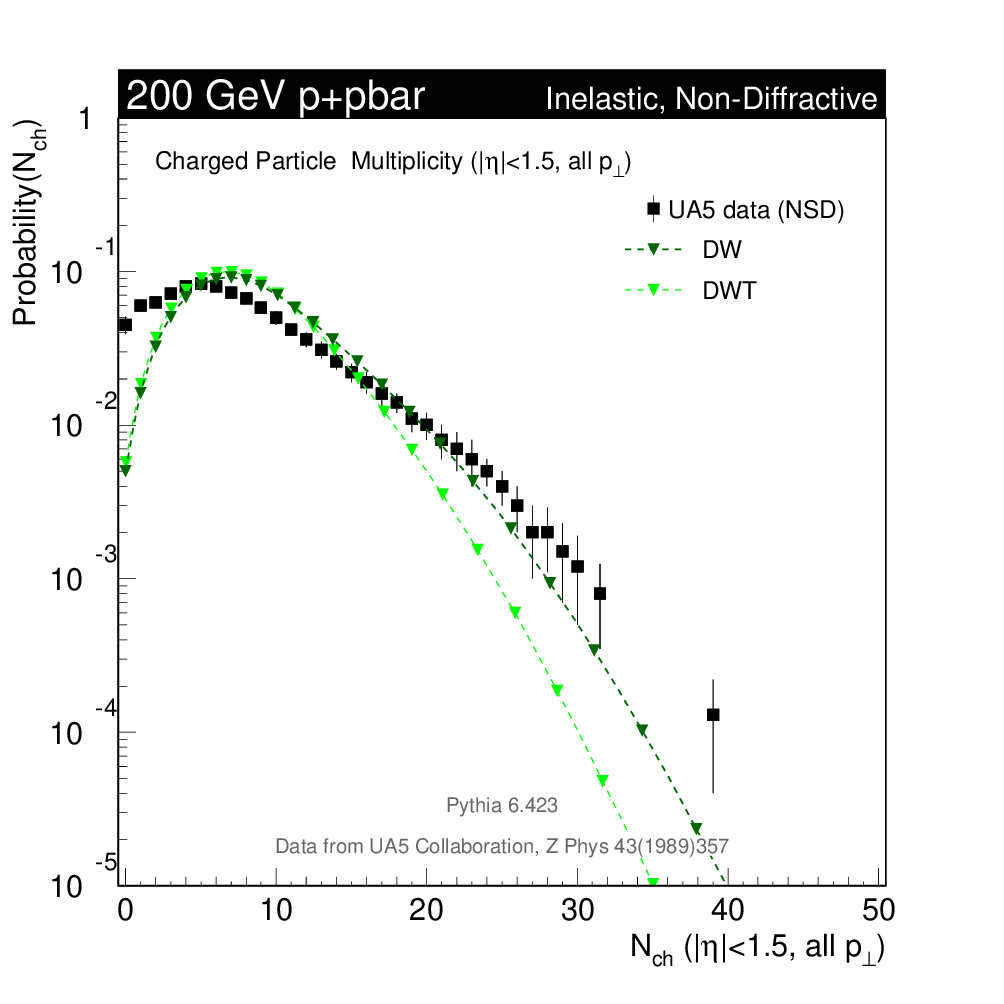}\hspace*{-5mm}
\includegraphics*[scale=0.34]{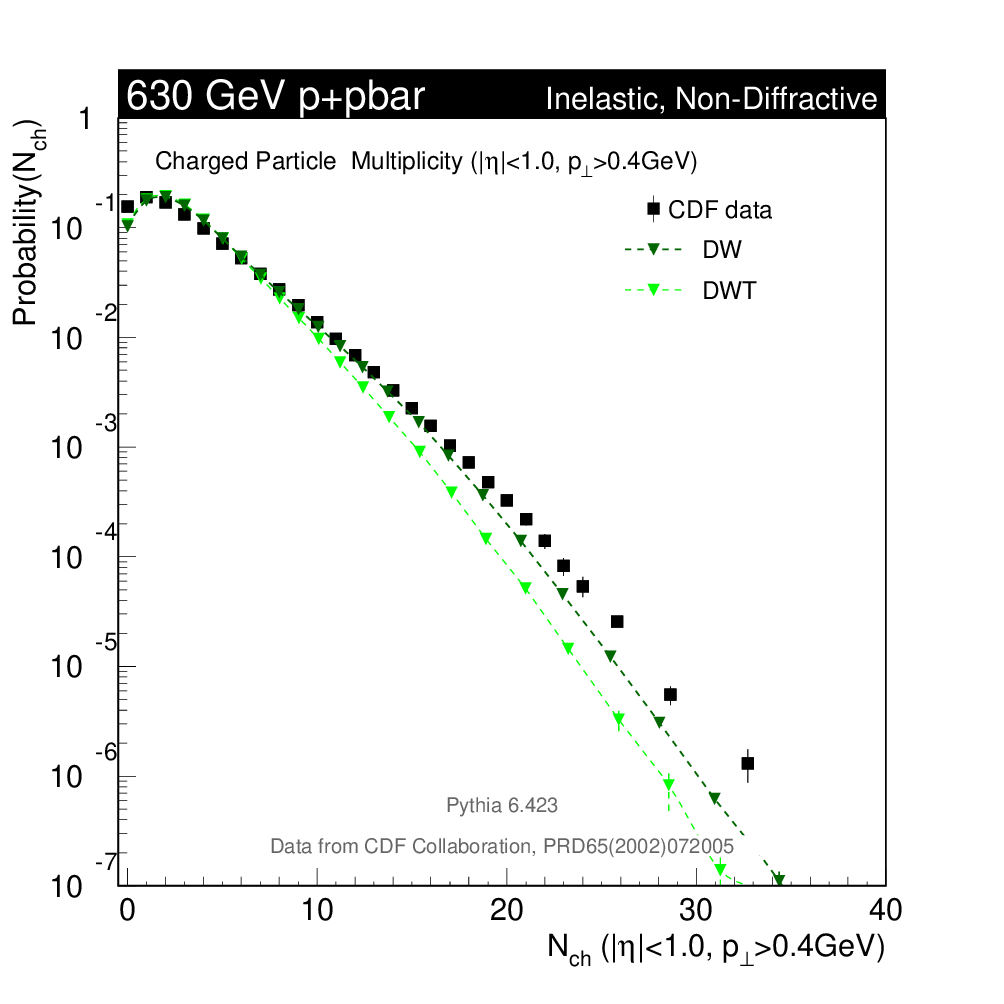}\hspace*{-5mm}
\includegraphics*[scale=0.34]{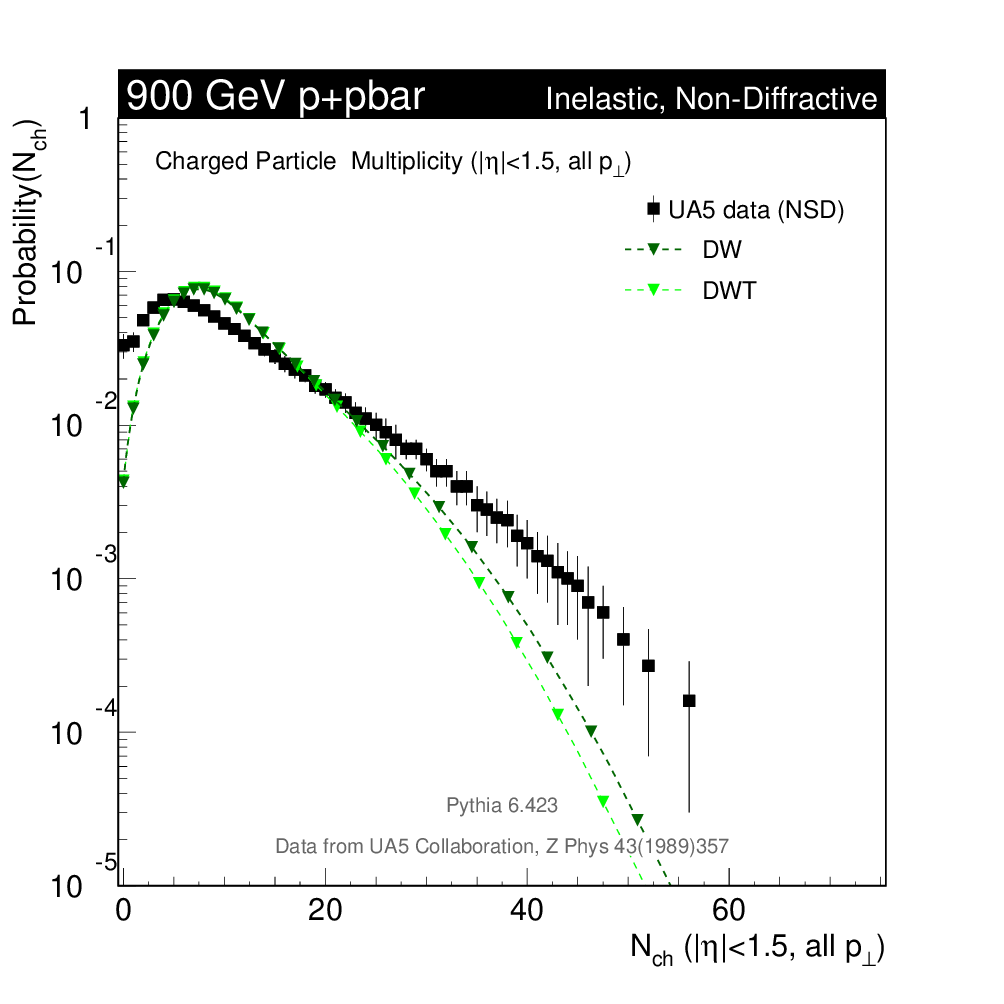}\hspace*{-8mm}\\[-5mm]
\includegraphics*[scale=0.34]{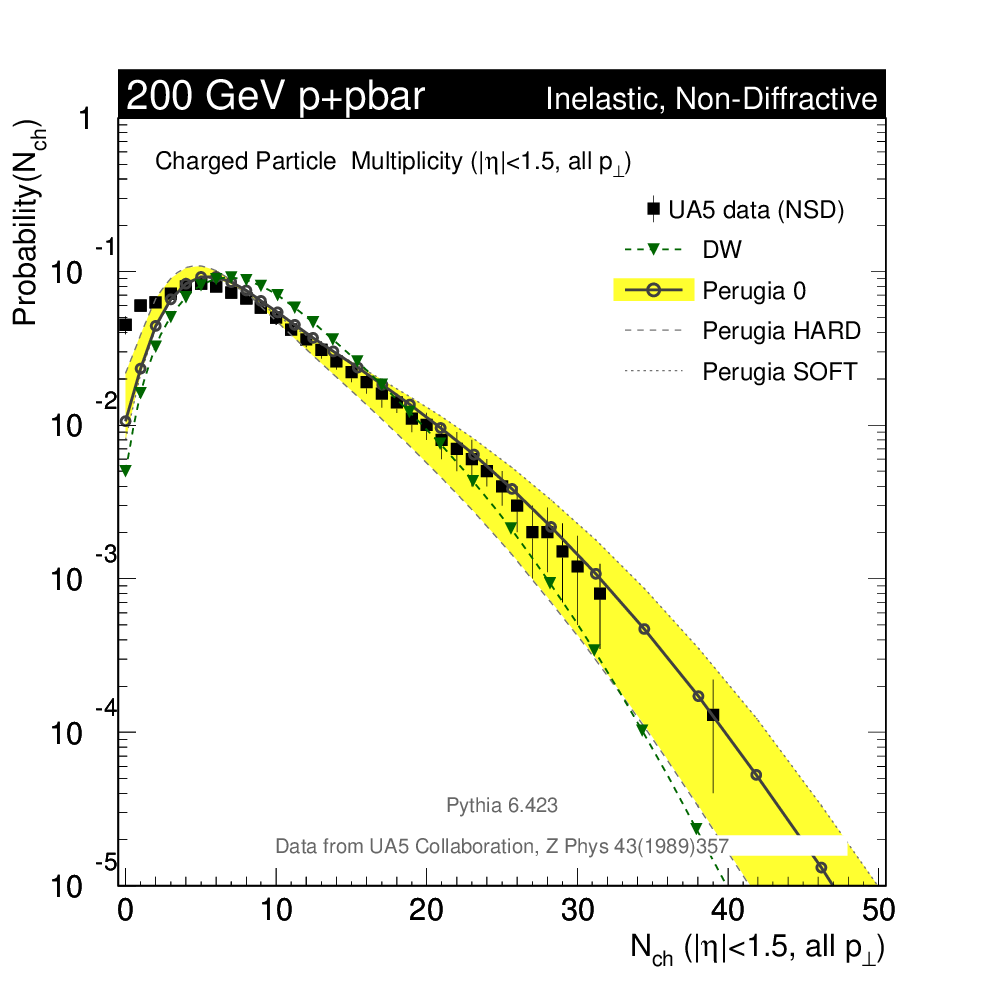}\hspace*{-5mm}
\includegraphics*[scale=0.34]{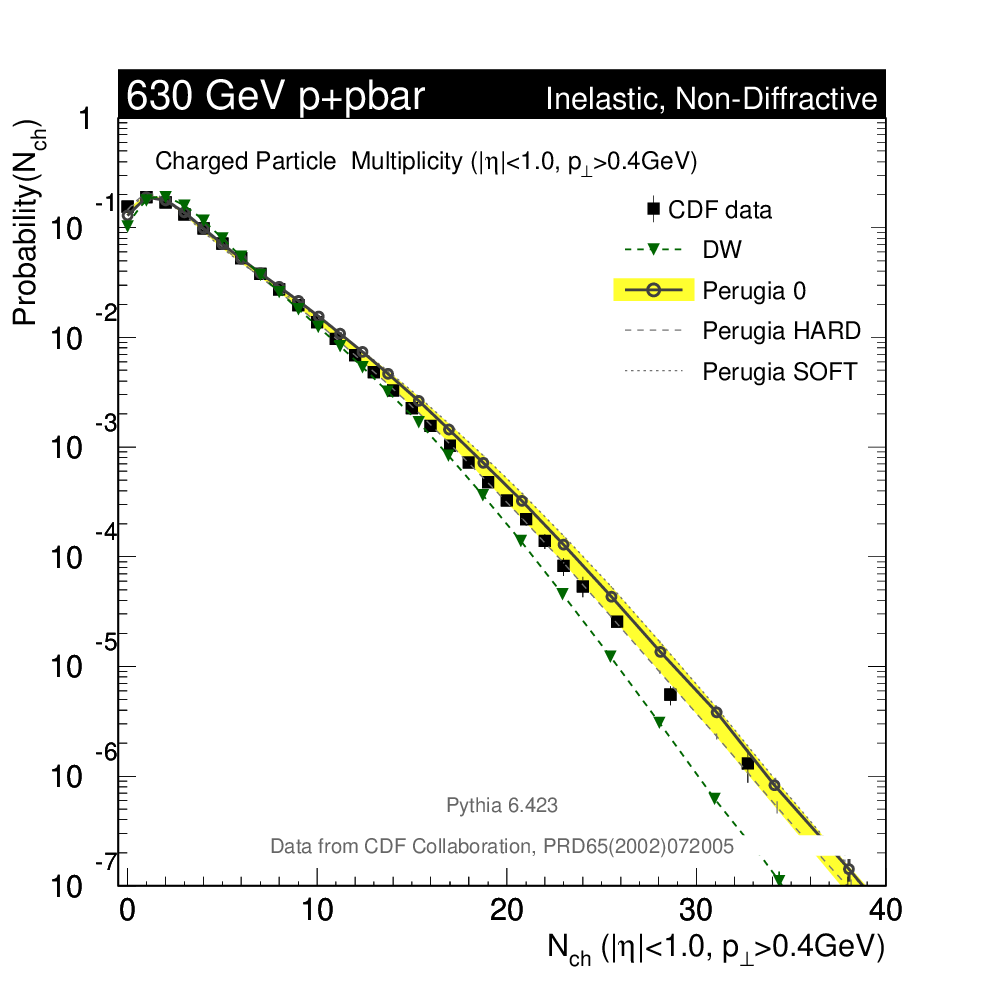}\hspace*{-5mm}
\includegraphics*[scale=0.34]{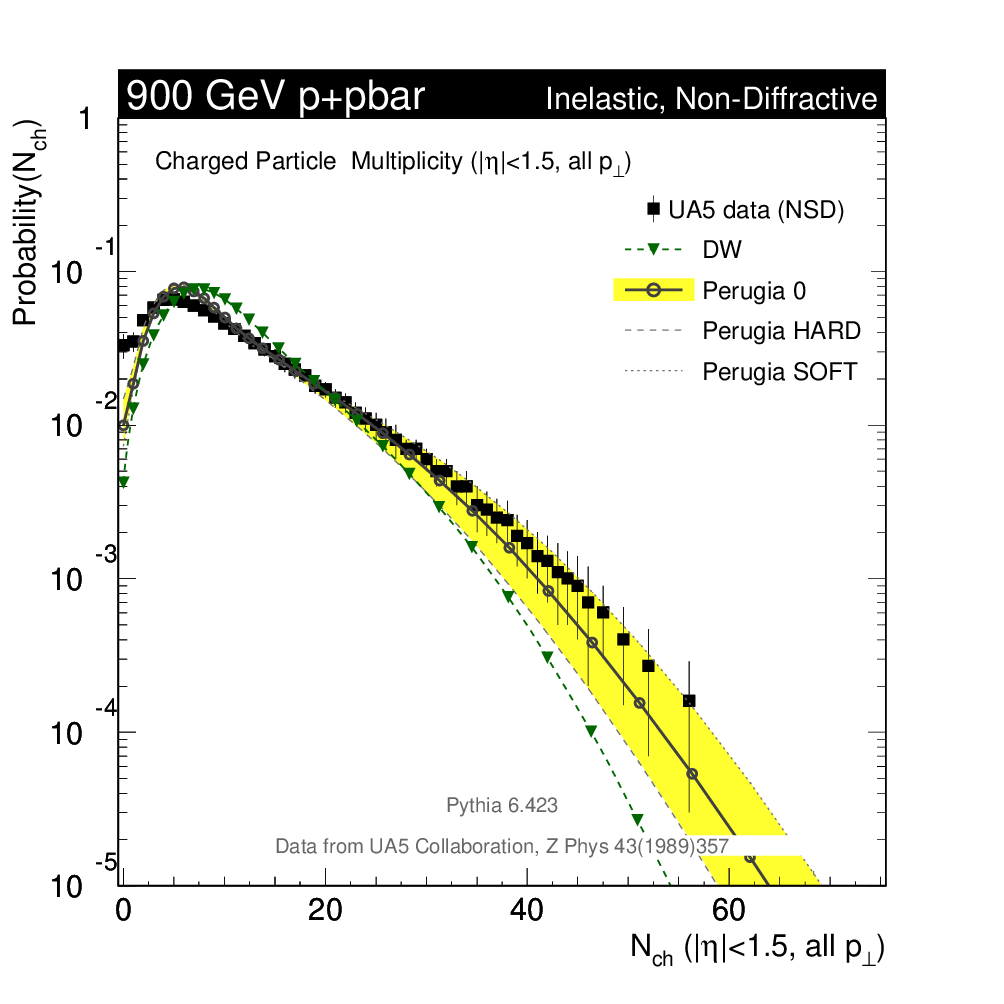}\hspace*{-8mm}\\[-3mm]
\caption{\small Comparisons to UA5 and CDF measurements of
 the charged track multiplicity in minimum-bias $p\bar{p}$ collisions 
at 200 GeV ({\sl left}), 630 GeV ({\sl middle}), and  900 GeV ({\sl right}).  
{\sl Top:} Rick Field's tunes DW and DWT. {\sl Bottom:} the  main
Perugia variations compared to DW. See
  \cite{lhplots} for 
  other tunes and collider energies. 
\label{fig:escaling}}
\end{center}
\end{figure}

The default value for the scaling power in \textsc{Pythia} 6.2 was 
$P_{90}=0.16$, motivated \cite{Sjostrand:1987su}
by relating it to the scaling of the total cross section, which grows
like  $\propto E_\mrm{cm}^{0.16}$. When comparing to Tevatron data
at 630 GeV, Rick Field found that this resulted in too little activity
at that energy, as  illustrated in the 
top row of fig.~\ref{fig:escaling}, where tune DWT uses the old
default scaling away from the Tevatron and DW uses Rick Field's 
value of $P_{90}=0.25$.
(The total cross section is still obtained from a Donnachie-Landshoff
fit \cite{Donnachie:1992ny} and is not affected by this change.)  
Note that the lowest-multiplicity bins of
the UA5 data in particular 
and the first bin of the CDF data were ignored for our comparisons
here, since these
contain a large diffractive component, which has not been simulated in
the model comparisons.  

For the Perugia tunes, the main variations of which are 
shown in the bottom row of fig.~\ref{fig:escaling}, we find that a
large range of values, between $0.22$ and $0.32$, can be
accommodated without ruining the agreement with the available data, 
with Perugia 0 using $0.26$. 

The
energy scaling is therefore still a matter of large uncertainty, and
the possibility of getting good additional constraints from the early LHC
data is encouraging. The message so far appears to be contradictory,
however, with early ATLAS results at 900 GeV \cite{Aad:2010rd} 
appearing to confirm the tendency of the
current tunes to undershoot the high-multiplicity tail at 900 GeV (see
the right-hand column of fig.~\ref{fig:escaling}), which
would indicate a slower scaling between 900 and 1800 GeV than what is
generated by the models (since they all fit well at 1800 GeV) but
preliminary CMS results on the average multiplicity 
at 2360 GeV  \cite{Collaboration:2010xs} indicate the opposite, 
that the pace of evolution in the models is actually too
slow. Furthermore, the CDF data at 630 GeV and the UA5 data at 200 GeV
provide additional constraints at lower energies 
which have made it difficult for us to
increase the tail at 900 GeV without coming into conflict with at
least one of these other data sets. In view of these tensions, we
strongly recommend future studies to include comparisons at different
energies. 

One issue that can be clearly separated out in this discussion, however, 
is that the average multiplicity is sensitive to  ``contamination''
from events of diffractive origin, while the high-multiplicity tail
is not, and hence a different scaling behaviour with energy (or just a different
relative fraction?) of diffractive vs.\ non-diffractive
events may well generate differences between the scaling behaviour of
each individual moment of the multiplicity distribution. Attempting to
pin down the scaling behaviour moment by moment would therefore also be
an interesting possible study. Since the \Py\ 6
modelling of diffraction is relatively crude, however, we did not attempt to
pursue this question further in the present study, but note that 
a discussion of whether these tendencies could be
given other meaningful physical 
interpretations, e.g., in terms of low-$x$, saturation, and/or
unitarization effects, would be interesting to follow up on.

It should be safe to conclude, however, that there is clearly a need
for more systematic examinations of the energy scaling behavior, both
theoretically and experimentally, for both diffractive and
non-diffractively enhanced event topologies separately. It would also
be interesting, for instance, to attempt to separately determine the
scaling behaviours for low-activity/peripheral events and for
active/central events, e.g., by considering the scaling of the various
moments of the multiplicity distribution and by other observables
weighted by powers of the event multiplicity. 

\section{The Perugia Tunes: Tune by Tune}

The starting point for all the Perugia tunes, apart from Perugia NOCR,
was S0A-Pro, i.e., the original tune ``S0''
\cite{Sjostrand:2004pf,Sjostrand:2004ef,Sandhoff:2005jh,Skands:2007zg},
with the Tune A energy scaling (S0A), revamped to
include the Professor tuning of flavour and fragmentation parameters
to LEP data \cite{Buckley:2009vk,Buckley:2009bj} (S0A-Pro). 
The starting point for Perugia NOCR was 
NOCR-Pro. From these starting points, the main hadron collider parameters
were retuned to better describe the data sets described above. 

As in previous versions, each tune is associated with a
3-digit number which can be given in \ttt{MSTP(5)} as a convenient
shortcut. A complete overview of the Perugia tune parameters is given
in appendix \ref{sec:tables} and a list of all the predefined tunes
that are included with \textsc{Pythia} version 6.423 can be found in 
appendix \ref{sec:tunes}. 

\begin{figure}[t]
\begin{center}\vspace*{-5mm}
\scalebox{1.}{
\includegraphics*[scale=0.34]{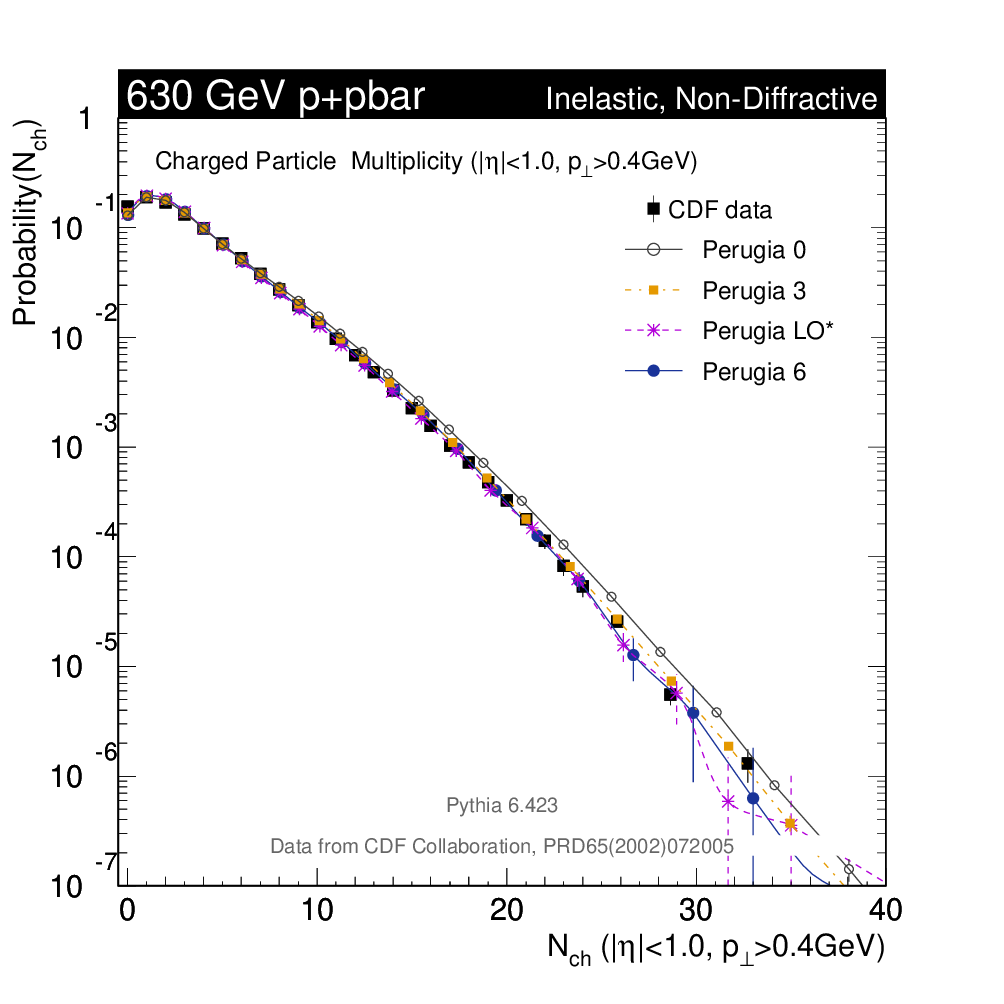}\hspace*{-3mm}
\includegraphics*[scale=0.34]{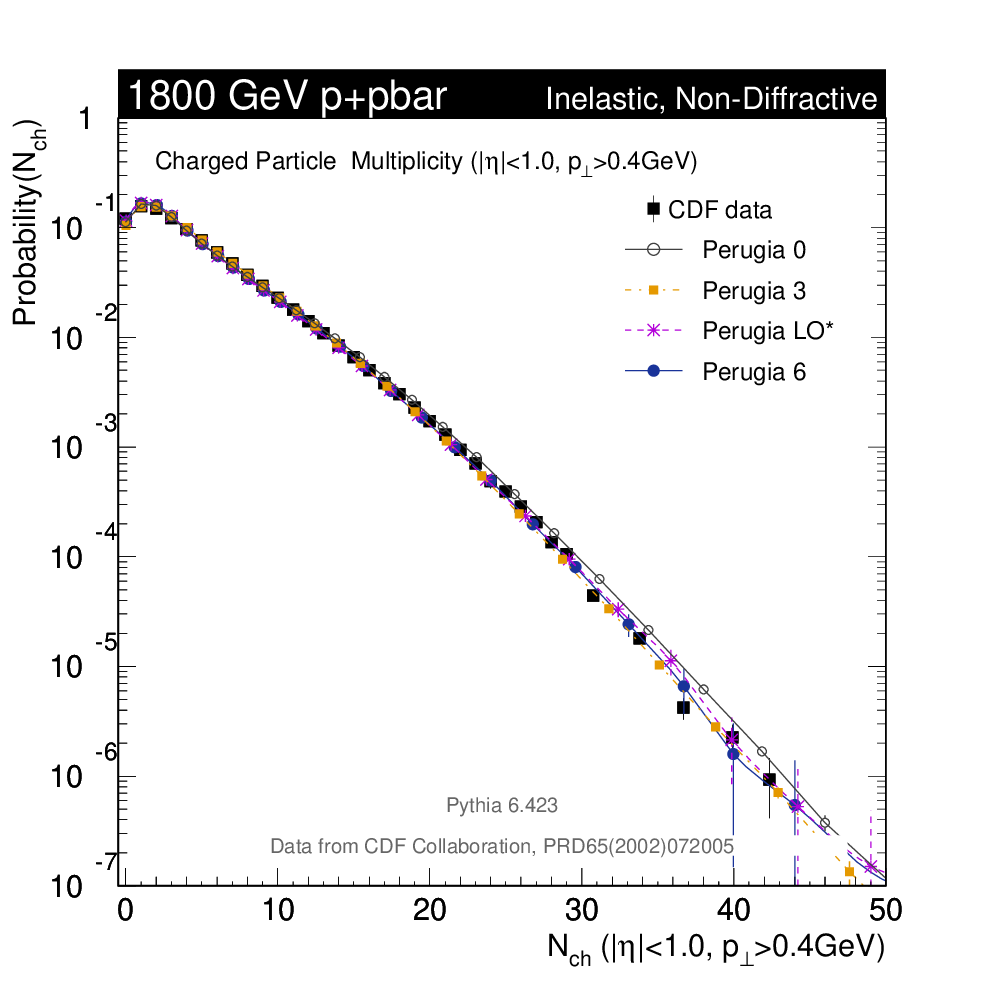}\hspace*{-3mm}
\includegraphics*[scale=0.34]{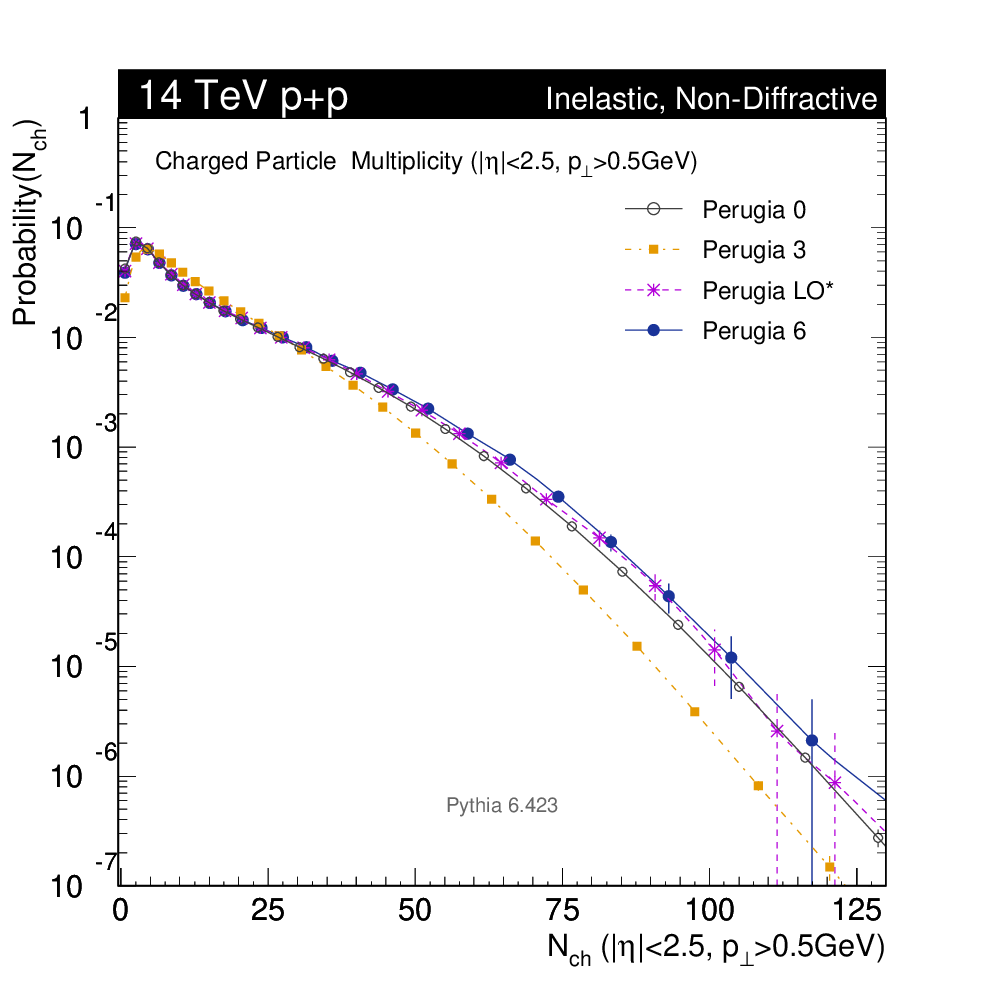}}\vspace*{-3mm}\\
\scalebox{1.}{
\includegraphics*[scale=0.34]{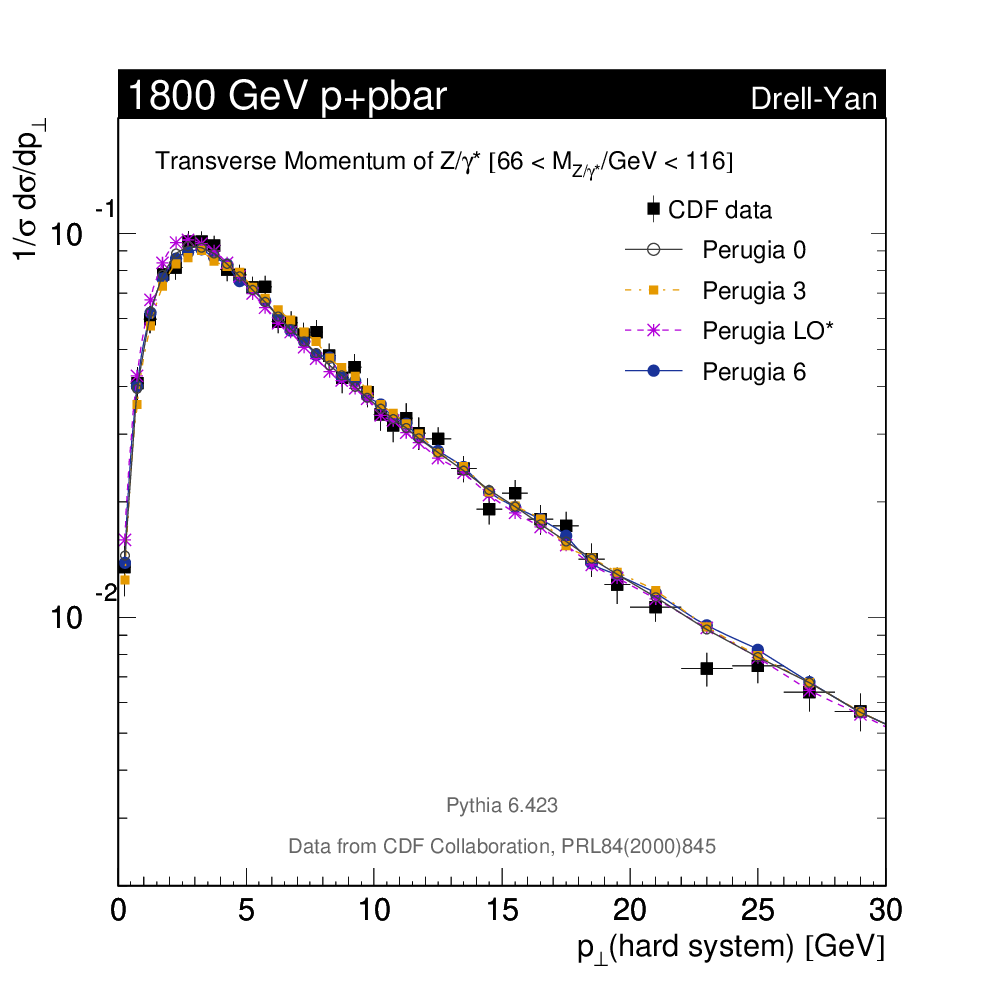}\hspace*{-3mm}
\includegraphics*[scale=0.34]{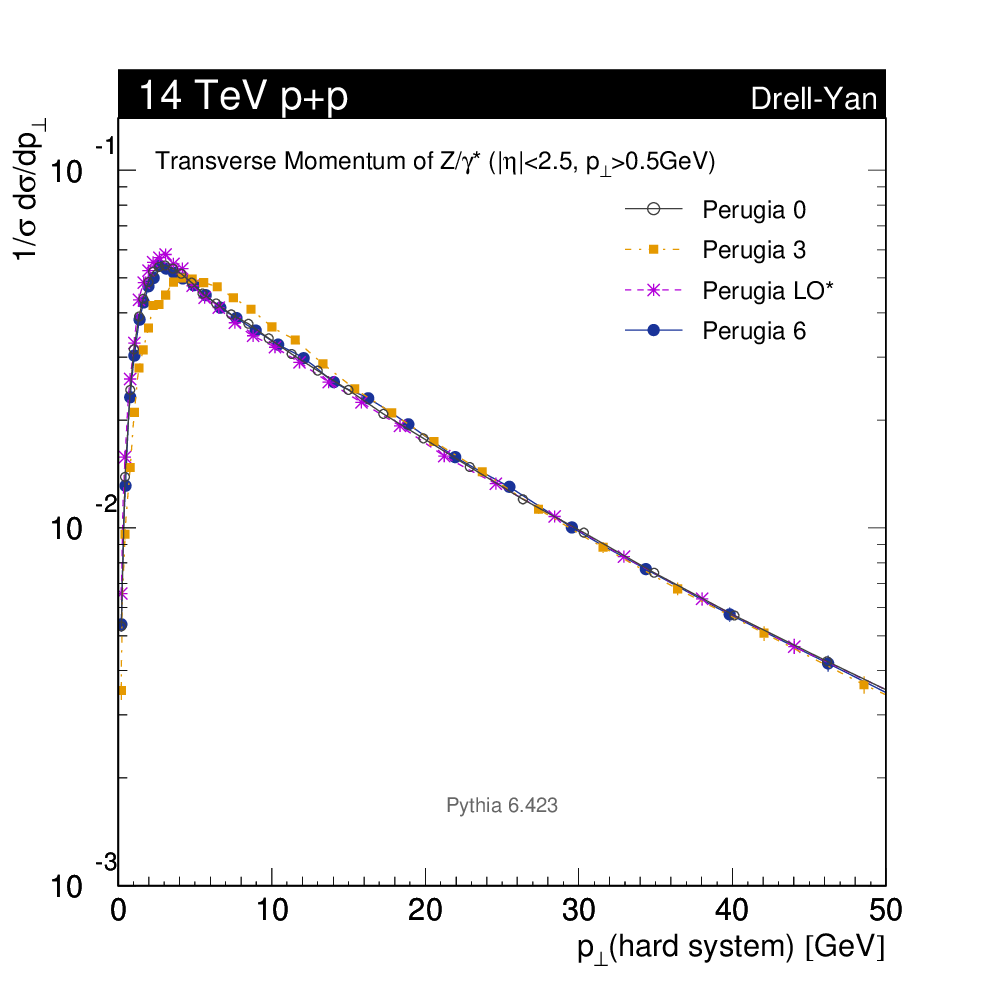}}\vspace*{-3mm}
\caption{\small Charged particle multiplicity and 
Drell-Yan $p_\perp$ spectra at the Tevatron 
  ({\sl left}) and at the LHC at 14 TeV  ({\sl right}) for
the Perugia 0, 3, LO*, and 6 tunes.  In particular, the Perugia 3
curve on the lower right-hand plot illustrates the consequences of
choosing a different regularization procedure for ISR in the infrared,
which shifts the position of the infrared peak of the Drell-Yan
$p_\perp$ spectrum without affecting the
tail of the distribution.
\label{fig:tune323}}
\end{center}
\end{figure}

\paragraph{Perugia 0 (320): } Uses CTEQ5L parton distributions
\cite{Lai:1999wy} (the default in \textsc{Pythia} and the most recent
set available in the standalone version --- see below for Perugia
variations using external CTEQ6L1 and MRST LO* distributions). 
Uses $\Lambda_\mrm{CMW}$ \cite{Catani:1990rr} 
instead of $\Lambda_{\overline{\mrm{MS}}}$,
which results in near-perfect agreement with the Drell-Yan $p_\perp$
spectrum, both in the tail and in the peak, cf.~fig.~\ref{fig:tevatronDY}.  
Also has slightly less colour reconnections than S0(A), especially among 
high-$\pT{}$ string pieces, which improves the agreement both with 
the $\left<\pT{}\right>(N_\mrm{ch})$ distribution and with the
high-$\pT{}$ tail 
of charged particle $\pT{}$ spectra, cf~\cite[dN/dpT (tail)]{lhplots}). 
Slightly more beam-remnant breakup than S0(A) (more baryon number transport),
mostly in order to explore this possibility than due to any necessity
of tuning at this point. 
Without further changes, these modifications would lead to a greatly
increased average multiplicity as well as larger multiplicity
fluctuations. To keep the total multiplicity unchanged, relative to
S0A-Pro, the
changes above were accompanied by an increase in the MPI infrared
cutoff, \pT{0}, which decreases the overall MPI-associated activity, and 
by a slightly smoother proton mass profile, which decreases the
fluctuations. Finally, the energy scaling is closer to that of Tune A
(and S0A) than to the old default scaling that was used for S0.
\paragraph{Perugia HARD (321): } A variant of Perugia 0 which has a
higher amount of activity from perturbative physics and
counter-balances that partly by having less particle production from
nonperturbative sources. Thus, the $\Lambda_{{\mrm{CMW}}}$ 
value is used for ISR, together with a
renormalization scale for ISR of $\mu_R=\frac12\pT{}$, yielding a
comparatively hard Drell-Yan $\pT{}$ spectrum,
cf.~the dashed curve labeled ``HARD'' in the 
right pane of fig.~\ref{fig:tevatronDY}. It also has a
slightly larger phase space 
for both ISR and FSR, uses higher-than-nominal values for
FSR, and has a slightly harder hadronization. To partly counter-balance
these choices, it has less ``primordial $k_T$'', a higher IR
cutoff for the MPI, and more active colour reconnections,
yielding a comparatively high curve for
$\left<\pT{}\right>(N_\mrm{ch})$,
cf.~fig.~\ref{fig:tevatronAVG}. Warning: this tune has more
ISR but also more FSR. The final number of reconstructed jets may
therefore not appear to change very much, and if the number of ISR jets is
held fixed (e.g., by matching), this tune may even produce
\emph{fewer} events, due to the increased broadening. For a full
ISR/FSR systematics study, the amount of ISR and FSR should be changed
independently. 
\paragraph{Perugia SOFT (322): } A variant of Perugia 0 which has a
lower amount of activity from perturbative physics and makes up for it
partly by adding more particle production from nonperturbative sources. 
Thus, the $\Lambda_{\overline{\mrm{MS}}}$ value is used for ISR, 
together with a
renormalization scale of $\mu_R=\sqrt{2}\pT{}$, yielding a
comparatively soft Drell-Yan $\pT{}$ spectrum,
cf.~the dotted curve labeled ``SOFT'' in the right pane of 
fig.~\ref{fig:tevatronDY}. It also has a slightly smaller phase space
for both ISR and FSR, uses lower-than-nominal values for
FSR, and has a slightly softer hadronization. To partly counter-balance
these choices, it has a more 
sharply peaked proton mass distribution, a more active beam remnant
fragmentation, a slightly lower IR
cutoff for the MPI, and slightly less active colour reconnections,
yielding a comparatively low curve for
$\left<\pT{}\right>(N_\mrm{ch})$,
cf.~fig.~\ref{fig:tevatronAVG}. Again, a more complete variation would
be to vary the amount of ISR and FSR independently, at the price of
introducing two more variations (see  above). 
We encourage users that desire a complete ISR/FSR systematics study
to make these additional variations on their own. 
\paragraph{Perugia 3 (323): } A variant of Perugia 0 which has a
different balance between MPI and ISR and a different energy
scaling. Instead of a smooth dampening of ISR all the way to zero
$\pT{}$, this tune uses a sharp cutoff at 1.25 GeV, which produces
a slightly harder ISR spectrum. The additional ISR activity is
counter-balanced by a higher infrared MPI cutoff. Since the ISR cutoff
is independent of the collider CM energy in this tune, 
the multiplicity would nominally evolve very rapidly with energy. To 
offset this, the MPI cutoff itself must scale very quickly, hence
this tune has a very large value of the scaling power of that
cutoff. This leads to an interesting systematic difference in the
scaling behavior, with ISR becoming an increasingly more important 
source of particle production as the energy increases in this tune,
relative to Perugia 0. This is illustrated in fig.~\ref{fig:tune323},
where we show the scaling of the min-bias charged multiplicity
distribution and the Drell-Yan $p_\perp$ spectrum
between the Tevatron (left) and the LHC at 14 TeV (right).
One sees that, while the overall multiplicity grows less fast with
energy in Perugia 3, the position of the soft peak in Drell-Yan
becomes harder, reflecting the relative increase in ISR, despite the
decrease in MPI. 
\paragraph{Perugia NOCR (324): } An update of NOCR-Pro that attempts
to fit the data sets as well as possible, without invoking any
explicit colour reconnections. Can reach an acceptable agreement with
most distributions, except for the $\left<\pT{}\right>(N_\mrm{ch})$
one, cf.~fig.~\ref{fig:tevatronAVG}. Since there is a large amount of
``colour disturbance'' in the remnant, this tune gives rise to a very
large amount of baryon number transport, even greater than for 
the SOFT variant above. 
\paragraph{Perugia X (325): } A Variant of Perugia 0 which uses the
MRST LO* PDF set \cite{Sherstnev:2007nd}. 
Due to the increased gluon densities, a slightly 
lower ISR renormalization scale and a higher MPI cutoff than for
Perugia 0 is used. Note that, since we are not yet sure the
implications of using LO* for the MPI interactions have been fully
understood, this tune should be considered experimental for the time
being. In fig.~\ref{fig:tune323}, we see that the choice of PDF does
not greatly affect neither the min-bias multiplicity nor the Drell-Yan
$p_\perp$ distribution, once the slight retuning has been done. Thus,
this tune is not intended to differ significantly from Perugia 0, but
only to allow people to explore the LO* set of PDFs without ruining
the tuning. See \cite[Perugia PDFs]{lhplots} for more distributions.
\paragraph{Perugia 6 (326): } A Variant of Perugia 0 which uses the
CTEQ6L1 PDF set \cite{Pumplin:2002vw}. 
Identical to Perugia 0 in all other respects, except
for a slightly lower MPI infrared cutoff at the Tevatron and a
lower scaling power of the MPI infrared cutoff (in other words, the
CTEQ6L1 distributions are slightly lower than the CTEQ5L ones, on
average, and hence a lower regularization scale can be tolerated). 
The predictions
obtained are similar to those of Perugia 0, cf., e.g.,
fig.~\ref{fig:tune323} and \cite{lhplots}.  
\paragraph{Perugia 2010 (327): }  
A variant of Perugia 0 with the amount of FSR outside resonance decays
increased to agree with the level inside them 
(specifically the Perugia-0 value for hadronic $Z$ decays at LEP is
used for FSR also outside $Z$ decays in Perugia 2010, where Perugia
0 uses the lower $\alpha_s$ value derived from the PDFs instead), 
in an attempt to bracket the description of hadronic event shapes 
relative to the comparison of Perugia 0 to NLO+NLL resummations in
\cite{Banfi:2010xy} and also to improve the description of jet shapes 
\cite{Acosta:2005ix}. The total strangeness yield has also 
been increased, since the original
parameters, tuned by Professor, were obtained for the $Q^2$-ordered
shower and small changes were observed when going to the
$\pT{}$-ordered ones. High-$z$ fragmentation has been modified by a 
slightly larger infrared cutoff, which hardens the fragmentation
spectrum slightly. 
The amount of baryon number transport has been increased slightly, 
mostly in order to explore the consequences of the junction
fragmentation framework better\footnote{Although there is room in the model to
increase the baryon 
asymmetry further, this would also increase the frequency of 
multi-junction-junction strings in $p\bar{p}$ events, which \textsc{Pythia} 6 
is currently not equipped to deal with, and hence the strength of this
effect was left at an intermediate level (cf.\ \ttt{PARP(80)} in table
\ref{tab:uebrcr} in appendix \ref{sec:tables}).}, and the colour
reconnection model has been changed to the newest one, \ttt{MSTP(95)=8}. 
See \cite{lhplots} for plots using this tune.  
\paragraph{Perugia K (328): } 
A variant of Perugia 2010 that introduces a ``$K$'' factor on the QCD
$2\to 2$ scattering cross sections used in the
multiple-parton-interaction framework. The $K$-factor applied is set
to a constant value of 1.5. This should make the underlying
event more ``jetty'' and pushes the underlying-event activity towards higher
$\pT{}$. To compensate for the increased activity at higher \pT{}, the
infrared regularization scale is larger for this tune, cf.~table \ref{tab:uebrcr}
in appendix \ref{sec:tables}. It does not give an extremely good
central fit to all data, but represents a theoretically interesting
variation to explore. 

\paragraph{The Perugia 2011 Tunes (350-359):} The 2011 updates of the
Perugia tunes were not included in the original published 
version of this manuscript. For reference, a description of them 
has been included in Appendix \ref{app:2011} of this updated preprint.

\paragraph{The Perugia 2012 Tunes (370-383):} The 2012 updates of the
Perugia tunes were not included in the original published 
version of this manuscript. For reference, a description of them 
has been included in Appendix \ref{app:2012} of this updated preprint.

\section{Extrapolation to the LHC}
\paragraph{``Predictions''}
Part of the motivation for updating the S0 family of tunes was
specifically to improve the constraints on the 
energy scaling to come up with tunes
that extrapolate more reliably to the LHC. 
This is not to
say that the uncertainty is still not large, but as mentioned above,
it does seem that, e.g., the default \textsc{Pythia} scaling is not
able to account for the scaling between 
the lower-energy data sets, and so this is naturally 
reflected in the updated parameters. 

\begin{figure}[t]
\begin{center}\hspace*{-2mm}
\includegraphics*[scale=0.34]{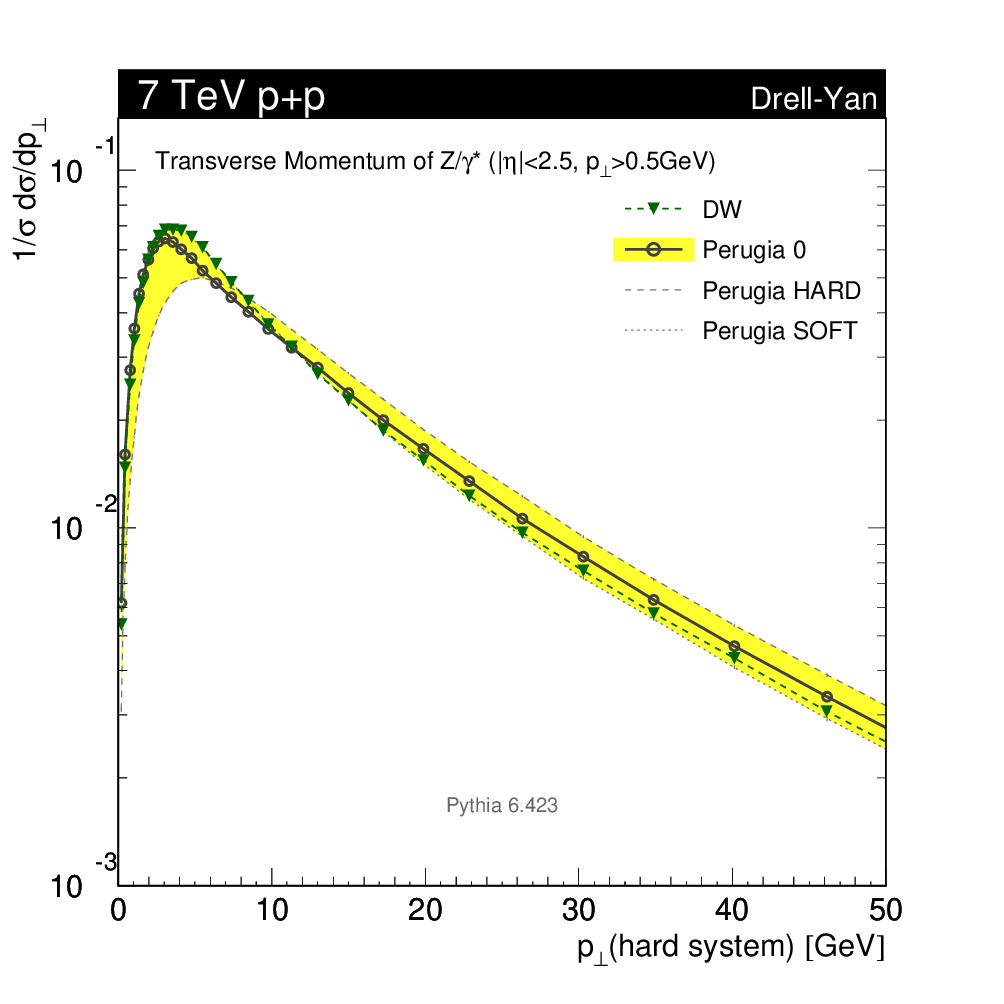}\hspace*{-5mm}
\includegraphics*[scale=0.34]{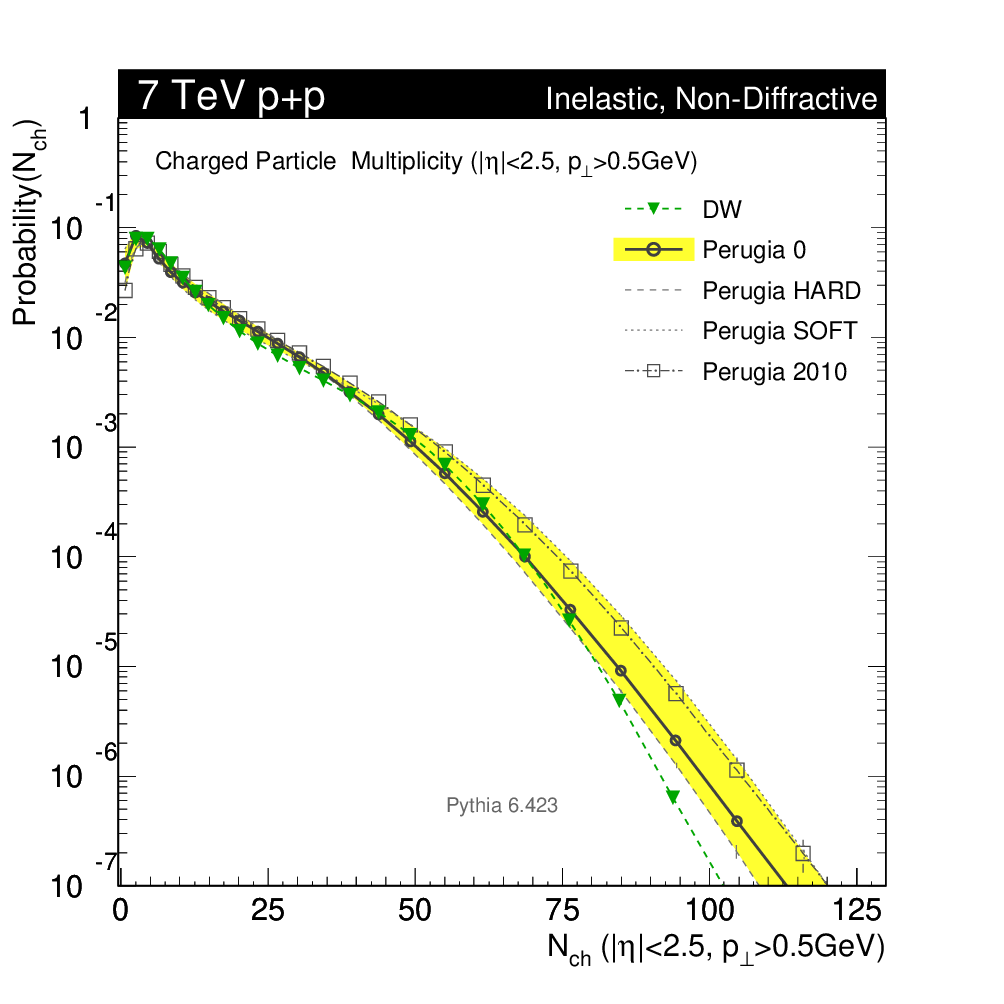}\hspace*{-5mm}
\includegraphics*[scale=0.34]{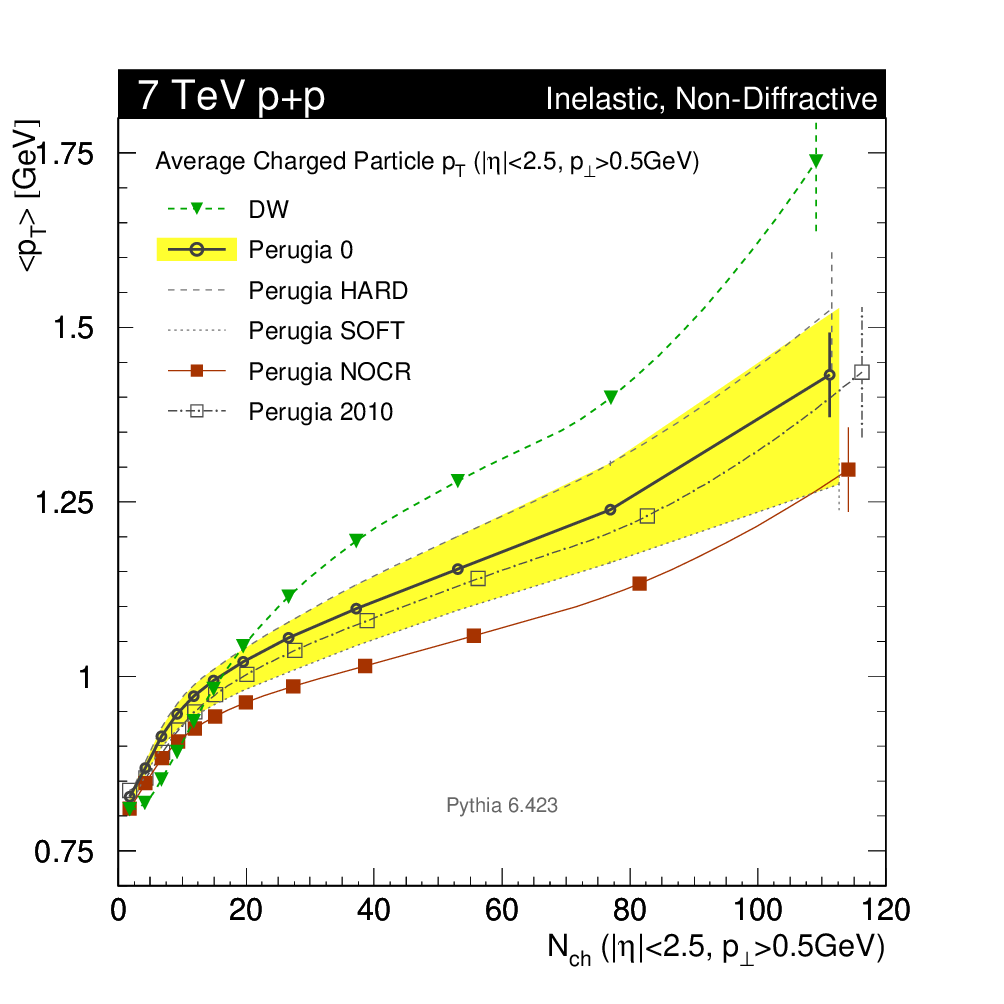}\hspace*{-8mm}\\[-4mm]
\caption{\small Perugia ``predictions'' for the $\pT{}$ of Drell-Yan pairs (left),
  the charged track multiplicity in min-bias (center), and the average
  track $\pT{}$ vs.~$N_{\mrm{ch}}$ in min-bias (right) at the LHC at 7
  TeV. 
See
  \cite{lhplots} for other tunes and collider energies. 
\label{fig:LHC}}
\end{center}
\end{figure}
In fig.~\ref{fig:LHC}, we compare the main Perugia variations to Rick
Field's Tune DW on the Drell-Yan
$\pT{}$ distribution (using the CDF cuts), the charged track
multiplicity distribution in (inelastic, non-diffractive) 
minimum-bias collisions, and the average
track $\pT{}$ as a function of multiplicity at the initial LHC
center-of-mass energy of 7 TeV. We
hope this helps to give a feeling for the kind of ranges spanned by
the Perugia tunes (the PDF variations give almost identical results to
Perugia 0 for these distributions and are not shown. The Perugia 2010
variation gives the same Drell-Yan $p_\perp$ spectrum and is therefore
not shown in the left-hand pane). 
A full set of plots including also the 14 TeV center-of-mass energy, 
for both the central region, $|\eta|<2.5$, and the region 
$1.8<\eta<4.9$ covered by LHCb, can be found on the web
\cite{lhplots}. 

However, in addition to these plots, 
we thought it would be interesting to make at
least one set of numerical predictions for an infrared sensitive
quantity that could be tested with the very earliest high-energy LHC data. 
We therefore used
the Perugia variations to get an estimate for the mean 
multiplicity of charged tracks in (inelastic, non-diffractive) 
minimum-bias $pp$ collisions at center-of-mass energies of 0.9, 2.36,
7, 10, and 14 TeV, as shown in table \ref{tab:predictions}. In order to facilitate
  comparison with data sets that may include diffraction in the
  first few multiplicity bins, we recomputed the means with up to the
  first 4 bins excluded, and model uncertainties were inflated
  slightly for the first two bins. The uncertainty estimates 
  correspond to roughly twice the largest difference between
  individual models and only drop below 10\% near the collider
  energies used to constrain the models and then only when the
  lowest-multiplicity bins are excluded. Note also, however, that the
  uncertainties nowhere become larger than 20\%. This presumably still
  underestimates the full theoretical uncertainty, due to intrinsic
  limitations in our ability to vary the models, but we hope
  nonetheless that it furnishes a useful first estimate. 

\begin{table}[t]
\begin{center}
\begin{tabular}{lrrrrrr}
\multicolumn{6}{c}{\sl Predictions for Mean Densities of Charged
  Tracks (Inelastic, Non-Diffractive Events)} \\[1mm]\hline\\[-3mm]
           & $\displaystyle\frac{\left<N_\mrm{ch}\right>|_{N_{\mrm{ch}}\ge 0}}{\Delta\eta\Delta\phi}$ 
           & $\displaystyle\frac{\left<N_\mrm{ch}\right>|_{N_{\mrm{ch}}\ge 1}}{\Delta\eta\Delta\phi}$
           & $\displaystyle\frac{\left<N_\mrm{ch}\right>|_{N_{\mrm{ch}}\ge 2}}{\Delta\eta\Delta\phi}$
           & $\displaystyle\frac{\left<N_\mrm{ch}\right>|_{N_{\mrm{ch}}\ge 3}}{\Delta\eta\Delta\phi}$
           & $\displaystyle\frac{\left<N_\mrm{ch}\right>|_{N_{\mrm{ch}}\ge 4}}{\Delta\eta\Delta\phi}$
\\[4mm]LHC 0.9 TeV & $0.21 \pm 0.03$ & $0.22 \pm 0.03$ & $ 0.24\pm 0.02$ &
$0.26 \pm 0.02$ & $0.30 \pm 0.02$
\\[1mm]LHC 2.36 TeV & $0.27\pm 0.03$ & $0.28 \pm 0.03$ & $0.30 \pm
0.02$ & $0.33\pm 0.02$ & $0.36 \pm 0.02$
\\[1mm]LHC 7 TeV & $0.36\pm 0.04$ & $0.37\pm 0.04$ & $0.39\pm 0.04$ &
$0.42\pm 0.05$ & $0.46\pm 0.04$
\\[1mm]LHC 10 TeV & $0.40 \pm 0.05$ & $0.41 \pm 0.05$ & $ 0.43\pm 0.05$ &
$0.46\pm 0.06$ & $0.50\pm 0.06$
\\[1mm]LHC 14 TeV & $0.44 \pm 0.06$ & $0.45 \pm 0.06$ & $0.47 \pm 0.06$ 
& $0.51 \pm 0.06$ & $0.54 \pm 0.07$
 \\[1mm]\hline
\end{tabular}
\caption{\small Best-guess predictions for the mean density of charged
  tracks for min-bias $pp$ collisions at several different LHC energies. These
  numbers should be compared to data corrected to 100\% track finding
  efficiency for tracks with $|\eta|<2.5$ and $\pT{}>0.5$ GeV
  and 0\% efficiency outside that region. The definition of a stable
  particle was set at $c\tau \ge 10$mm (e.g., the two tracks
  from a $\Lambda^0\to p^+\pi^-$ decay were not counted). 
  The $\pm$ values represent the estimated uncertainty,
  based on the Perugia tunes. No simulation of diffraction was
  included in these numbers.
\label{tab:predictions}}
\end{center}
\end{table}

\begin{figure}[t]
\begin{center}\hspace*{-2mm}
\includegraphics*[scale=0.34]{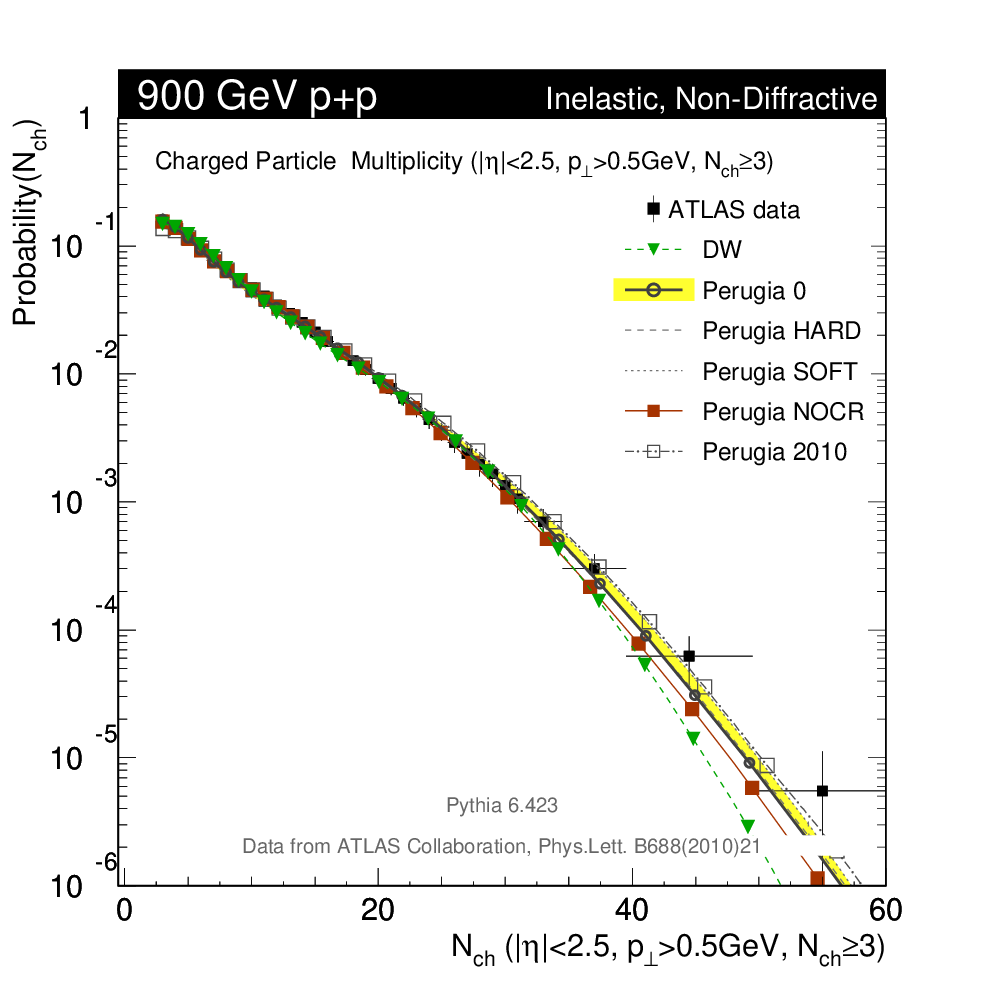}\hspace*{-5mm}
\includegraphics*[scale=0.34]{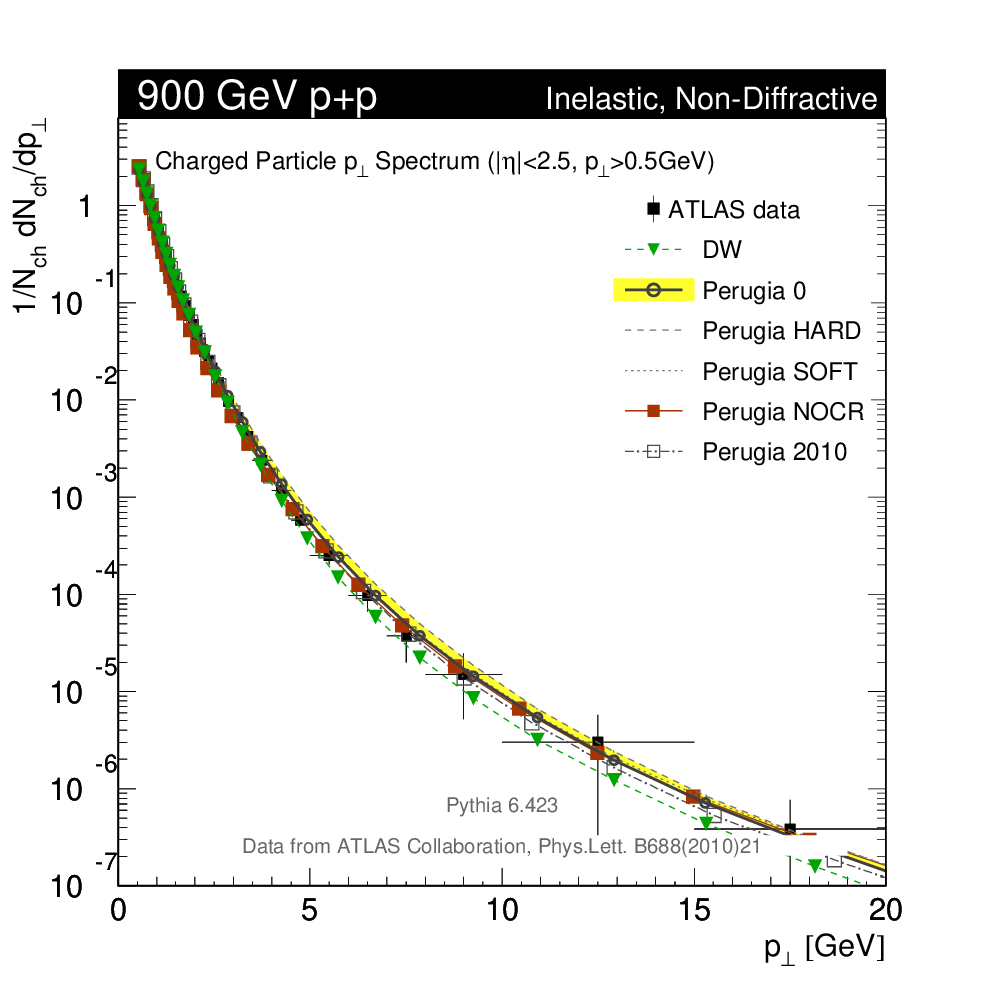}\hspace*{-5mm}
\includegraphics*[scale=0.34]{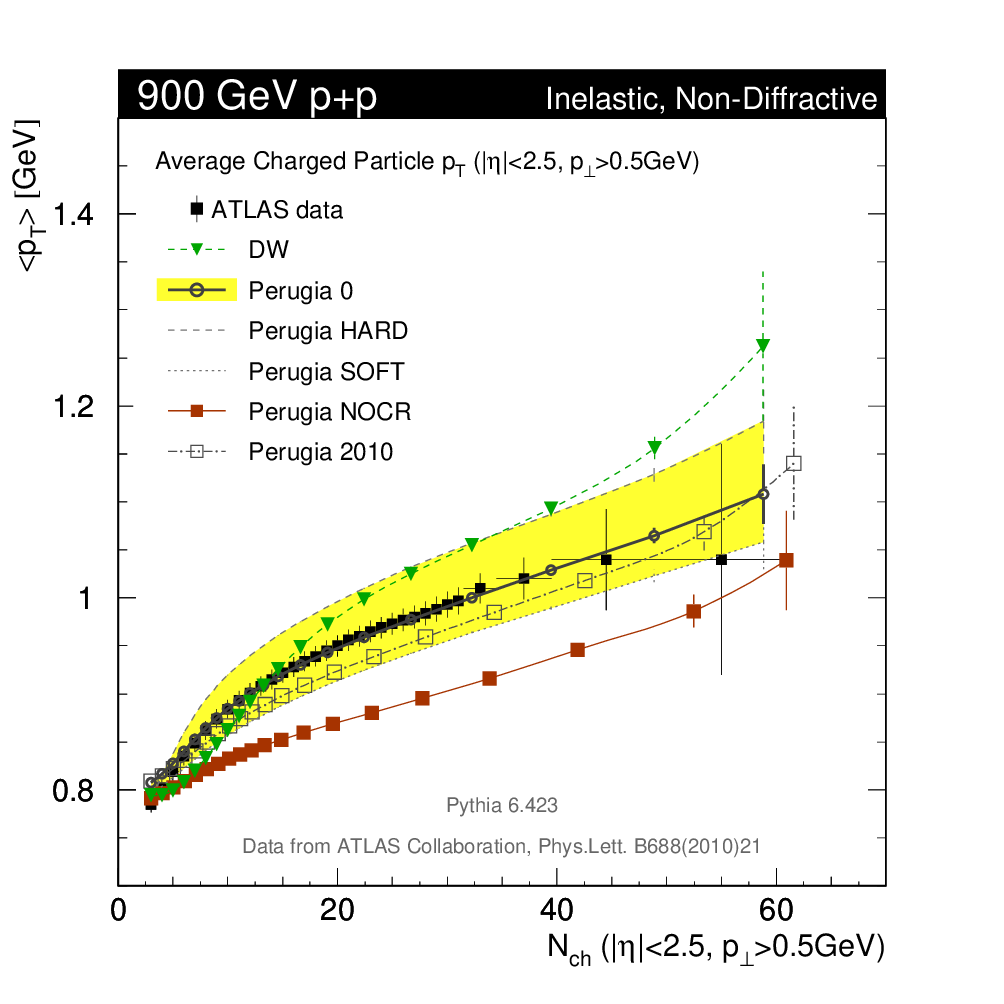}\hspace*{-8mm}\\[-4mm]
\caption{\small Perugia ``predictions'' for the charged multiplicity (left),
  $\pT{}$ (center), and $\left<\pT{}\right>(N_{\mrm{ch}})$
  distributions in
  inelastic, non-diffractive $pp$ collisions at 900 GeV, compared with
  ATLAS data. 
See
  \cite{lhplots} for 
  other tunes and collider energies. 
\label{fig:atlas}}
\end{center}
\end{figure}

\paragraph{Comparison to the Current LHC Data}
At a late stage while preparing this article, data from the initial
LHC runs at 900 GeV became available in the HepDATA web repository.
We were therefore able to include a comparison of Perugia 0 and a few
main variations to the 900 GeV ATLAS data \cite{Aad:2010rd}. We 
here explicitly omit bins with $N_{\mrm{ch}}<3$ in
the multiplicity and $\left<\pT{}\right>(N_{\mrm{ch}})$ distributions 
since we did not include diffractive events in the simulation. The
resulting  comparisons are shown in fig.~\ref{fig:atlas}. 

The overall agreement between the models and the data 
is good, which is not surprising given that the 900 GeV beam energy
lies well within the energy span inside which the models were
tuned. One point that may be worth remarking on is that the models
appear to be undershooting the tail of the multiplicity distribution
slightly (left). This confirms the tendency already observed in the
comparison to the UA5 data, cf.~fig.~\ref{fig:escaling} while the
models had a tendency to overshoot the tails of the Tevatron
distributions, cf.~figs.~\ref{fig:tevatronNCH} and
\ref{fig:escaling}. Combined with early indications at 7 TeV from
ALICE \cite{Aamodt:2010pp} 
and CMS \cite{Collaboration:2010us} that, likewise, confirm an
undershooting by the models of the  high-multiplicity tail, 
we observe that it may be particularly difficult to
describe both the Tevatron and LHC data sets simultaneously and that
more work in this direction would be fruitful. One way of getting
closer to an apples-to-apples comparison in a study of this particular
issue would be to perform an LHC measurement applying 
the same cuts as those used by the CDF min-bias analysis. 

\section{Conclusions}
We have presented a set of updated parameter sets (tunes) 
for the interleaved
$\pT{}$-ordered shower and underlying-event model in \textsc{Pythia}
6.4. These parameter sets include the revisions to the fragmentation and flavour
parameters obtained by the Professor group 
\cite{Buckley:2009vk,Buckley:2009bj}. The new sets further include more
Tevatron data and more data from different collider CM energies in an
attempt to simultaneously improve the overall description of the Tevatron
data while also improving the reliability of the extrapolations to the
LHC. We have also attempted to deliver a first set of 
``theoretical uncertainty bands'', by including alternative tunes with
systematically different parameter choices. The new tunes are
available from Pythia version 6.4.23, via the routine \ttt{PYTUNE} or,
alternatively, via the switch \ttt{MSTP(5)}. 

Our conclusions are that reasonably good overall fits can be obtained,
at the 10-20\% level, but that the contribution of diffractive
processes and the  scaling of the overall activity with
collider energy are still highly uncertain. 
Other interesting questions
to pursue concern the spectrum of ultra-hard single hadrons
with momenta above 30 GeV
\cite{Aaltonen:2009ne,Albino:2010em,Arleo:2010kw,Cacciari:2010yd,Yoon:2010fa},
the 
(possibly connected) question of collective effects in $pp$ and the
dynamics driving  such effects, the contribution and
properties of diffractive interactions, tests of jet universality by constraining 
fragmentation models better \emph{in situ} at hadron colliders 
as compared to constraints coming from LEP and HERA, 
and the question of
the relative balance between different particle production mechanisms
with different characteristics; e.g., between soft beam remnant fragmentation,
multiple parton interactions, and traditional parton-shower /
radiative corrections to the fundamental scattering processes. 

We note that these tunes still only included LEP, Drell-Yan, 
and minimum-bias data directly, and that the lowest-multiplicity bins of 
the latter were ignored due to their relatively stronger sensitivity
to diffractive physics which we deemed it beyond the scope of this
analysis to attack. Furthermore, only one Drell-Yan distribution was
used, the inclusive \pT{} spectrum.
Leading-jet, $V/\gamma$+jet(s), underlying-event and jet structure 
observables were  not considered explicitly. 
We wish to emphasize that such studies  furnish additional important inputs 
both to tuning and to jet  calibration efforts through such
observables as jet rates, jet pedestals,  jet masses, jet-jet masses (and
inter-jet distances), jet profiles, and dedicated jet substructure 
variables. 

We hope these tunes will be useful to the RHIC, Tevatron, and 
LHC communities. 

\subsubsection*{Acknowledgments}
The Perugia tunes derive their names from the Perugia MPI Workshop in 2008,
which brought people from different communities 
together, and helped us take some steps towards finding a common
language. We thank the Fermilab computing division, S.~Timm in
particular, and the Fermilab theory group for providing and maintaining 
excellent dedicated computing resources without which
the large runs necessary for this tuning effort would have been
impossible. We acknowledge many fruitful interactions with the RHIC,
Tevatron, and LHC experimental communities, 
and are particularly grateful to B.~Cooper, L.~Galtieri,
B.~Heinemann, G.~Hesketh, D.~Kar, P.~Lenzi, A.~Messina, 
and L.~Tomkins for detailed counter-checks and feedback. 
In combination with the writeup of this article, the author 
agrees to owing Lisa Randall a bottle of champagne if 
the first published measurement at 10 or 14 TeV of any number in 
table \ref{tab:predictions} 
is outside the range given in the table, and vice versa.

This work was supported in part by the Marie Curie research training
network ``MCnet'' (contract number MRTN-CT-2006-035606) 
and by the U.S.~Department of Energy under contract No. DE-AC02-07CH11359. 

\appendix
\section{Parameters for the Perugia Tunes \label{sec:tables}}
The following tables give an overview of the parameter settings in
\textsc{Pythia} corresponding to the Perugia tunes described in this
paper. The settings for the 
previous ``best'' tune of the \pT{}-ordered model, Tune S0A-Pro, 
are included for reference. 
\begin{table}[h]
\begin{center}
\scalebox{0.825}{
\begin{tabular}{lc|r|rrrrrrrrr}
Parameter & Type & S0A$_\mrm{Pro}$ & P$_0$ & P$_{\mrm{HARD}}$ &
P$_{\mrm{SOFT}}$ & P$_{\mrm{3}}$ & P$_{\mrm{NOCR}}$ & P$_{\mrm{LO*}}$
& P$_6$ & P$_{\mrm{2010}}$ & P$_K$ \\
\hline
\ttt{MSTP(5)} & Tune & 310 & 320 & 321 & 322 & 323 & 324 & 325 & 326 &
327 & 328\\
\hline
\ttt{PARJ(81)} & FSR &0.257&0.257&   0.3  &   0.2 & 0.257 & 0.257 &
0.257 &  0.257 & 0.26 & 0.26\\
\ttt{PARJ(82)} & FSR & 0.8 & 0.8 &   0.8 &    0.8 &  0.8 &    0.8 & 
0.8 &   0.8 & 1.0 & 1.0\\
\hline
\ttt{MSTJ(11)} & HAD &   5 &   5 &     5  &      5 &    5 &     5 &
5     &     5 & 5 & 5\\
\ttt{PARJ(1)} & HAD & 0.073 & 0.073 & 0.073 & 0.073 & 0.073 & 0.073 &
0.073 & 0.073 & 0.08 & 0.08 \\
\ttt{PARJ(2)} & HAD &  0.2 & 0.2 & 0.2 & 0.2 & 0.2 & 0.2 & 0.2 & 0.2 &
0.21 & 0.21\\
\ttt{PARJ(3)} & HAD & 0.94 & 0.94 & 0.94 & 0.94 & 0.94 & 0.94 & 0.94 &
0.94 & 0.94 & 0.94\\
\ttt{PARJ(4)} & HAD & 0.032 & 0.032 & 0.032 & 0.032 & 0.032 & 0.032 & 0.032
& 0.032 & 0.04 & 0.04\\
\ttt{PARJ(11)} & HAD & 0.31 & 0.31 & 0.31 & 0.31 & 0.31 & 0.31 & 0.31
& 0.31 & 0.35 & 0.35\\
\ttt{PARJ(12)} & HAD & 0.4 & 0.4 & 0.4 & 0.4 & 0.4 & 0.4 & 0.4 & 0.4 &
0.35 & 0.35\\
\ttt{PARJ(13)} & HAD & 0.54 & 0.54 & 0.54 & 0.54 & 0.54 & 0.54 & 0.54
& 0.54 & 0.54 & 0.54\\
\ttt{PARJ(21)} & HAD &0.313&0.313 &  0.34 &   0.28 & 0.313& 0.313 &
0.313 &  0.313 & 0.36 & 0.36\\
\ttt{PARJ(25)} & HAD & 0.63 & 0.63 & 0.63 & 0.63 & 0.63 & 0.63 & 0.63
& 0.63 & 0.63 & 0.63\\
\ttt{PARJ(26)} & HAD & 0.12 & 0.12 & 0.12 & 0.12 & 0.12 & 0.12 & 0.12
& 0.12 & 0.12 & 0.12\\
\ttt{PARJ(41)} & HAD & 0.49& 0.49 &  0.49 &   0.49 & 0.49 &  0.49 &
0.49  &   0.49 & 0.35 & 0.35\\
\ttt{PARJ(42)} & HAD & 1.2& 1.2 &  1.2   &   1.2 & 1.2 &  1.2 &
1.2 &   1.2 & 0.9 & 0.9\\
\ttt{PARJ(46)} & HAD & 1.0& 1.0 &  1.0   &   1.0 & 1.0 &  1.0 &
1.0 &   1.0 & 1.0 & 1.0\\  
\ttt{PARJ(47)} & HAD & 1.0& 1.0 &  1.0   &   1.0 & 1.0 &  1.0 &
1.0 &   1.0 & 1.0 & 1.0\\
\hline
\end{tabular}}
\caption{\small Final-State Radiation and Hadronization Parameters of the
  Perugia tunes compared to S0A-Pro. For more information on each
  parameter, see \cite{Sjostrand:2006za}. \label{tab:fsrhad}}
\end{center}
\end{table}

\begin{table}[h!]
\begin{center}
\scalebox{0.825}{
\begin{tabular}{lc|r|rrrrrrrrr}
Parameter & Type & S0A$_\mrm{Pro}$ & P$_0$ & P$_{\mrm{HARD}}$ &
P$_{\mrm{SOFT}}$ & P$_{\mrm{3}}$ & P$_{\mrm{NOCR}}$ & P$_{\mrm{LO*}}$
& P$_6$ & P$_{\mrm{2010}}$ & P$_K$\\
\hline
\ttt{MSTP(5)} & Tune & 310 & 320 & 321 & 322 & 323 & 324 & 325 & 326 &
327 & 328 \\
\hline
\ttt{MSTP(51)} & PDF &   7 &   7 &      7 &      7 &    7 &      7 &
20650 & 10042 & 7 & 7\\
\ttt{MSTP(52)} & PDF &   1 &   1 &      1 &      1 &    1 &      1 &   
2     &     2 & 1 & 1 \\
\ttt{MSTP(3)} &  $\Lambda$ &    2 &   2 &      2 &      2 &    2 &      2 &
2 &        2 & 1 & 1\\
\ttt{MSTU(112)} & $\Lambda$ &    - &   - &      - &      - &    - &      - &
- &        - & 4 & 4\\
\ttt{PARU(112)} & $\Lambda$ &    - &   - &      - &      - &    - &      - &
- &        - & 0.192 & 0.192\\
\ttt{PARP(1)} & ME &    - &   - &      - &      - &    - &      - &
- &        - & 0.192 & 0.192\\
\hline
\ttt{PARP(61)} & ISR &   - &   - &      - &      - &    - &      - & 
- &        - & 0.192 & 0.192\\ 
\ttt{PARP(72)} & IFSR & - & - &  -  &   - & - & - &
- &  - & 0.26 & 0.26\\
\ttt{MSTP(64)} & ISR &   2 &   3 &      3 &      2 &    3 &      3 &    
3     &     3 & 3 & 3\\
\ttt{PARP(64)} & ISR & 1.0 & 1.0 &   0.25 &    2.0 &  1.0 &    1.0 & 
2.0   &    1.0 & 1.0 & 1.0\\
\ttt{MSTP(67)} & ISR &   2 &   2 &      2 &      2 &    2 &      2 &
2     &      2 & 2 & 2\\
\ttt{PARP(67)} & ISR & 4.0 & 1.0 &    4.0 &    0.25 &  1.0 &    1.0 & 
1.0   &    1.0 & 1.0 & 1.0\\
\ttt{PARP(71)} & IFSR & 4.0 & 2.0 &    4.0 &    1.0 &  2.0 &    2.0 &
2.0   &    2.0 & 2.0 & 2.0\\
\ttt{MSTP(70)} & ISR &   2 &   2 &      0 &      1 &    0 &      2 &
2     &      2 & 2 & 2\\
\ttt{PARP(62)} & ISR &   - &   - &   1.25 &      - & 1.25 &      - &
-     &      - & - & -\\
\ttt{PARP(81)} & ISR &   - &   - &      - &    1.5 &    - &      - &
-     &      - & -& -\\
\ttt{MSTP(72)} & ISR &   0 &   1 &      1 &      0 &    2 &      1 &
1     &      1 & 2 & 2\\
\hline
\ttt{MSTP(91)} &  BR &   1 &   1 &      1 &      1 &    1 &      1 &
1     &     1 & 1& 1\\
\ttt{PARP(91)} &  BR & 2.0 & 2.0 &    1.0 &    2.0 &  1.5 &    2.0 &
2.0   &   2.0 & 2.0& 2.0\\
\ttt{PARP(93)} &  BR &10.0 & 10.0&   10.0 &   10.0 & 10.0 &  10.0 &
10.0  &  10.0 & 10.0 & 10.0\\
\hline
\end{tabular}}
\caption{\small Parton-Density, 
Initial-State Radiation, and Primordial $k_T$ parameters of the
  Perugia tunes compared to S0A-Pro. For more information on each
  parameter, see \cite{Sjostrand:2006za}. \label{tab:isrkt}}
\end{center}
\end{table}

\begin{table}[h!]
\begin{center}
\scalebox{0.825}{
\begin{tabular}{lc|r|rrrrrrrrr}
Parameter & Type & S0A$_\mrm{Pro}$ & P$_0$ & P$_{\mrm{HARD}}$ &
P$_{\mrm{SOFT}}$ & P$_{\mrm{3}}$ & P$_{\mrm{NOCR}}$ & P$_{\mrm{LO*}}$
& P$_6$ & P$_{\mrm{2010}}$ & P$_{K}$ \\
\hline
\ttt{MSTP(5)} & Tune & 310 & 320 & 321 & 322 & 323 & 324 & 325 & 326 &
327 & 328\\
\hline
\ttt{MSTP(81)} &  UE &  21 &  21 &     21 &     21 &   21 &     21 &
21    &     21 & 21 & 21\\
\ttt{PARP(82)} &  UE &1.85 & 2.0 &    2.3 &    1.9 &  2.2 &   1.95 &   
2.2   &   1.95 &  2.05  &  2.45\\
\ttt{PARP(89)} &  UE &1800 &1800 &   1800 &   1800 & 1800 &   1800 &
1800  &   1800 & 1800 & 1800\\  
\ttt{PARP(90)} &  UE &0.25 &0.26 &   0.30 &   0.24 & 0.32 &   0.24 &
0.23  &  0.22 & 0.26 & 0.26\\
\ttt{MSTP(82)} &  UE &   5 &   5 &      5 &      5 &    5 &      5 &
5     &     5 & 5 & 5\\
\ttt{PARP(83)} &  UE & 1.6 & 1.7 &    1.7 &    1.5 &  1.7 &    1.8 &
1.7   &   1.7 & 1.5 & 1.5\\
\ttt{PARP(84)} &  UE &   - &   - &      - &      - &    - &      - &
  -   & - &     - &   -\\
\ttt{MSTP(33)} & ``K'' & 0 & 0 & 0 & 0 & 0 & 0 & 0  & 0 & 0  & 10 \\
\ttt{PARP(32)} & ``K'' & - & - & - & - & - & - & - &  - & -  & 1.5\\ 
\hline
\ttt{MSTP(88)} &  BR &   0 &   0 &      0 &      0 &    0 &      0 &
0     &     0 & 0 & 0\\
\ttt{PARP(79)} &  BR & 2.0 & 2.0 &    2.0 &    2.0 &  2.0 &    2.0 &
2.0   &   2.0 & 2.0 & 2.0\\
\ttt{MSTP(89)} &  BR &   1 &   1 &      1 &      0 &    1 &      2 &
1     &     1 & 0 & 0\\
\ttt{PARP(80)} &  BR &0.01 &0.05 &   0.01 &   0.05 & 0.03 &   0.01 &
0.05  &  0.05 & 0.1 & 0.1\\
\hline
\ttt{MSTP(95)} &  CR &   6 &   6 &      6 &      6 &    6 &     6 &
6     &     6 & 8 & 8\\
\ttt{PARP(78)} &  CR & 0.2 &0.33 &   0.37 &    0.15& 0.35 &   0.0 &
0.33  &  0.33 & 0.035 & 0.033\\ 
\ttt{PARP(77)} &  CR & 0.0 & 0.9 &   0.4  &    0.5 & 0.6  &   0.0 &
0.9   &   0.9 & 1.0 & 1.0\\
\hline
\end{tabular}}
\caption{\small Underlying-Event, Beam-Remnant, and Colour-Reconnection parameters of the
  Perugia tunes compared to S0A-Pro. For more information on each
  parameter, see \cite{Sjostrand:2006za}. \label{tab:uebrcr}}
\end{center}
\end{table}

\clearpage 
\section{The Perugia 2011 Tunes \label{app:2011}}

An update of the Perugia tunes was prepared in the Spring of 2011,
with the following main goals, 
\begin{itemize}
\item Use the same value of $\Lambda_{\mrm{QCD}}$ for all shower
  activity (ISR and FSR), in particular to simplify matching applications.
\begin{itemize}
\item The common value is taken from a \textsc{Professor} 
  fit to LEP event shapes and jet
  rates \cite{Buckley:2009vk,Buckley:2009bj} 
  and ignores the value given by the PDF set. 
\item A variant is provided which also uses this $\Lambda_\mrm{QCD}$ 
  value for the MPI cross sections in the underlying event. This 
  increases the rate of semi-hard mini-jets produced by MPI relative
  to the central Perugia 2011 tunes. 
\item Due to a slightly increased level of soft ISR, the Perugia 2011 tunes
  only need 1 GeV of primordial $k_T$ to describe the CDF and D$\O$
  Drell-Yan $\pT{}$ spectra, as compared to 2 GeV in the previous tunes.
\end{itemize}
\item Take into account some of the early lessons of LHC minimum-bias
  and underlying-event data at 900 and 7000 GeV:
\begin{itemize}
\item Faster scaling of multiplicities with energy, motivated, e.g.,
  by the ALICE \cite{Aamodt:2010pp} and ATLAS \cite{atlas:2010ir}
  min-bias charged multiplicity measurements. 
\item Slightly larger underlying event as compared to Perugia 2010,
  motivated by the ATLAS UE measurement \cite{Aad:2010fh}, see also
  \cite{Karneyeu:2013aha}. 
\item Increased baryon production, especially of strange baryons
  (larger $\Lambda/K$ ratio), motivated by identified-particle
  measurements by the ALICE 
  \cite{Aamodt:2011zz,Aamodt:2011zj} and CMS
  \cite{Khachatryan:2011tm} experiments 
   and by the $p/\pi$ ratio measured by STAR \cite{Adams:2006nd}. The 
  total amount of baryon production (dominated by protons and
  neutrons) now appears to be at the upper limit of the range allowed
  by LEP~\cite{Karneyeu:2013aha}.
\item Increased baryon transport from the beam remnant (though still less than
  the Perugia SOFT tune), motivated by $\bar{p}/p$ and
  $\bar{\Lambda}/{\Lambda}$ measurements performed by the ALICE
  \cite{Aamodt:2010dx} and LHCb experiments \cite{lhcb-inprep}.
\item Slightly softer LEP fragmentation functions than in the Perugia 2010
  tune, since the plots on \cite{Karneyeu:2013aha} indicated this was
  previously slightly too hard. A slight \emph{additional}
  softening of baryon fragmentation functions was made to
  improve the agreement with baryon $x$ distributions at LEP~\cite{Karneyeu:2013aha}.
\item The default suppression of strangeness in association
  with popcorn mesons (\ttt{PARJ(6)} and \ttt{PARJ(7)})
  was removed to help improve $\Xi$ and $\Omega$ yields at LEP
  \cite{Karneyeu:2013aha}. (Note, however, the consequences of this on particle-particle
  correlations have not been checked.)
\item Slightly larger $K^*/K$ ratio, motivated by comparisons of
  Perugia 2010 to LEP data \cite{Karneyeu:2013aha}. 
\item Lower color-reconnection strength than the AMBT1 tune, in order
  to lower $\left<\pT{}\right>(N_\mrm{ch})$, cf., e.g., \cite{atlas:2010ir}.
\end{itemize}
\end{itemize}
In total, ten tune variations are included in the ``Perugia 2011''
set. The starting point was in all cases Perugia 2010, with
modifications as documented in the tables below.
\begin{center}
\begin{tabular}{clp{9.5cm}}
\multicolumn{3}{c}{\textbf{Perugia 2011 Tune Set}}\\
\hline
(350) & Perugia 2011 & Central Perugia 2011 tune (CTEQ5L)\\
(351) & Perugia 2011 radHi & Variation using $\alpha_s(\frac12\pT{})$
  for ISR and FSR \\
(352) & Perugia 2011 radLo & Variation using $\alpha_s(2\pT{})$
  for ISR and FSR \\
(353) & Perugia 2011 mpiHi & Variation using
  $\Lambda_{\mrm{QCD}}=0.26\,\mrm{GeV}$ also for MPI\\
(354) & Perugia 2011 noCR & Variation without color reconnections \\
(355) & Perugia 2011 M & Variation using MRST LO** PDFs\\
(356) & Perugia 2011 C & Variation using CTEQ 6L1 PDFs \\
(357) & Perugia 2011 T16 & Variation using \ttt{PARP(90)=0.16} scaling
  away from 7 TeV\\
(358) & Perugia 2011 T32 & Variation using \ttt{PARP(90)=0.32} scaling
  away from 7 TeV\\
(359) & Perugia 2011 Tevatron & Variation optimized for Tevatron \\
\hline
\end{tabular}
\end{center}
Note that these
variations do not explicitly include variations of the
non-perturbative hadronization parameters, cf.\ table
\ref{tab:p2011had}, hence those parameters 
would still have to be varied independently (i.e., manually) to
estimate uncertainties associated specifically 
with the hadronization process. 

Though updated plots showing the 2011
Perugia tunes are not provided in this writeup, a complete set of such
plots can now be found at the \ttt{mcplots.cern.ch} web site~\cite{Karneyeu:2013aha}, under the
``PYTHIA 6'' tab. 
Tables containing explicit parameter values for the Perugia 2011 tunes,
and comparing them to those of the Perugia 0 and Perugia 2010 ones, are
provided in Appendix \ref{app:params11}.

\section{The Perugia 2012 Tunes \label{app:2012}}
A further update of the Perugia tunes was prepared during 2012, with
the following main goals:
\begin{itemize}
\item Change to using CTEQ6L1 as the baseline PDF choice instead of
  the CTEQ5L one that was used for the previous Perugia tune sets. A
  variation using MRST LO** PDFs is still included (379), along with a new
  variation using the MSTW 2008 LO PDF set (378). Note that we use a
  slightly smaller $\Lambda_{\mrm{QCD}}$ (\ttt{PARP(1)} and
  \ttt{PARU(112)}) for the LO** variation (0.14 rather than 0.16), in
  order to slightly reduce the extremely large inclusive jet cross section
  one otherwise obtains with this PDF set, compare e.g.\ the LO** variations
  between the 2011 and 2012 Perugia tune sets on the
  $d\sigma_{\mrm{jet}}/dp_\perp$ distribution on the MCPLOTS web
  site~\cite{Karneyeu:2013aha}. 
\item Slighly increase strangeness production with respect to Perugia
  2011 (by about 5\%).
\item Soften the hard tail of the momentum spectrum of baryons, by
  using a significantly larger value for \ttt{PARJ(45)}, motivated mainly by
  the $\Lambda^0$ spectrum at LEP.
\item Include an additional CR variation, with ``low'' rather than
  just ``no'' color reconnections. This ``loCR'' variation (374)
  both uses a slightly lower CR strength than the default variations,
  combined with a slightly different CR algorithm. It should still 
  at least be in 
  borderline agreement with the min-bias data, though still erring on the low
  side, and should provide a more aggressive precision target for
  uncertainties related to CR. The
  full dynamics of CR is not yet understood (for recent discussions,
  see~\cite{Sjostrand:2013sma,Sjostrand:2013cya}), hence this variation
  cannot be guaranteed to be conservative, wherefore a ``noCR''
  variation is also still provided (375), but at least the loCR
  variation may give an indication of how much CR uncertainties could
  be reduced by improved physics modeling in the future.
\item Include genuine high/low variations of the underlying-event
  activity. Previously, this was only represented by the ``mpiHi''
  variation. However, what that variation really does is shift
  the UE activity to be produced by slightly higher-pT
  ``minijets'', making the UE more ``lumpy'', while the average UE
  level does not necessarily change much. Therefore, two genuine
  variations of the MPI pT0 parameter have now been included, which
  affect the average UE level, called ``ueHi'' (381) and ``ueLo''
  (382). These are complementary to the ``mpiHi'' variation (373). As
  mentioned above, the latter increases the amount of semi-hard MPI scatterings 
    (i.e., the amount of MPI minijets) by increasing the alphaS associated 
    with MPI. The ``ueHi'' and ``ueLo'' variations adjust the soft pT0 scale
    and hence affect  
    the amount of soft MPI produced, without changing the rate of hard MPI. 
    The energy-scaling of the new variations has also been chosen
    conservatively, so that the activity of the ``ueLo'' variation scales 
    slower with CM energy (i.e., pT0 scales up faster) than for
    ``ueHi''.  
\item Include variations of the hadronization parameters. Three such
  variations are now included, altering the fragmentation process to
  be more in the longitudinal direction (376; smaller Lund $a$
  parameters and smaller non-perturbative $p_\perp$ in string breaks), 
  more transverse (377; smaller Lund $b$ parameter and larger
  non-perturbative $p_\perp$ in string breaks),
  and replacing the baseline Perugia fragmentation parameters by
  complementary ones obtained independently by the Innsbruck group
  (383)~\cite{Firdous:2013noa}. Note that the Innsbruck (IBK)
  parameter set is quite different from the baseline Perugia ones; we
  have not here made any independent validations of how well the Perugia-IBK
  combination works in practice; cross checks will be made available
  in a future update of the MCPLOTS web site~\cite{Karneyeu:2013aha}. 
  Note also that nine full-fledged Innsbruck
  tunes~\cite{Firdous:2013noa} have also separately been included in \Py,
  starting from version 6.4.29, with numbers 390-398. (Thanks to
  N.~Firdous and G.~Rudolph for providing these.)
\item Include a variation exploring the ambiguity between $q\bar{q}$ and
  $gg$ scatterings at low transverse momentum scales. By default, a
  fraction of the generated $gg$ scatterings are replaced by
  $q\bar{q}$ ones at low $p_\perp$, in order to account for an assumed
  dominance of valence quarks at low scales. In the new ``mb2'' variation
  (380), the proportions are instead taken directly from the PDFs, with no
  enhancement of the $q\bar{q}$ component.  The ``mb2'' variation
  appears to have a slightly improved  
    behavior at very low minimum-bias multiplicities.
\end{itemize}
The first 10 tune variations (370-379) were made available starting from
\Py~6.4.27. The last 4 (380-383) are available starting from 6.4.28. 

Though updated plots showing the 2012
Perugia tunes are not provided in this writeup, a complete set of such
plots will soon be available at the \ttt{mcplots.cern.ch} web
site~\cite{Karneyeu:2013aha}, under the ``PYTHIA 6'' tab. Tables
containing explicit parameter values for the Perugia 2012 tunes, are
provided in Appendix \ref{app:params11}.
 
\begin{center}
\begin{tabular}{clp{9.5cm}}
\multicolumn{3}{c}{\textbf{Perugia 2012 Tune Set}}\\
\hline
(370) & Perugia 2012 & Central Perugia 2012 tune (CTEQ6L1)\\
(371) & Perugia 2012 radHi & Variation using $\alpha_s(\frac12\pT{})$
  for ISR and FSR \\
(372) & Perugia 2012 radLo & Variation using $\alpha_s(2\pT{})$
  for ISR and FSR \\
(373) & Perugia 2012 mpiHi & Variation using
  $\Lambda_{\mrm{QCD}}=0.26\,\mrm{GeV}$ also for MPI\\
(374) & Perugia 2012 loCR & Variation with less color reconnections \\
(375) & Perugia 2012 noCR & Variation with no color reconnections \\
(376) & Perugia 2012 FL & Variation with more longitudinal fragmentation \\
(377) & Perugia 2012 FT & Variation with more transverse fragmentation \\
(378) & Perugia 2012 M8LO & Variation using MSTW 2008 LO PDFs \\
(379) & Perugia 2012 LO** & Variation using MRST LO** PDFs \\
(380) & Perugia 2012 mb2 & Same as Perugia 2012, with PARP(87)=0D0 \\
(381) & Perugia 2012 ueHi & Variation with higher UE (lower pT0) \\
(382) & Perugia 2012 ueLo & Variation with lower UE (higher pT0) \\
(383) & Perugia 2012 IBK & Variation using Innsbruck hadronization parameters\\
\hline
\end{tabular}
\vspace*{0.5cm}
\end{center} 

\section{Parameters of the Perugia 2011 and 2012 Tunes \label{app:params11}}

\begin{table}[ht!]
\begin{center}
\scalebox{0.85}{
\begin{tabular}{lc|rr|rrrrrrrrrrr}
Parameter & Type & Perugia 0 & Perugia 2010 &  Perugia 2011 (All) &
Perugia 2012 & FL & FT & IBK\\
\hline
\ttt{MSTP(5)} & Tune & 310 & 327 & 350 --- 359 & 370--375,378-382 & 376 & 377
& 383\\
\hline
\ttt{MSTJ(11)} & HAD &   5 &  5 & 5 & 5 &   &   & 5\\
\ttt{PARJ(1)} & HAD & 0.073 & 0.08 & 0.087 & 0.085 & & & 0.128\\
\ttt{PARJ(2)} & HAD &  0.2 & 0.21 &  0.19  & 0.20 & & & 0.268 \\
\ttt{PARJ(3)} & HAD & 0.94 &  0.94 & 0.95  & 0.92 & & & 0.772 \\
\ttt{PARJ(4)} & HAD & 0.032 & 0.04 & 0.043 & 0.043 & & & 0.05 \\
\textcolor{blue}{\ttt{PARJ(6)}} & HAD & 0.5 & 0.5 & 1.0 & 1.0 &  &
 & 0.5\\
\textcolor{blue}{\ttt{PARJ(7)}} & HAD & 0.5 & 0.5 & 1.0 & 1.0 &  &
 & 0.5\\
\ttt{PARJ(11)} & HAD & 0.31 & 0.35 & 0.35 & 0.35 & & & 0.549\\
\ttt{PARJ(12)} & HAD & 0.4 & 0.35 &  0.40 & 0.40 & & & 0.450\\
\ttt{PARJ(13)} & HAD & 0.54 & 0.54 & 0.54 & 0.54 & & & 0.500\\
\ttt{PARJ(21)} & HAD &0.313& 0.36 & 0.33 & 0.33 & 0.30 & 0.36 & 0.329\\
\ttt{PARJ(25)} & HAD & 0.63 & 0.63 & 0.63 & 0.70 & & & 1.0\\
\ttt{PARJ(26)} & HAD & 0.12 & 0.12 & 0.12 & 0.135 & & & 0.245\\
\ttt{PARJ(41)} & HAD & 0.49& 0.35 & 0.35 & 0.45 & 0.36 & 0.45 & 0.425\\
\ttt{PARJ(42)} & HAD & 1.2& 0.9 & 0.80 & 1.0 & 1.0 & 0.75 & 1.65\\
\textcolor{blue}{\ttt{PARJ(45)}} & HAD & 0.5& 0.5 & 0.55 & 0.86 & 0.75
& 0.90 & 0.50\\  
\ttt{PARJ(46)} & HAD & 1.0& 1.0 & 1.0 & 1.0 & & & 1.42 \\  
\ttt{PARJ(47)} & HAD & 1.0& 1.0 & 1.0 & 1.0 & & & 0.975\\
\hline
\end{tabular}}
\caption{\small Hadronization Parameters of the
  Perugia 2011 and 2012 tunes compared to Perugia 0 and Perugia
  2010. Parameters that were not explicitly part of the Perugia 0 and
  Perugia 2010 tuning but were included in Perugia 2011 are
  highlighted in blue. Note that the IBK
  variation~\cite{Firdous:2013noa} includes a few additional
  paremeters, not shown here, related to $L=1$ mesons. 
  For more information on each
  parameter, see \cite{Sjostrand:2006za}. \label{tab:p2011had}}
\end{center}
\end{table}

\begin{table}[p]
\centering
\scalebox{0.81}{
\begin{tabular}{lc|rr|rrrrrrrrrr}
Parameter & Type & P$_0$ & P$_{2010}$ & P$_{11}$ & 
rad$_\mrm{Hi}$ & rad$_\mrm{Lo}$ & 
mpi$_\mrm{Hi}$ & no$_\mrm{CR}$ 
& M$_\mrm{LO^{**}}$ & C$_{\mrm{6L1}}$ & T$_{0.16}$ 
& T$_{0.32}$ & TeV \\
\hline
\ttt{MSTP(5)} & Tune & 320 & 327 & 350 & 351 & 352 & 353 & 354 & 355 &
356 & 357 & 358 & 359\\
\hline
\ttt{MSTP(51)} & PDF &   7 &  7 &7&7&7&7&7& \textcolor{red}{20651} & \textcolor{red}{10042} &7&7&7\\
\ttt{MSTP(52)} & PDF &   1 &  1 &1&1&1&1&1&\textcolor{red}{2}&\textcolor{red}{2}&1&1&1\\
\hline
\ttt{MSTP(3)}  & $\Lambda$ &  2 &  1   &    1  &    1 & 1 & 1 & 1 & 1 & 1 & 1 &
1 & 1\\
\ttt{MSTP(64)} & $\Lambda$ &  3   & 3  & 2& 2& 2& 2& 2& 2& 2& 2& 2&2  \\
\ttt{MSTU(112)}& $\Lambda$ & -    &    4  &    5  &5&5&5&5&5&5&5&5&5\\
\ttt{PARP(61)} & ISR  &  -   & 0.192 & 0.26& \textcolor{red}{0.52}& \textcolor{red}{0.13}& 0.26& 0.26& 0.26& 0.26& 0.26& 0.26& 0.26  \\ 
\ttt{PARP(72)} & IFSR & -    & 0.26  & 0.26& \textcolor{red}{0.52}& \textcolor{red}{0.13}& 0.26& 0.26& 0.26& 0.26& 0.26& 0.26& 0.26  \\
\ttt{PARJ(81)} & FSR &0.257  & 0.26  & 0.26& \textcolor{red}{0.52}& \textcolor{red}{0.13}& 0.26& 0.26& 0.26& 0.26& 0.26& 0.26& 0.26  \\
\hline
\ttt{PARP(1)}  & ME   & -    & 0.192 & 0.16& 0.16& 0.16& \textcolor{red}{0.26}& 0.16& 0.16& 0.16& 0.16& 0.16& 0.16  \\
\ttt{PARU(112)}& ME   & -    & 0.192 & 0.16& 0.16& 0.16& \textcolor{red}{0.26}& 0.16& 0.16& 0.16& 0.16& 0.16& 0.16   \\
\ttt{PARP(64)} & ISR & 1.0 & 1.0 & 1.0 & 1.0& 1.0& 1.0& 1.0& 1.0& 1.0& 1.0& 1.0 &1.0\\
\ttt{MSTP(67)} & ISR &   2 &   2 &   2& 2& 2& 2& 2& 2& 2& 2& 2& 2\\
\ttt{PARP(67)} & ISR & 1.0 & 1.0 & 1.0& 1.0& 1.0& 1.0& 1.0& 1.0& 1.0& 1.0& 1.0& 1.0\\
\ttt{PARP(71)} & IFSR & 2.0 & 2.0& 1.0& 1.0& 1.0& 1.0& 1.0& 1.0& 1.0& 1.0& 1.0& 1.0\\
\ttt{MSTP(70)} & ISR &   2 &  2  &   0& 0& 0& 0& 0& 0& 0& 0& 0& 0\\
\ttt{MSTP(72)} & ISR &   1 &   2 &   2& 2& 2& 2& 2& 2& 2& 2& 2& 2\\
\ttt{PARP(62)} & ISR &   - &   - & 1.5&1.75&1.0&1.5&1.5&1.5&1.5&1.5&1.5&1.5\\
\ttt{PARJ(82)} & FSR & 0.8 &  1.0& 1.0&1.75&0.75&1.0&1.0&1.0&1.0&1.0&1.0&1.0\\
\hline
\ttt{MSTP(91)} &  BR &   1 &   1  & 1 & 1 & 1 & 1& 1& 1& 1& 1& 1 &1\\
\ttt{PARP(91)} &  BR & 2.0 &  2.0 & 1.0 & 1.0 & 1.0 & 1.0 & 1.0 & 1.0
& 1.0 & 1.0 & 1.0&1.0\\
\ttt{PARP(93)} &  BR &10.0 & 10.0 &10.0 & 10.0 &10.0 & 10.0 &10.0 & 10.0 &10.0 & 10.0 &10.0 &10.0 \\
\hline
\end{tabular}}\\[2ex]
\scalebox{0.81}{
\begin{tabular}{l|rrrrrrrrrrrrrr}
Parameter & P$_{2012}$ & 
rad$_\mrm{Hi}$ & rad$_\mrm{Lo}$ & 
mpi$_\mrm{Hi}$ & \hspace*{-1mm}lo$_\mrm{CR}$, no$_\mrm{CR}$ 
& F$_L$, F$_T$ 
& M$_\mrm{8LO}$ & M$_\mrm{LO^{**}}$ 
& mb$_2$ & ue$_\mrm{Hi}$ & ue$_\mrm{Lo}$ 
& I$_{\mrm{BK}}$\\
\hline
\ttt{MSTP(5)} &  370 & 371 & 372 & 373 & 374, 375 & 376, 377 & 378 &
379 &
380 & 381 & 382 & 383\\
\hline
\ttt{MSTP(51)} & 10042 & 10042 & 10042 & 10042 & 10042 & 10042 & \textcolor{red}{21000}
& \textcolor{red}{20651} & 10042 & 10042 & 10042 & 10042 \\
\ttt{MSTP(52)} &   2 &  2 &2&2&2&2&2&2 & 2&2&2&2\\
\hline
\ttt{MSTP(3)}  &  1 &  1   &    1  &    1 & 1 & 1 & 1 & 1 & 1 & 1 &
1 & 1\\
\ttt{MSTP(64)} & 2   & 2  & 2& 2& 2& 2& 2& 2& 2& 2& 2&2  \\
\ttt{MSTU(112)}& 5    &   5  &    5  &5&5&5&5&5&5&5&5&5\\
\ttt{PARP(61)} & 0.26& \textcolor{red}{0.52}& \textcolor{red}{0.13}&
0.26& 0.26& 0.26& 0.26& 0.26& 0.26& 0.26 & 0.26 & 0.26 \\ 
\ttt{PARP(72)} & 0.26& \textcolor{red}{0.52}& \textcolor{red}{0.13}&
0.26& 0.26& 0.26& 0.26& 0.26& 0.26& 0.26 & 0.26 & 0.26  \\
\ttt{PARJ(81)} & 0.26& \textcolor{red}{0.52}& \textcolor{red}{0.13}&
0.26& 0.26& 0.26& 0.26& 0.26& 0.26& 0.26 & 0.26 &  0.261 \\
\hline
\ttt{PARP(1)}  & 0.16& 0.16& 0.16& \textcolor{red}{0.26}& 0.16& 0.16&
0.16& \textcolor{red}{0.14}& 0.16& 0.16 & 0.16 & 0.16 \\
\ttt{PARU(112)}& 0.16& 0.16& 0.16& \textcolor{red}{0.26}& 0.16& 0.16&
0.16& \textcolor{red}{0.14}& 0.16& 0.16 & 0.16 & 0.16   \\
\ttt{PARP(64)} &  1.0 & 1.0 & 1.0 & 1.0& 1.0& 1.0& 1.0& 1.0& 1.0& 1.0& 1.0 &1.0\\
\ttt{MSTP(67)} &   2 &   2 &   2& 2& 2& 2& 2& 2& 2& 2& 2& 2\\
\ttt{PARP(67)} &  1.0 & 1.0 & 1.0& 1.0& 1.0& 1.0& 1.0& 1.0& 1.0& 1.0& 1.0& 1.0\\
\ttt{PARP(71)} &  1.0 & 1.0 & 1.0& 1.0& 1.0& 1.0& 1.0& 1.0& 1.0& 1.0& 1.0& 1.0\\
\ttt{MSTP(70)} &   0 & 0 &  0& 0& 0& 0& 0& 0& 0& 0& 0& 0\\
\ttt{MSTP(72)} &  2 &   2 &   2& 2& 2& 2& 2& 2& 2& 2& 2& 2\\
\ttt{PARP(62)} &  1.5& 1.75&1.0&1.5&1.5&1.5&1.5&1.5&1.5&1.5 & 1.5 & 1.5\\
\ttt{PARJ(82)} &  1.0& 1.75&0.75&1.0&1.0&1.0&1.0&1.0&1.0&1.0 & 1.0 & 0.90\\
\hline
\ttt{MSTP(91)} &   1 &   1  & 1 & 1 & 1 & 1& 1& 1& 1& 1& 1 &1\\
\ttt{PARP(91)} &  1.0 & 1.0 & 1.0 & 1.0 & 1.0 & 1.0
& 1.0 & 1.0 & 1.0&1.0 & 1.0 & 1.0\\
\ttt{PARP(93)} &  10.0 & 10.0 &10.0 & 10.0 &10.0 & 10.0 &10.0 & 10.0 &10.0 & 10.0 &10.0 &10.0 \\
\hline
\end{tabular}}
\caption{\small Parton-Density, 
Initial-State Radiation, and Primordial $k_T$ parameters of the
  Perugia 2011 and 2012 tunes compared to Perugia 0 and Perugia 2010. The main
  distinguishing features of each variation are highlighted in red.
  For more information on each
  parameter, see \cite{Sjostrand:2006za}. \label{tab:isrkt11}}
\end{table}

\begin{table}[p]
\centering
\scalebox{0.81}{
\begin{tabular}{lc|rr|rrrrrrrrrr}
Parameter & Type &  P$_0$ & P$_{2010}$ & P$_{11}$ & 
rad$_\mrm{Hi}$ & rad$_\mrm{Lo}$ & 
mpi$_\mrm{Hi}$ & no$_\mrm{CR}$ 
& M$_\mrm{LO^{**}}$ & C$_{\mrm{6L1}}$ & T$_{0.16}$ 
& T$_{0.32}$ & TeV \\
\hline
\ttt{MSTP(5)} & Tune & 320 & 327 & 350 & 351 & 352 & 353 & 354 & 355 &
356 & 357 & 358 & 359\\
\hline
\ttt{MSTP(81)} &  UE &  21 &  21 &     21 &     21 &   21 &     21 &
21    &     21 & 21 & 21 & 21 & 21\\
\ttt{PARP(82)} &  UE & 2.0 & 2.05  & 2.93 & 3.0 & 2.95 & \textcolor{red}{3.35} & 3.05 & 3.4 &
2.65 & 2.93 & 2.93 & \textcolor{red}{2.1}\\
\ttt{PARP(89)} &  UE &1800 &  1800 & 7000 & 7000 & 7000 & 7000 & 7000 & 7000 & 7000 & 7000 & 7000 & \textcolor{red}{1800} \\  
\ttt{PARP(90)} &  UE &0.26 &  0.26 & 0.265 & 0.28&0.24&0.26&0.265&0.23&0.22&\textcolor{red}{0.16}&\textcolor{red}{0.32}&0.28\\
\ttt{MSTP(82)} &  UE &   5 &   5 & 3& 3& 3& 3& 3& 3& 3& 3& 3& 3\\
\ttt{PARP(83)} &  UE &1.7 &  1.5 & -& -& -& -& -& -& -& -& -&-\\
\ttt{PARP(84)} &  UE    - &    - & -& -& -& -& -& -& -& -& -& -&-\\
\ttt{MSTP(33)} & ``K'' & 0 & 0 & 0 & 0 & 0 & 0 & 0  & 0 & 0  & 0 & 0& 0\\
\ttt{PARP(32)} & ``K'' & - & - & - & - & - & - & - &  - & -  & -& -& -\\ 
\hline
\ttt{MSTP(88)} &  BR &   0 &   0 & 0& 0& 0& 0& 0& 0& 0& 0& 0&0\\
\ttt{PARP(79)} &  BR & 2.0 & 2.0 &    2.0 &    2.0 &  2.0 &    2.0 &
2.0   &   2.0 & 2.0 & 2.0& 2.0& 2.0\\
\ttt{MSTP(89)} &  BR &   1 &   0 & 0& 0& 0& 0& 0& 0& 0& 0& 0& 0 \\
\ttt{PARP(80)} &  BR &0.05 & 0.1 &0.015&0.015&0.015&0.015&0.015&0.015&0.015&0.015&0.015&0.015 \\
\textcolor{blue}{\ttt{PARP(87)}} & BR & 0.7 & 0.7 & 0.7 & 0.7 & 0.7 &
0.7 & 0.7 & 0.7 & 0.7 & 0.7 & 0.7 & 0.7\\ 
\hline
\ttt{MSTP(95)} &  CR &   6 & 8 & 8& 8& 8& 8& \textcolor{red}{0}& 8& 8& 8& 8& 8\\
\ttt{PARP(78)} &  CR & 0.33 & 0.035 & 0.036 & 0.036 & 0.036 & 0.036 &
- & 0.034 & 0.036 & 0.036 & 0.036 & \textcolor{red}{0.05}\\ 
\ttt{PARP(77)} &  CR & 0.9 &  1.0 &1.0 &1.0 &1.0 &1.0 & - &1.0 &1.0 &1.0 &1.0&1.0 \\
\hline
\end{tabular}}\\[2ex]
\scalebox{0.81}{
\begin{tabular}{l|rrrrrrrrrrrrr}
Parameter & P$_{2012}$ & 
rad$_\mrm{Hi}$ & rad$_\mrm{Lo}$ & 
mpi$_\mrm{Hi}$ & \hspace*{-1mm}lo$_\mrm{CR}$ & no$_\mrm{CR}$ 
& F$_L$, F$_T$ 
& M$_\mrm{8LO}$ & M$_\mrm{LO^{**}}$ 
& mb$_2$ & ue$_\mrm{Hi}$ & ue$_\mrm{Lo}$ 
& I$_{\mrm{BK}}$\\
\hline
\ttt{MSTP(5)} &  370 & 371 & 372 & 373 & 374 & 375 & 376, 377 & 378 &
379 &
380 & 381 & 382 & 383\\
\hline
\ttt{MSTP(81)} & 21 &  21 &     21 &     21 &   21 &     21 &
21    &     21 & 21 & 21 & 21 & 21 & 21\\
\ttt{PARP(82)} &  2.65 & 2.725 & 2.6 & \textcolor{red}{3.0} & 2.7 & 2.8 &
2.65 & 2.90 & 3.25 & 2.65 & \textcolor{red}{2.46} &
\textcolor{red}{2.92} & 2.65\\
\ttt{PARP(89)} & 7000 & 7000 & 7000 & 7000 & 7000 & 7000 & 7000 & 7000
& 7000 & 7000 & 7000 & 7000 & 7000\\  
\ttt{PARP(90)} & 0.24 & 0.25 & 0.23 & 0.24 & 0.24 & 0.24 & 0.24 & 0.29
& 0.25 & 0.245 & \textcolor{red}{0.23} & \textcolor{red}{0.26} & 0.24\\
\ttt{MSTP(82)} & 3& 3& 3& 3& 3& 3& 3& 3& 3& 3 & 3 & 3 & 3\\
\hline
\ttt{MSTP(88)} &  0& 0& 0& 0& 0& 0& 0& 0& 0&0 & 0 & 0 & 0\\
\ttt{PARP(79)} &    2.0 &    2.0 &  2.0 &    2.0 &
2.0   &   2.0 & 2.0 & 2.0& 2.0& 2.0 & 2.0 & 2.0 & 2.0\\
\ttt{MSTP(89)} &  0& 0& 0& 0& 0& 0& 0& 0& 0& 0 & 0 & 0 & 0\\
\ttt{PARP(80)} &
0.015&0.015&0.015&0.015&0.015&0.015&0.015&0.015&0.015&0.015 & 0.015 &
0.015 & 0.015\\
\textcolor{blue}{\ttt{PARP(87)}} & 0.7 & 0.7 & 0.7 & 0.7 & 0.7 & 0.7 &
0.7 & 0.7 & 0.7 & \textcolor{red}{0.0} & 0.7 & 0.7 & 0.7\\ 
\hline
\ttt{MSTP(95)} & 8& 8& 8& 8& \textcolor{red}{6}& \textcolor{red}{0} &
8& 8& 8& 8 & 8 & 8 & 8\\
\ttt{PARP(78)} &  0.035 & 0.035 & 0.035 & 0.035 & 0.25 &
- & 0.035 & 0.035 & 0.034 & 0.035 & 0.035 & 0.035 & 0.035\\ 
\ttt{PARP(77)} & 1.0 &1.0 &1.0 &1.0 & 1.0& - &1.0 &1.0 &1.0 &1.0&1.0 &
1.0 & 1.0\\
\hline
\end{tabular}}
\caption{\small Underlying-Event, Beam-Remnant, and
  Colour-Reconnection parameters of the 
  Perugia 2011 and 2012 tunes compared to Perugia 0 and Perugia 2010. The main
  distinguishing features of each variation are highlighted in
  red. Parameters that were only explicitly included as part of 
  the Perugia 2012 tuning variations are highlighted in blue. For
  more information on each 
  parameter, see \cite{Sjostrand:2006za}. \label{tab:uebrcr11}}
\end{table}

\clearpage

\section{Overview of Tunes included in PYTHIA \label{sec:tunes}}
The following three tables give an overview of the tune presets that
have so far been implemented in \textsc{Pythia}, as of version
6.4.23 (see additionally the table in Appendix \ref{app:2012} for the
Perugia 2012 set, introduced in versions 6.4.27 and 6.4.28). They can
be obtained either by setting \ttt{MSTP(5) = NNN}, 
where \ttt{NNN} is the tune number, or by calling \ttt{PYTUNE(NNN)}
before the call to \ttt{PYINIT}. It is not advisable to do both. Note
that, when \ttt{MSTP(5)} is used, \ttt{PYINIT} calls
\ttt{PYTUNE}, and the tune parameters will then overwrite any previous
user modifications. Also consult the output of \ttt{PYTUNE} which
informs you about useful references for each
tune, its parameters, and a brief description of their meaning. 

\noindent{\small
\begin{tabular}{rp{2.1cm}p{8.5cm}r}
\multicolumn{4}{l}{\colorbox{black}{\textcolor{white}{\large \bf{100+:
        $Q^2$-ordered shower and ``old'' underlying-event model}}}}\\
\hline
\ttt{MSTP(5)} & Name & Description & Date\\
\hline
\multicolumn{4}{l}{     1st generation: Rick Field's CDF tunes and a few more}\\
100 & A & :  Rick Field's CDF Tune A & (Oct 2002) \\
     101  &  AW &:  Rick Field's CDF Tune AW                 &   (Apr 2006)\\
     102  &    BW& :  Rick Field's CDF Tune BW                &    (Apr 2006)\\
     103  &   DW& :  Rick Field's CDF Tune DW                  &  (Apr 2006)\\
     104  &   DWT& :  As DW but with the old default ECM-scaling &  (Apr 2006)\\
     105  &    QW& :  Rick Field's CDF Tune QW using CTEQ6.1M    &    \\
     106 &ATLAS-DC2&: Arthur Moraes' (old) ATLAS tune (``Rome'')    &  \\
     107  &   ACR& :  Tune A modified with new CR model           &(Mar 2007)\\
     108   &   D6& :  Rick Field's CDF Tune D6 using CTEQ6L1       &  \\
     109    & D6T& :  Rick Field's CDF Tune D6T using CTEQ6L1       & \\
\hline
\multicolumn{4}{l}{     2nd generation: The same, but with Professor's LEP parameters}\\
     110 &  A-Pro& :  Tune A, but with Professor's LEP parameters     &   (Oct 2008)\\
     111 & AW-Pro& :  Tune AW, but with Professor's LEP parameters                             &   (Oct 2008)\\
     112 & BW-Pro& :  Tune BW, but with Professor's LEP parameters                            &   (Oct 2008)\\
     113 & DW-Pro& :  Tune DW,    but with Professor's LEP parameters                         &   (Oct 2008)\\
     114 &DWT-Pro& :  Tune DWT, but with Professor's LEP parameters                            &   (Oct 2008)\\
     115  &QW-Pro& :  Tune QW,but with Professor's LEP parameters                              &   (Oct 2008)\\
     116 &ATLAS-DC2-Pro&: ATLAS-DC2/Rome, but with Professor's LEP parameters                &  (Oct 2008)\\
     117 &ACR-Pro& :  Tune ACR, but with Professor's LEP parameters                            &   (Oct 2008)\\
     118 & D6-Pro& :  Tune D6, but with Professor's LEP parameters                            &   (Oct 2008)\\
     119 &D6T-Pro& :  Tune D6T, but with Professor's LEP parameters                            &   (Oct 2008)\\
\hline
\multicolumn{4}{l}{     3rd generation: Complete Q2-ordered Tune by Professor }\\    
     129 & Pro-Q2O & :  Professor Q2-ordered tune             &      (Feb 2009)\\
\hline
 \end{tabular}}

{\small\noindent\begin{tabular}{rp{2.1cm}p{8.5cm}r}
\multicolumn{4}{l}{\colorbox{black}{\textcolor{white}{\large \bf{200+:
        Intermediate and hybrid models}}}}\\\hline
\ttt{MSTP(5)} & Name & Description & Date\\
\hline
     200 &   IM 1 &: Intermediate model: new UE, Q2-ord. showers, new
     CR& \\
     201 &    APT &: Tune A w. pT-ordered FSR                    &(Mar 2007)\\
     211 &APT-Pro &: Tune APT, with LEP tune from Professor      &(Oct 2008)\\
     221 &Perugia APT & : "Perugia" update of APT-Pro            &(Feb 2009)\\
     226 &Perugia APT6 &: "Perugia" update of APT-Pro w. CTEQ6L1 &(Feb 2009)\\
\hline
 \end{tabular}}

{\small\noindent
\begin{tabular}{rp{2.64cm}p{8.4cm}r}
\multicolumn{4}{l}{
\colorbox{black}{\textcolor{white}{\large \bf{300+: 
        $\pT{}^2$-ordered shower and interleaved underlying-event model}}}}\\
\hline
\ttt{MSTP(5)} & Name & Description & Date\\
\hline
\multicolumn{4}{l}{     
     1st generation: Sandhoff-Skands CDF Min-Bias tunes and a few more}\\
     300  &    S0 &: Sandhoff-Skands Tune using the S0 CR model & (Apr 2006)\\
     301  &    S1 &: Sandhoff-Skands Tune using the S1 CR model & (Apr 2006)\\
     302  &    S2 &: Sandhoff-Skands Tune using the S2 CR model & (Apr 2006)\\
     303  &   S0A &: S0 with ``Tune A'' UE energy scaling        &  (Apr 2006)\\
     304  &  NOCR &:  ``best try'' without CR       &  (Apr 2006)\\
     305  &   Old &:  Original (primitive) CR model &  (Aug 2004)\\
     306 &ATLAS-CSC&: Arthur Moraes' $\pT{}$-ordered ATLAS tune w.~CTEQ6L1   & \\
\hline
\multicolumn{4}{l}{     
     2nd generation : The same, but with Professor's LEP parameters}\\
     310 &  S0-Pro &: S0, but with Professor's LEP parameters   & (Oct 2008)\\
     311 &  S1-Pro &: S1, but with Professor's LEP parameters                                     & (Oct 2008)\\
     312 &  S2-Pro &: S2, but with Professor's LEP parameters                                    & (Oct 2008)\\
     313 & S0A-Pro &: S0A, but with Professor's LEP parameters                                 & (Oct 2008)\\
     314 &NOCR-Pro &: NOCR, but with Professor's LEP parameters                                  & (Oct 2008)\\
     315 & Old-Pro &: Old, but with Professor's LEP parameters                                    & (Oct 2008)\\
\hline
\multicolumn{4}{l}{     
     3rd generation : The Perugia, Professor, and ATLAS MC09 pT-ordered Tunes}\\
     320 &Perugia 0 &: "Perugia" update of S0-Pro                &(Feb 2009)\\
     321 &Perugia HARD &: More ISR, More FSR, Less MPI, Less BR, Less HAD&(Feb 2009)\\
     322 &Perugia SOFT &: Less ISR, Less FSR, More MPI, More BR, More HAD&(Feb 2009)\\
     323 &Perugia 3 &: Alternative to Perugia 0, with different ISR/MPI
                     balance \& different scaling to LHC \& RHIC &(Feb 2009)\\
     324 &Perugia NOCR &: "Perugia" update of NOCR-Pro           &(Feb 2009) \\
     325 &Perugia * &: "Perugia" Tune w. (external) MRSTLO* PDFs &(Feb 2009) \\
     326 &Perugia 6 &: "Perugia" Tune w. (external) CTEQ6L1 PDFs &(Feb 2009) \\
     327 & Perugia 2010 & :  Perugia 0 with more FSR off
     ISR and more $s$ & (Mar 2010) \\
     328 & Perugia K & :  Perugia 2010 with a ``$K$'' factor on 
     $\sigma_{\mrm{MPI}}$  & (Mar 2010) \\
     329 &Pro-pT0  & : Professor pT-ordered tune w.~S0 CR model  &(Feb 2009)\\
     330 &MC09    &  :  ATLAS MC09 tune  with (external) LO* PDFs                         &  (2009)\\
   335 &Pro-pT*  & : Professor Tune with (external) LO* PDFs             &   (Mar 2009)\\
   336 &Pro-pT6  & : Professor Tune with (external) CTEQ6L1 PDFs           &   (Mar 2009)\\
   339 &Pro-pT** & : Professor Tune with (external) LO** PDFs       &   (Mar 2009)\\
\hline
\multicolumn{4}{l}{4th generation : Tunes after LHC 7 TeV data}\\
   340 & AMBT1 & :  ATLAS Min-Bias tune     &  \\
   341 & Z1 & :  Underlying-Event tune based on AMBT1      &  \\
   342 & Z1-Pro & :  As Z1, but with Professor's LEP tune & \\
   343 & Z2 & :  Underlying-Event tune based on AMBT1      &  \\
   344 & Z2-Pro & : As Z1, but with Professor's LEP tune & \\
   350 & Perugia 11 & : Central Perugia 2011 tune (with CTEQ5L) & (Mar
   2011)\\ 
   351 & Perugia 11 radHi & : Using $\alpha_s(\frac12\pT{})$ for ISR
   and FSR & (Mar 2011)\\ 
   352 & Perugia 11 radLo & : Using $\alpha_s(2\pT{})$ for ISR and FSR& (Mar 2011)\\ 
   353 & Perugia 11 mpiHi& : Using $\Lambda_{\mrm{QCD}}=0.26$ also for
   MPI& (Mar 2011)\\ 
   354 & Perugia 11 noCR& : Best try without color reconnections& (Mar 2011)\\ 
   355 & Perugia 11 M & : Using MRST LO** PDFs& (Mar 2011)\\ 
   356 & Perugia 11 C & : Using CTEQ6L1 PDFs& (Mar 2011)\\ 
   357 & Perugia 11 T16 & : \ttt{PARP(90)=0.16} away
   from 7 TeV& (Mar 2011)\\ 
   358 & Perugia 11 T32 & : \ttt{PARP(90)=0.32} away
   from 7 TeV& (Mar 2011)\\ 
   359 & Perugia 11 TeV & : Optimized for Tevatron& (Mar 2011)\\ 
\hline
 \end{tabular}}

\bibliography{perugia6}

\begin{thebibliography}{100}

\bibitem{Skands:2012ts}
P.~Skands,
\newblock (2012), 1207.2389.

\bibitem{Snigirev:2003cq}
A.~M. Snigirev,
\newblock Phys. Rev. {\bf D68}, 114012 (2003), hep-ph/0304172.

\bibitem{Korotkikh:2004bz}
V.~L. Korotkikh and A.~M. Snigirev,
\newblock Phys. Lett. {\bf B594}, 171 (2004), hep-ph/0404155.

\bibitem{Akesson:1986iv}
AFS, T.~Akesson {\em et~al.},
\newblock Z. Phys. {\bf C34}, 163 (1987).

\bibitem{ua1minijets}
UA1, C.-E. Wulz,
\newblock in proceedings of the 22nd Rencontres de Moriond, Les Arcs, France,
  15-21 March 1987.

\bibitem{Alitti:1991rd}
UA2, J.~Alitti {\em et~al.},
\newblock Phys. Lett. {\bf B268}, 145 (1991).

\bibitem{Abe:1993rv}
CDF, F.~Abe {\em et~al.},
\newblock Phys. Rev. {\bf D47}, 4857 (1993).

\bibitem{Abe:1997bp}
CDF, F.~Abe {\em et~al.},
\newblock Phys. Rev. Lett. {\bf 79}, 584 (1997).

\bibitem{Abe:1997xk}
CDF, F.~Abe {\em et~al.},
\newblock Phys. Rev. {\bf D56}, 3811 (1997).

\bibitem{Abazov:2002mr}
D{\O}, V.~M. Abazov {\em et~al.},
\newblock Phys. Rev. {\bf D67}, 052001 (2003), hep-ex/0207046.

\bibitem{Abazov:2009gc}
D{\O}, V.~M. Abazov {\em et~al.},
\newblock (2009), 0912.5104.

\bibitem{Gwenlan:2002st}
ZEUS, C.~Gwenlan {\em et~al.},
\newblock Acta Phys. Polon. {\bf B33}, 3123 (2002).

\bibitem{Sjostrand:2004ef}
T.~Sj{\"o}strand and P.~Z. Skands,
\newblock Eur. Phys. J. {\bf C39}, 129 (2005), hep-ph/0408302.

\bibitem{Sjostrand:2004pf}
T.~Sj{\"o}strand and P.~Z. Skands,
\newblock JHEP {\bf 03}, 053 (2004), hep-ph/0402078.

\bibitem{Sjostrand:2006za}
T.~Sj{\"o}strand, S.~Mrenna, and P.~Skands,
\newblock JHEP {\bf 05}, 026 (2006), hep-ph/0603175.

\bibitem{Sjostrand:2007gs}
T.~Sj{\"o}strand, S.~Mrenna, and P.~Skands,
\newblock Comput. Phys. Commun. {\bf 178}, 852 (2008), 0710.3820.

\bibitem{Corke:2009tk}
R.~Corke and T.~Sj{\"o}strand,
\newblock JHEP {\bf 01}, 035 (2010), 0911.1909.

\bibitem{Azimov:1984np}
Y.~I. Azimov, Y.~L. Dokshitzer, V.~A. Khoze, and S.~I. Troyan,
\newblock Z. Phys. {\bf C27}, 65 (1985).

\bibitem{Koba:1972ng}
Z.~Koba, H.~B. Nielsen, and P.~Olesen,
\newblock Nucl. Phys. {\bf B40}, 317 (1972).

\bibitem{GrosseOetringhaus:2009kz}
J.~F. Grosse-Oetringhaus and K.~Reygers,
\newblock (2009), 0912.0023.

\bibitem{Field:2000dy}
CDF, R.~Field,
\newblock Int. J. Mod. Phys. {\bf A16S1A}, 250 (2001).

\bibitem{Affolder:2001xt}
CDF, A.~A. Affolder {\em et~al.},
\newblock Phys. Rev. {\bf D65}, 092002 (2002).

\bibitem{Field:2002vt}
CDF, R.~D. Field,
\newblock (2002), hep-ph/0201192.

\bibitem{Acosta:2004wqa}
CDF, D.~E. Acosta {\em et~al.},
\newblock Phys. Rev. {\bf D70}, 072002 (2004), hep-ex/0404004.

\bibitem{Field:2005qt}
CDF, R.~Field,
\newblock Acta Phys. Polon. {\bf B36}, 167 (2005).

\bibitem{Kar:2009kc}
CDF, D.~Kar,
\newblock (2009), 0905.2323.

\bibitem{Werner:2008zza}
K.~Werner,
\newblock Nucl. Phys. Proc. Suppl. {\bf 175-176}, 81 (2008).

\bibitem{Apel:2009sv}
KASCADE, W.~D. Apel {\em et~al.},
\newblock J. Phys. {\bf G36}, 035201 (2009), 0901.4650.

\bibitem{Pierog:2009zt}
T.~Pierog and K.~Werner,
\newblock Nucl. Phys. Proc. Suppl. {\bf 196}, 102 (2009), 0905.1198.

\bibitem{Buckley:2009bj}
A.~Buckley, H.~Hoeth, H.~Lacker, H.~Schulz, and J.~E. von Seggern,
\newblock Eur. Phys. J. {\bf C65}, 331 (2010), 0907.2973.

\bibitem{Bacchetta:2010hh}
A.~Bacchetta, H.~Jung, A.~Knutsson, K.~Kutak, and F.~von Samson-Himmelstjerna,
\newblock (2010), 1001.4675.

\bibitem{Giele:2007di}
W.~T. Giele, D.~A. Kosower, and P.~Z. Skands,
\newblock Phys. Rev. {\bf D78}, 014026 (2008), 0707.3652.

\bibitem{Giele:2011cb}
W.~Giele, D.~Kosower, and P.~Skands,
\newblock Phys.Rev. {\bf D84}, 054003 (2011), 1102.2126.

\bibitem{Sandhoff:2005jh}
M.~Sandhoff and P.~Skands,
\newblock presented at Les Houches Workshop on Physics at TeV Colliders, Les
  Houches, France, 2-20 May 2005, in hep-ph/0604120.

\bibitem{Skands:2007zg}
P.~Skands and D.~Wicke,
\newblock Eur. Phys. J. {\bf C52}, 133 (2007), hep-ph/0703081.

\bibitem{Wicke:2008iz}
D.~Wicke and P.~Z. Skands,
\newblock (2008), 0807.3248.

\bibitem{Skands:2007zz}
P.~Z. Skands,
\newblock {Some interesting min-bias distributions for early LHC runs},
\newblock FERMILAB-CONF-07-706-T, in C. Buttar et al., arXiv:0803.0678
  [hep-ph].

\bibitem{Alekhin:2005dx}
S.~Alekhin {\em et~al.},
\newblock (2005), hep-ph/0601012.

\bibitem{Albrow:2006rt}
TeV4LHC QCD Working Group, M.~G. Albrow {\em et~al.},
\newblock (2006), hep-ph/0610012.

\bibitem{Buttar:2008jx}
C.~Buttar {\em et~al.},
\newblock (2008), 0803.0678.

\bibitem{Bartalini:2008zz}
P.~Bartalini {\em et~al.},
\newblock In *Hamburg 2008, Multiparticle dynamics (ISMD08)* 406-411.

\bibitem{updatenotes}
T.~Sj{\"o}strand, S.~Mrenna, and P.~Skands,
\newblock {PYTHIA update notes},
\newblock available from {\\\small
  \texttt{http://projects.hepforge.org/pythia6/}}.

\bibitem{Skands:2009zm}
P.~Z. Skands,
\newblock (2009), 0905.3418.

\bibitem{lhplots}
P.~Skands,
\newblock Peter's pythia plots,
\newblock see
  {\small\\\texttt{http://home.fnal.gov/$\sim$skands/leshouches-plots/}}.

\bibitem{Karneyeu:2013aha}
A.~Karneyeu, L.~Mijovic, S.~Prestel, and P.~Skands,
\newblock Eur.Phys.J. {\bf C74}, 2714 (2014), 1306.3436,
\newblock see {\texttt{http://mcplots.cern.ch}}.

\bibitem{Buckley:2009vk}
A.~Buckley, H.~Hoeth, H.~Lacker, H.~Schulz, and E.~von Seggern,
\newblock (2009), 0906.0075.

\bibitem{Ackerstaff:1998hz}
OPAL, K.~Ackerstaff {\em et~al.},
\newblock Eur. Phys. J. {\bf C7}, 369 (1999), hep-ex/9807004.

\bibitem{Amsler:2008zzb}
Particle Data Group, C.~Amsler {\em et~al.},
\newblock Phys. Lett. {\bf B667}, 1 (2008).

\bibitem{Adams:2006nd}
STAR, J.~Adams {\em et~al.},
\newblock Phys.Lett. {\bf B637}, 161 (2006), nucl-ex/0601033.

\bibitem{Abelev:2006cs}
STAR, B.~I. Abelev {\em et~al.},
\newblock Phys. Rev. {\bf C75}, 064901 (2007), nucl-ex/0607033.

\bibitem{Acosta:2005ix}
CDF, D.~E. Acosta {\em et~al.},
\newblock Phys. Rev. {\bf D71}, 112002 (2005), hep-ex/0505013.

\bibitem{Banfi:2010xy}
A.~Banfi, G.~P. Salam, and G.~Zanderighi,
\newblock (2010), 1001.4082.

\bibitem{Affolder:1999jh}
CDF, A.~A. Affolder {\em et~al.},
\newblock Phys. Rev. Lett. {\bf 84}, 845 (2000), hep-ex/0001021.

\bibitem{:2007nt}
D{\O}, V.~M. Abazov {\em et~al.},
\newblock Phys. Rev. Lett. {\bf 100}, 102002 (2008), 0712.0803.

\bibitem{hesketh}
A.~Buckley {\em et~al.},
\newblock Effect of {QED} {FSR} on measurements of {$Z/\gamma^*$} and {$W$}
  leptonic final states at hadron colliders,
\newblock in Tools and Monte Carlo Working Group: Summary Report, Les Houches,
  France, 2009, arXiv:1003.1643.

\bibitem{Acosta:2001rm}
CDF, D.~E. Acosta {\em et~al.},
\newblock Phys. Rev. {\bf D65}, 072005 (2002).

\bibitem{moggi}
N.~Moggi, M.~Mussini, and F.~Rimondi,
\newblock {CDF Public Note 9936},
\newblock see{\\\tt\small http://www-cdf.fnal.gov/physics/new/qcd/QCD.html}.

\bibitem{Abe:1988yu}
CDF, F.~Abe {\em et~al.},
\newblock Phys. Rev. Lett. {\bf 61}, 1819 (1988).

\bibitem{Aaltonen:2009ne}
CDF, T.~Aaltonen {\em et~al.},
\newblock Phys. Rev. {\bf D79}, 112005 (2009), 0904.1098.

\bibitem{Alexopoulos:1998bi}
T.~Alexopoulos {\em et~al.},
\newblock Phys. Lett. {\bf B435}, 453 (1998).

\bibitem{Alner:1987wb}
UA5, G.~J. Alner {\em et~al.},
\newblock Phys. Rept. {\bf 154}, 247 (1987).

\bibitem{Ansorge:1988kn}
UA5, R.~E. Ansorge {\em et~al.},
\newblock Z. Phys. {\bf C43}, 357 (1989).

\bibitem{Collaboration:2009dt}
ALICE,
\newblock Eur. Phys. J. {\bf C65}, 111 (2010), 0911.5430.

\bibitem{Collaboration:2010xs}
CMS,
\newblock JHEP {\bf 02}, 041 (2010), 1002.0621.

\bibitem{Aad:2010rd}
ATLAS, G.~Aad {\em et~al.},
\newblock Phys. Lett. {\bf B688}, 21 (2010), 1003.3124.

\bibitem{Field:1976ve}
R.~D. Field and R.~P. Feynman,
\newblock Phys. Rev. {\bf D15}, 2590 (1977).

\bibitem{tunea}
R.~D. Field,
\newblock CDF Note 6403, in hep-ph/0201192; further recent talks available from
  webpage
  {\\\small\texttt{http://www.phys.ufl.edu/}$\sim$\texttt{rfield/cdf/}}.

\bibitem{Banfi:2004nk}
A.~Banfi, G.~P. Salam, and G.~Zanderighi,
\newblock JHEP {\bf 08}, 062 (2004), hep-ph/0407287.

\bibitem{D0jets}
{D\O}, V.~M. Abazov {\em et~al.},
\newblock Phys. Rev. Lett. {\bf 94}, 221801 (2005), hep-ex/0409040.

\bibitem{Field:2005sa}
R.~Field and R.~C. Group,
\newblock (2005), hep-ph/0510198.

\bibitem{Field:2005yw}
CDF, R.~Field,
\newblock AIP Conf. Proc. {\bf 828}, 163 (2006).

\bibitem{Aaltonen:2008yn}
CDF, T.~Aaltonen {\em et~al.},
\newblock Phys. Rev. Lett. {\bf 102}, 232002 (2009), 0811.2820.

\bibitem{Cacciari:2009dp}
M.~Cacciari, G.~P. Salam, and S.~Sapeta,
\newblock (2009), 0912.4926.

\bibitem{Aaltonen:2010rm}
CDF, T.~Aaltonen {\em et~al.},
\newblock (2010), 1003.3146.

\bibitem{Kar:2008zza}
D.~Kar,
\newblock {\em {Using Drell-Yan to probe the underlying event in Run II at
  CDF}},
\newblock PhD thesis,
\newblock FERMILAB-THESIS-2008-54.

\bibitem{Carli:2010cg}
T.~Carli, T.~Gehrmann, and S.~Hoeche,
\newblock (2010), 0912.3715.

\bibitem{atlasmc09}
ATLAS,
\newblock {ATLAS Monte Carlo Tunes for MC09},
\newblock ATL-PHYS-PUB-2010-002, 2010.

\bibitem{LundFrag}
B.~Andersson, G.~Gustafson, and B.~S{\"o}derberg,
\newblock Z. Phys. {\bf C20}, 317 (1983).

\bibitem{Bowler}
M.~G. Bowler,
\newblock Z. Phys. {\bf C11}, 169 (1981).

\bibitem{Nagy:2009vg}
Z.~Nagy and D.~E. Soper,
\newblock (2009), 0912.4534.

\bibitem{Catani:1990rr}
S.~Catani, B.~R. Webber, and G.~Marchesini,
\newblock Nucl. Phys. {\bf B349}, 635 (1991).

\bibitem{Corcella:2000bw}
G.~Corcella {\em et~al.},
\newblock JHEP {\bf 01}, 010 (2001), hep-ph/0011363.

\bibitem{Bahr:2008pv}
M.~B{\"a}hr {\em et~al.},
\newblock Eur. Phys. J. {\bf C58}, 639 (2008), 0803.0883.

\bibitem{Sherstnev:2007nd}
A.~Sherstnev and R.~S. Thorne,
\newblock Eur. Phys. J. {\bf C55}, 553 (2008), 0711.2473.

\bibitem{Kasemets:2010bx}
T.~Kasemets,
\newblock (2010), 1002.4376.

\bibitem{Bengtsson:1986hr}
M.~Bengtsson and T.~Sj{\"o}strand,
\newblock Phys. Lett. {\bf B185}, 435 (1987).

\bibitem{Bengtsson:1986et}
M.~Bengtsson and T.~Sj{\"o}strand,
\newblock Nucl. Phys. {\bf B289}, 810 (1987).

\bibitem{Plehn:2005cq}
T.~Plehn, D.~Rainwater, and P.~Z. Skands,
\newblock Phys. Lett. {\bf B645}, 217 (2007), hep-ph/0510144.

\bibitem{Skands:2005bj}
P.~Z. Skands, T.~Plehn, and D.~Rainwater,
\newblock ECONF {\bf C0508141}, ALCPG0417 (2005), hep-ph/0511306.

\bibitem{Corke:2010zj}
R.~Corke and T.~Sj{\"o}strand,
\newblock Eur.Phys.J. {\bf C69}, 1 (2010), 1003.2384.

\bibitem{Alwall:2008qv}
J.~Alwall, S.~de~Visscher, and F.~Maltoni,
\newblock JHEP {\bf 02}, 017 (2009), 0810.5350.

\bibitem{Arleo:2010kw}
F.~Arleo, D.~d'Enterria, and A.~S. Yoon,
\newblock (2010), 1003.2963.

\bibitem{Albino:2010em}
S.~Albino, B.~A. Kniehl, and G.~Kramer,
\newblock (2010), 1003.1854.

\bibitem{Cacciari:2010yd}
M.~Cacciari, G.~P. Salam, and M.~J. Strassler,
\newblock (2010), 1003.3433.

\bibitem{Yoon:2010fa}
A.~S. Yoon, E.~Wenger, and G.~Roland,
\newblock (2010), 1003.5928.

\bibitem{Sjostrand:1987su}
T.~Sj{\"o}strand and M.~van Zijl,
\newblock Phys. Rev. {\bf D36}, 2019 (1987).

\bibitem{Campanelli:2009hc}
M.~Campanelli and J.~W. Monk,
\newblock (2009), 0910.5108.

\bibitem{Alexopoulos:1995ft}
E735, T.~Alexopoulos {\em et~al.},
\newblock Phys. Lett. {\bf B353}, 155 (1995).

\bibitem{Donnachie:1992ny}
A.~Donnachie and P.~V. Landshoff,
\newblock Phys. Lett. {\bf B296}, 227 (1992), hep-ph/9209205.

\bibitem{Lai:1999wy}
CTEQ, H.~L. Lai {\em et~al.},
\newblock Eur. Phys. J. {\bf C12}, 375 (2000), hep-ph/9903282.

\bibitem{Pumplin:2002vw}
J.~Pumplin {\em et~al.},
\newblock JHEP {\bf 07}, 012 (2002), hep-ph/0201195.

\bibitem{Aamodt:2010pp}
ALICE, K.~Aamodt {\em et~al.},
\newblock Eur. Phys. J. {\bf C68}, 345 (2010), 1004.3514.

\bibitem{Collaboration:2010us}
CMS, V.~Khachatryan {\em et~al.},
\newblock Phys.Rev.Lett. {\bf 105}, 022002 (2010), 1005.3299.

\bibitem{atlas:2010ir}
ATLAS, G.~Aad {\em et~al.},
\newblock (2010), 1012.5104.

\bibitem{Aad:2010fh}
ATLAS, G.~Aad {\em et~al.},
\newblock (2010), 1012.0791.

\bibitem{Aamodt:2011zz}
ALICE, K.~Aamodt {\em et~al.},
\newblock Eur.Phys.J. {\bf C71}, 1594 (2011), 1012.3257.

\bibitem{Aamodt:2011zj}
ALICE, K.~Aamodt {\em et~al.},
\newblock (2011), 1101.4110.

\bibitem{Khachatryan:2011tm}
CMS, V.~Khachatryan {\em et~al.},
\newblock (2011), 1102.4282.

\bibitem{Aamodt:2010dx}
ALICE, A.~K. Aamodt {\em et~al.},
\newblock Phys. Rev. Lett. {\bf 105}, 072002 (2010), 1006.5432.

\bibitem{lhcb-inprep}
LHCb,
\newblock in preparation,
\newblock 2011.

\bibitem{Sjostrand:2013sma}
T.~Sj{\"o}strand,
\newblock Phys.Scripta {\bf T158}, 014002 (2013), 1309.6747.

\bibitem{Sjostrand:2013cya}
T.~Sj{\"o}strand,
\newblock (2013), 1310.8073.

\bibitem{Firdous:2013noa}
N.~Firdous and G.~Rudolph,
\newblock EPJ Web Conf. {\bf 60}, 20056 (2013).

\end{thebibliography}

\end{document}